\documentclass[fleqn,usenatbib]{mnras}

\usepackage{newtxtext,newtxmath}
\usepackage[T1]{fontenc}
\usepackage{ae,aecompl,times}

\usepackage{graphbox,graphicx}	% Including figure files
\usepackage{amsmath}	% Advanced maths commands
\usepackage{multirow}
\usepackage[normalem]{ulem}
\usepackage{xcolor}
\usepackage{bm}
\usepackage{lipsum}

\newcommand{\Gpch}{\,h^{-1}{\rm Gpc}}
\newcommand{\Mpch}{\,h^{-1}{\rm Mpc}}
\newcommand{\kms}{{\rm km}\,{\rm s}^{-1}}
\newcommand{\kpch}{\,h^{-1}{\rm kpc}}
\newcommand{\Mpchc}{\,h^{-3}{\rm Mpc}^3}
\newcommand{\Gpchc}{\,h^{-3}{\rm Gpc}^3}
\newcommand{\hMpc}{\,h\,{\rm Mpc}^{-1}}
\newcommand{\hMpcc}{\,h^3{\rm Mpc}^{-3}}
\newcommand{\Msh}{\,h^{-1}{\rm M}_\odot}

\title[MillenniumTNG -- Matter clustering and halo statistics]{The MillenniumTNG Project: High-precision predictions for matter clustering and halo statistics}

\author[C.~Hern\'andez-Aguayo et al.]{%
\parbox{0.9\textwidth}{
C\'esar Hern\'andez-Aguayo$^{1,2}$\thanks{E-mail: 
cesarhdz@mpa-garching.mpg.de (CH-A)},
Volker Springel$^{1}$,
R\"udiger Pakmor$^{1}$,
Monica Barrera$^{1}$,
Fulvio Ferlito$^{1}$,
Simon D. M. White$^{1}$,
Lars Hernquist$^{3}$, 
Boryana Hadzhiyska$^{3,4,5}$,
Ana Maria Delgado$^{3}$, 
Rahul Kannan$^{3}$,
Sownak Bose$^{6}$
and Carlos Frenk$^{6}$
}
\\%
\\%
$^{1}$Max-Planck-Institut f\"ur Astrophysik, Karl-Schwarzschild-Str. 1, D-85748, Garching, Germany\\%
$^{2}$Excellence Cluster ORIGINS, Boltzmannstrasse 2, D-85748 Garching, Germany\\%
$^{3}$Harvard-Smithsonian Center for Astrophysics, 60 Garden St, Cambridge, MA 02138, USA\\%
$^{4}$Miller Institute for Basic Research in Science, University of California, Berkeley, CA, 94720, USA\\%
$^{5}$Physics Division, Lawrence Berkeley National Laboratory, Berkeley, CA 94720, USA\\%
$^{6}$Institute for Computational Cosmology, Department of Physics, Durham University, South Road, Durham, DH1 3LE, UK
}

\date{Accepted XXX. Received YYY; in original form ZZZ}

\pubyear{2022}

\begin{document}
\label{firstpage}
\pagerange{\pageref{firstpage}--\pageref{lastpage}}
\maketitle

\begin{abstract} 
Cosmological inference with large galaxy surveys requires theoretical models that combine precise predictions for large-scale structure with robust and flexible galaxy formation modelling throughout a sufficiently large cosmic volume. Here, we introduce the {\sc MillenniumTNG} (MTNG) project which combines the hydrodynamical galaxy formation model of {\sc IllustrisTNG} with the large volume of the {\sc Millennium} simulation. Our largest hydrodynamic simulation, covering $(500 \Mpch)^3 \simeq (740\, {\rm Mpc})^3$, is complemented by a suite of dark-matter-only simulations with up to $4320^3$ dark matter particles (a mass resolution of $1.32\times 10^8 \Msh$) using the fixed-and-paired technique to reduce large-scale cosmic variance.  The hydro simulation adds $4320^3$ gas cells, achieving a baryonic mass resolution of $2\times 10^7 \Msh$. High time-resolution merger trees and direct lightcone outputs facilitate the construction of a new generation of semi-analytic galaxy formation models that can be calibrated against both the hydro simulation and observation, and then applied to even larger volumes -- MTNG includes a flagship simulation with 1.1 trillion dark matter particles and massive neutrinos in a volume of $(3000\,{\rm Mpc})^3$. In this introductory analysis we carry out convergence tests on basic measures of non-linear clustering such as the matter power spectrum, the halo mass function and halo clustering, and we compare simulation predictions to those from current cosmological emulators. We also use our simulations to study matter and halo statistics, such as  halo bias and clustering at the baryonic acoustic oscillation scale. Finally we measure the impact of baryonic physics on the matter and halo distributions.  
\end{abstract}


%---------------------------------------------------------------
\section{Introduction}
\label{sec:intro}
%---------------------------------------------------------------
The amazing progress in observational cosmology over the last decades has brought many surprises. Perhaps the most stunning is that we live in a Universe where most of the matter is comprised of yet unidentified collisionless dark matter particles, while ordinary baryons produced in the Big Bang make up only a subdominant part \citep{Planck:2018vyg}. Furthermore, in the last 5 billion years or so, a dark energy component has progressively become stronger and begun to overwhelm the matter density, driving an accelerated expansion of the Universe. The real physical nature of dark energy, the identity of the dark matter particles, as well as the mass of the neutrinos which contribute a tiny admixture of hot dark matter, are profound and fundamental open questions in physics.

To make further progress, the firmly established $\Lambda$CDM standard cosmological model will be subjected to precision tests in the coming years that are far more sensitive than anything done thus far. Forthcoming cosmological mega galaxy surveys carried out by space missions such as Euclid \citep{Laureijs:2011gra} and Roman \citep{wfirst}, as well as new powerful telescopes on Earth such as DESI \citep{DESI:2016zmz}, PFS \citep{PFS2014} and Rubin \citep{LSST:2009}, will map out billions of galaxies through extremely large regions of space. They will primarily use various measures of galaxy clustering and weak gravitational lensing to carry out meticulous tests of the cosmological model. The primary goals are to search for deviations of dark energy from a cosmological constant, for non-gaussianities in the primordial fluctuation field, for signatures of a law of gravity different from general relativity, and to measure the mass of the light neutrino flavors.

It has widely been recognised that systematic uncertainties in our ability to compute very accurate theoretical predictions for $\Lambda$CDM, as well as neighbouring cosmological models, could become a limiting factor in making full use of the statistical power of the upcoming data. Simulation predictions need to cover much larger cosmological volumes than customary so far to match the statistical power of the new surveys. They also have to be able to predict the non-linear matter clustering highly accurately, ideally account for the impact of baryonic physics on matter clustering in a reliable fashion, and produce realistic galaxy properties and galaxy clustering signals. In addition, one would like to be able to create such predictions not only for one set of cosmological parameters, but for many model variants in a computationally inexpensive fashion. Reaching this goal is very demanding, and will likely require an innovative combination of ``ground-truth'' simulations based on full-hydrodynamical and $N$-body simulations \citep[see][for a review]{Vogelsberger2020}, approximate but fast simulation techniques \citep[e.g.][]{Feng2016}, rescaling techniques \citep{Angulo2010}, semi-analytic galaxy formation methods \citep[e.g.][SAM]{White1991}, more empirical methods such as  halo occupation distribution modelling \citep[e.g.][HOD]{Berlind2002} and subhalo abundance matching \citep[e.g.][SHAM]{Conroy2006}, plus data-driven approaches such as machine-learning techniques \citep[e.g.][]{Villaescusa-Navarro2021}.

A number of groups have in recent years produced very large simulation models as initial steps to tackle this problem. Among them are the {\sc Millennium-XXL} \citep{Angulo2012}, the Euclid {\sc Flagship} \citep{Euclid:FS1}, the {\sc OuterRim} \citep{Heitmann:2019ytn}, the {\sc AbacusSummit} \citep{Maksimova:2021ynf}, or the {\sc Uchuu} \citep{Ishiyama:2020vao} simulations. These simulations are characterised by their large cosmological volume $(\gtrsim 2^3 \Gpchc)$ and by their large number of resolution elements $(N_p \gtrsim 6000^3)$, making them ideally suited to match the statistical power of the large-scale galaxy surveys. However, most of these simulations do not employ a physically motivated galaxy formation model to produce their mock catalogues. Instead, those mocks are often created by empirical models, such as HOD, which do not take into account the impact of baryonic physics on the internal structure of haloes and the clustering of matter. Moreover, most of these large simulation projects do not provide detailed dark matter (sub)halo merger trees which are needed to create realistic mocks through semi-analytical models of galaxy formation.

In this paper, we introduce a new project along this line of research, which we have named ``{\sc MillenniumTNG}'' based on its close connections to two older simulation projects, the {\sc Millennium} simulation \citep{Springel:2005nw}, and The Next Generation Illustris Simulations \citep[{\sc IllustrisTNG};][]{Pillepich:2017fcc,Nelson:2017cxy,Springel:2017tpz,Marinacci:2017wew,Naiman:2018MNR,Nelson:2019jkf,Pillepich:2019bmb}. Both have managed to substantially advance galaxy formation modelling, Millennium by introducing the first 10 billion particle simulation in a $500\Mpch$ box that was augmented with subhalo merger trees that allowed the construction of detailed semi-analytic galaxy formation models, while IllustrisTNG excelled by combining an accurate moving-mesh hydrodynamical technique with a sophisticated model for galaxy formation physics, as well as the use of a set of different box size, TNG50, TNG100, and TNG300. Still, even TNG300 has a boxsize\footnote{Which is very close to $300\,{\rm Mpc}$ without the conventional $h^{-1}$, hence the name TNG300.} of only $205\Mpch$ -- clearly too small for the required precision on large cosmological scales.

In the {\sc MillenniumTNG} project we push the hydrodynamical modelling of TNG to a volume nearly 15 times larger, reaching the $500\Mpch\,(\simeq 740\,{\rm Mpc})$ on a side that could be done with $N$-body techniques in the Millennium, more than a decade ago. To achieve this we use an unprecedentedly large number of resolution elements for a high-resolution hydrodynamical simulation of galaxy formation, with a mass resolution that is still very close to that of TNG300 and thus sufficient to model the formation of large galaxies with reasonable accuracy. We accompany this simulation with a set of dark matter only simulations, in the same volume and with the same initial phases. While we use the same volume, our best mass resolution is nearly an order of magnitude better than that of the original Millennium simulation. We have also considerably refined subhalo finding and tracking, thereby supporting more accurate semi-analytic modelling. In addition, we carry out these simulations in pairs using a variance suppression technique in order to boost the effective volume even further. Finally, we augment our simulation set with additional $N$-body runs that take massive neutrinos explicitly into account. Here we push also the volume and particle number further, reaching more than a trillion particles, and a volume of $(3\,{\rm Gpc})^3$.
This design of the simulation suite allows us to probe baryonic effects in simulation boxes of 740 Mpc on a side, and the effects of different neutrino masses in boxes of 430 Mpc on a side. While we also have an extremely large simulation with neutrinos that is 3000 Mpc on a side, we presently do not have a run yet that explicitly includes both baryons and massive neutrinos.

Our overarching goal with this simulation set is to better link large-scale structure studies with non-linear galaxy formation simulations. We can achieve this by comparing the full-hydrodynamical simulations with semi-analytic models applied to the dark matter-only simulations, thereby assessing and improving the modelling uncertainty, and then by rolling out the semi-analytic model to larger volumes, as realised, for example, in our neutrino simulations. Furthermore, the simulation set of MTNG provides many opportunities to test the internal consistency of the simulation predictions, in particular through detailed convergence tests, tests of box size dependencies, as well as the ability to assess the impact of baryonic physics and finite neutrino masses on matter and galaxy clustering.

This paper is one of a set of introductory papers we have prepared for the {\sc MillenniumTNG} project. In the present work, we introduce the technical aspects of the simulations and present a high-level analysis of the matter and halo statistics. \citet{Pakmor2022} describe in  detail the full-physics MTNG simulation and give a first impression of cluster cosmology with MTNG. \citet{Barrera2022} introduce an updated version of the {\sc L-Galaxies} semi-analytic model of galaxy formation \citep{Henriques:2014sga} and its application to the MTNG lightcones. \citet{Ferlito2022} analyse weak-lensing convergence maps from both DM-only and full-physics runs, while \citet{Bose2022} present a galaxy clustering study based on colour-selected (blue and red) galaxy samples. \citet{Hadzhiyska2022a,Hadzhiyska2022b} present an improved halo occupation model (refining the one-halo and two-halo terms) of luminous-red and emission-line galaxies using the MTNG simulations, and \citet{Delgado2022} study intrinsic alignments of galaxy shapes and compare predictions between the dark matter only and full-physics simulations. \citet{Kannan2022} investigate properties of very high redshift galaxies ($z>8$) in the MTNG full-hydrodynamical run. Finally, \citet{Contreras2022} use the MTNG simulations and SAM catalogues to infer cosmological parameters of SDSS-like samples. These introductory papers cover a range of interesting astrophysical and cosmological topics that can be addressed with the simulations, however they are far from exhaustive. %We therefore plan to make the data fully publicly available for general use by the scientific community in due time, following the successful and productive examples of the {\sc Millennium} \citep{Lemson2006} and {\sc IllustrisTNG} \citep{Nelson2019} projects.

The paper is organised as follows. In Section~\ref{sec:mtng}, we describe the technical aspects and specifications of the MTNG simulations, while in Section~\ref{sec:products} we introduce the various data sets produced by the different simulations. In Section~\ref{sec:matter_halo_stats}, we present results for the foundational matter and halo statistics, while in Section~\ref{sec:matter_halo_Pk} we extend this to an analysis of large-scale (baryonic acoustic oscillations scale) matter and halo clustering, as well as scale-dependent halo bias. In Section~\ref{sec:baryonic_impact}, we briefly discuss the baryonic impact on basic matter and halo statistics and make a comparison with the results from the {\sc IllustrisTNG} simulations. Finally, in Section~\ref{sec:conc} we summarise the results and present our conclusions.

%---------------------------------------------------------------
\section{The Millennium-TNG simulations}
\label{sec:mtng}
%---------------------------------------------------------------
%------------Table----------------
\begin{table*}
\centering
\begin{tabular}{ccccccccccc}
\hline
Type    & Run name        & Series & Box size  & $N_{\rm cdm}$ & $N_{\rm gas}$ & $N_{\nu}$ & \multicolumn{1}{c}{$m_{\rm cdm}$} & $m_{\rm gas/\nu}$   & $\sum\,m_{\nu}$  & $\epsilon_{\rm cdm}$ \\
        &                &        & $[\Mpch]$ &            &  &               & \multicolumn{1}{c}{$[\Msh]$}     & $[\Msh]$   &   $[{\rm eV}]$    & $[\kpch]$  \\ \hline
DM only & MTNG740-DM(-1) & A/B    & 500       & $4320^3$   & $-$ &      $-$         & $1.32 \times 10^8$               &        $-$         & $-$ & 2.5         \\
        & MTNG740-DM-2 & A/B    & 500       & $2160^3$   & $-$ &        $-$      & $1.06\times 10^9$                &       $-$          & $-$  & 5         \\
        & MTNG740-DM-3 & A/B    & 500       & $1080^3$   & $-$ &      $-$         & $8.50\times 10^9$                &        $-$         & $-$ & 10         \\
        & MTNG740-DM-4  & A/B    & 500       & $540^4$    & $-$ &      $-$         & $6.80\times 10^{10}$             &        $-$         & $-$ & 20         \\
        & MTNG740-DM-5  & A/B    & 500       & $270^3$    & $-$ &       $-$        & $5.44\times 10^{11}$             &        $-$         & $-$ & 40        \\ 
        & MTNG185-DM & A    & 125       & $1080^3$   & $-$ &      $-$         & $1.32 \times 10^8$               &       $-$          & $-$ & 2.5         \\ \hline
 Hydro   & MTNG740 & A      & 500       & $4320^3$   &   $4320^3$ &  $-$   & $1.12 \times 10^8$                & $2.00 \times 10^7$ & $-$ & 2.5         \\
        & MTNG185 & A      & 125       & $1080^3$   &   $1080^3$ &  $-$   & $1.12 \times 10^8$                & $2.00 \times 10^7$ & $-$ & 2.5       \\ \hline
Neutrinos  & MTNG3000-DM-0.1$\nu$ & A & 2040       & $10240^3$  & $-$ &   $2560^3$            & $6.66 \times 10^8$                &  $3.26 \times 10^8$    & 0.1 & 4         \\
        &  MTNG1500-DM-0.1$\nu$ & A      & 1020       & $5120^3$  & $-$ &  $1280^3$    & $6.66 \times 10^8$                & $3.26 \times 10^8$  & 0.1 & 4       \\ 
        & MTNG630-DM-0.3$\nu$ & A/B      & 430       & $2160^3$  & $-$  &  $540^3$ & $6.54 \times 10^8$                & $9.76 \times 10^8$ & 0.3 & 4       \\  
        &  MTNG630-DM-0.1$\nu$ & A/B      & 430       & $2160^3$  & $-$  &  $540^3$ & $6.66 \times 10^8$                & $3.26 \times 10^8$ & 0.1 & 4       \\  
        &  MTNG630-DM-0.0$\nu$ & A/B      & 430       & $2160^3$  & $-$  &  $540^3$ & $6.66 \times 10^8$                & $-$ & 0.0 & 4       \\ \hline
\end{tabular}
\caption{Specifications of the simulations of the {\sc MillenniumTNG} project introduced in this paper. The fiducial MTNG runs cover a volume of $500^3\Mpchc \simeq 740^3\,{\rm Mpc}^3$ with resolution elements varying from $4320^3$ (highest, level 1) to $270^3$ (lowest, level 5), spaced by a factor of 8. Our naming convention uses the tag ``MTNG'' followed directly by the box size in Mpc, and optionally the identifier ``DM'' for dark matter-only runs followed by the resolution level, in analogy to the convention of {\sc IllustrisTNG}. Where needed for clarity, we append the letters ``A'' or ``B'' to distinguish the two different realisations run. The full hydrodynamical runs match the resolution of the MTNG740-DM simulations in two different volumes, $500^3\Mpchc$ and $125^3\Mpchc$. In addition, we report the mass of dark matter particles and the initial mass of gaseous cells used in the full-physics runs, as well as the gravitational softening length. The last block of rows shows the specifications of the neutrino runs with box sizes of $2040\Mpch$, $1020\Mpch$ and $430\Mpch$, with neutrinos of mass $\Sigma\,m_{\nu} = 100\,{\rm meV}$ (available for all box sizes), and $\Sigma\,m_{\nu} = 300\,{\rm meV}$ and $\Sigma\,m_{\nu} = 0\,{\rm meV}$ (for MTNG630 runs only).}
\label{tab:sims}
\end{table*}

The {\sc MillenniumTNG} (MTNG) simulations combine large dark matter-only and full-physics hydrodynamical computations with the goal to link predictions for the evolution of large-scale structure to non-linear galaxy formation, while at the same time offering sufficient volume to allow accurate cosmological inferences. In this section, we give an overview of technical specifications of the simulation set. 

The dark matter-only simulations of the project were run with a slightly customised version of the modern {\sc Gadget-4} code \citep{Springel:2020plp}. Our code extensions relative to the public version are primarily concerned with the inclusion of relativistic matter-energy components, such as massive neutrinos and a photon background. The hydrodynamical simulations have been carried out with the {\sc Arepo} code \citep{Springel:2009aa, Pakmor2016, Weinberger:2019tbd} instead, which has been augmented by us with the group finding and lightcone outputting routines of the {\sc Gadget-4} code, and has furthermore been substantially modified to yield better memory efficiency and improved scalability when using a very large number of processor cores.

The dark matter simulations consist of one series of runs which all use the same volume, i.e. $(500 \Mpch)^3 \simeq (740\, {\rm Mpc})^3$, but which vary  the number of DM particles systematically from $270^2$ to $4320^3$, spaced by a factor of 8 in mass resolution. The highest-resolution run (abbreviated MTNG740-DM in the following) improves the original {\sc Millennium} mass resolution thus by about a factor of 8, which translates to a DM particle mass of $1.32\times 10^8 \Msh$. Actually, the exact value of the ratio of the particle masses is not precisely 8 because the cosmological parameters we use have changed compared to constraints at the time of the Millennium simulation.

We employ the variance suppression technique of \cite{Angulo:2016hjd} and run two realisations of each of the DM setups, with mode amplitudes set in each case to the square root of the power spectrum, and with the second realisation having phases that are mirrored relative to those of the first one. We refer to these runs as the  A- and B-series of otherwise identical simulations. With such a pair of simulations we can boost the statistical precision of the dark matter only simulations by factors of $\sim 30-40$, without biasing any of the results \citep{Chuang:2018ega}. 

These simulation are complemented by full-physics hydrodynamical simulations with {\sc Arepo} using the {\sc IllustrisTNG} galaxy formation model \citep{Weinberger:2017MNRAS,Pillepich:2017jle}, employing two different box sizes: $L=500\Mpch = 738.1\,{\rm Mpc} \simeq 740\,{\rm Mpc}$ and $L=125\Mpch = 184.5\,{\rm Mpc} \simeq 185\,{\rm Mpc}$. The combined number of dark matter particles and gaseous cells for each box are $2\times4320^3$ and $2\times1080^3$, respectively. These simulations match the mass resolution of the MTNG740 dark matter-only run, i.e., the corresponding dark matter and baryonic mass resolutions are $1.12\times 10^8\Msh$ and $2\times 10^7\Msh$. This lies intermediate to the mass resolution of the pairs TNG300/TNG300-2 and TNG100-2/TNG100-3, respectively, while the volume of MTNG740 is enlarged by a factor of 14.5 relative to TNG300. All the galaxy formation physics parameters of the MTNG runs are kept exactly the same as in the original {\sc IllustrisTNG} simulations, thereby allowing a direct assessment of numerical convergence by comparing galaxy properties to the TNG100 and even TNG50 series, and where adequate, allowing a Richardson extrapolating to a fiducial infinite resolution. There are however two significant changes to the physics model: Due to severe memory pressure in fitting the largest simulations onto the supercomputer available to us, we were forced to disable magnetic fields and to simplify the tracking of metallicity in the MTNG hydro simulations. This has, however, only a rather minor influence on galaxy properties.  We refer to \citet{Pakmor2022} for a full description and a first in-depth analysis of the MTNG hydrodynamical simulations, and a comparison of its galaxy properties to TNG.

We adopt the cosmological parameters given by \cite{Ade:2015xua}: $\Omega_{\rm m} = \Omega_{\rm cdm} + \Omega_{\rm b} = 0.3089$, $\Omega_{\rm b} = 0.0486$, $\Omega_\Lambda = 0.6911$, $h = 0.6774$, $\sigma_8 = 0.8159$ and $n_s=0.9667$ for this set of simulations, which have been used previously by {\sc IllustrisTNG}, such that the comparison to TNG is unaffected by any modification in cosmological parameters. The initial conditions were generated at $z=63$ with second-order Lagrangian perturbation theory based on a new version of the {\sc NgenIC} algorithm implemented into {\sc Gadget-4}, using the same linear theory power spectrum as used in the original {\sc IllustrisTNG} simulations.

The high computational cost of the hydrodynamic simulation makes running a second realisation prohibitive, so we are here content with constraining only the mode amplitudes, which already gives a good fraction of the benefit of the variance suppression technique, in particular for simple second-order statistics. The hydro simulation is matched to the ``A'' version of the corresponding dark matter only simulations. Note that the paired DM simulations in any case allow us to judge how important a second realisation with mirrored phases is for this technique.

While our MTNG simulations constitute a transformative numerical model for studying galaxy formation on the largest scales, it is not yet fully sufficient to address the cosmological science questions that require multi-Gpc$^3$ simulation volumes. To address this latter need, we boost the scope of our results with an extremely large $N$-body simulation, carried out with a two times better mass resolution than the original {\sc Millennium} simulation $(m_p = 6.66\times10^8\Msh)$ but with a much larger box size of $ 2.04\Gpch = 3\,{\rm Gpc}$, plus the addition of modelling of massive neutrinos, considering a sum of the neutrino masses equal to $\Sigma\, m_\nu = 0.1\, {\rm eV}$. The dark matter particle number used in this simulation is $10240^3$, augmented with a further $2560^3$ simulation particles used to represent the neutrinos. For the latter, we implemented a variant of the $\delta f$ method proposed by \citet{Elbers:2020lbn} in the {\sc Gadget-4} code, which will be described in detail in \citet{paperII}.

The high computational cost of this run also precludes carrying out a further B version for the moment. However, we have done this for a set of three paired simulations of the same mass resolution but much smaller volume, and in which we have changed the neutrino mass, from $\Sigma\, m_\nu = 0.3\, {\rm eV}$, over $\Sigma\, m_\nu = 0.1\, {\rm eV}$, to $\Sigma\, m_\nu = 0$. For all of the neutrino simulations, we have also updated the cosmological model used to the newest DES-Y3 constraints \citep{DES:2021wwk}. 
Note that when we change the neutrino mass, we keep the total matter density at $z=0$ constant. Also, we include photons and relativistic degrees of freedom in the background evolution, requiring changes in the {\sc Gadget-4} code, and a more sophisticated approach to set the initial conditions. Note, in particular, that in this case there is no closed form integration formula any more for computing the linear growth factor.

The basic parameters and naming conventions of the MTNG simulations are listed in Table~\ref{tab:sims}. Analogous to {\sc IllustrisTNG} and for ease of comparison, we refer to our simulations with a name that is composed of the identifier ``MTNG'' followed directly by the box size in Mpc, slightly rounded where appropriate. Dark matter-only simulations additionally carry a designator ``DM'', and simulations with neutrinos are labelled with the summed neutrino mass and a designator ``$\nu$''. When different mass resolutions are available, they are distinguished at the end with a numerical identifier encoding the resolution level, with level ``1'' denoting the best available mass resolution. When we need to explicitly distinguish the two different realisations, we append the letters ``A'' or ``B'' to the name. The simulations of the {\sc MillenniumTNG} project have been carried out for the most part on the SuperMUC-NG supercomputer at the Leibniz Computing Center. The large volume neutrino run (MTNG3000-DM-0.1$\nu$) has been done on the Cosma8\footnote{\url{https://www.dur.ac.uk/icc/cosma/cosma8}} machine at Durham University, while a few of the smaller simulations have been carried out on machines operated the Max Planck Computing and Data Facility (MPCDF)\footnote{\url{https://www.mpcdf.mpg.de/services/supercomputing}}.

%--------- Figure --------------
\begin{figure*}
  \centering     % \resizebox{18cm}{!}{\includegraphics{fig/composite_pie_darkmatter.png}}\vspace*{-0.05cm}\\%
%\hspace*{-0.235cm}\resizebox{18.44cm}{!}{\includegraphics{fig/pie_axis.pdf}}\vspace*{-0.5cm}\\%
\includegraphics[width=1\textwidth]{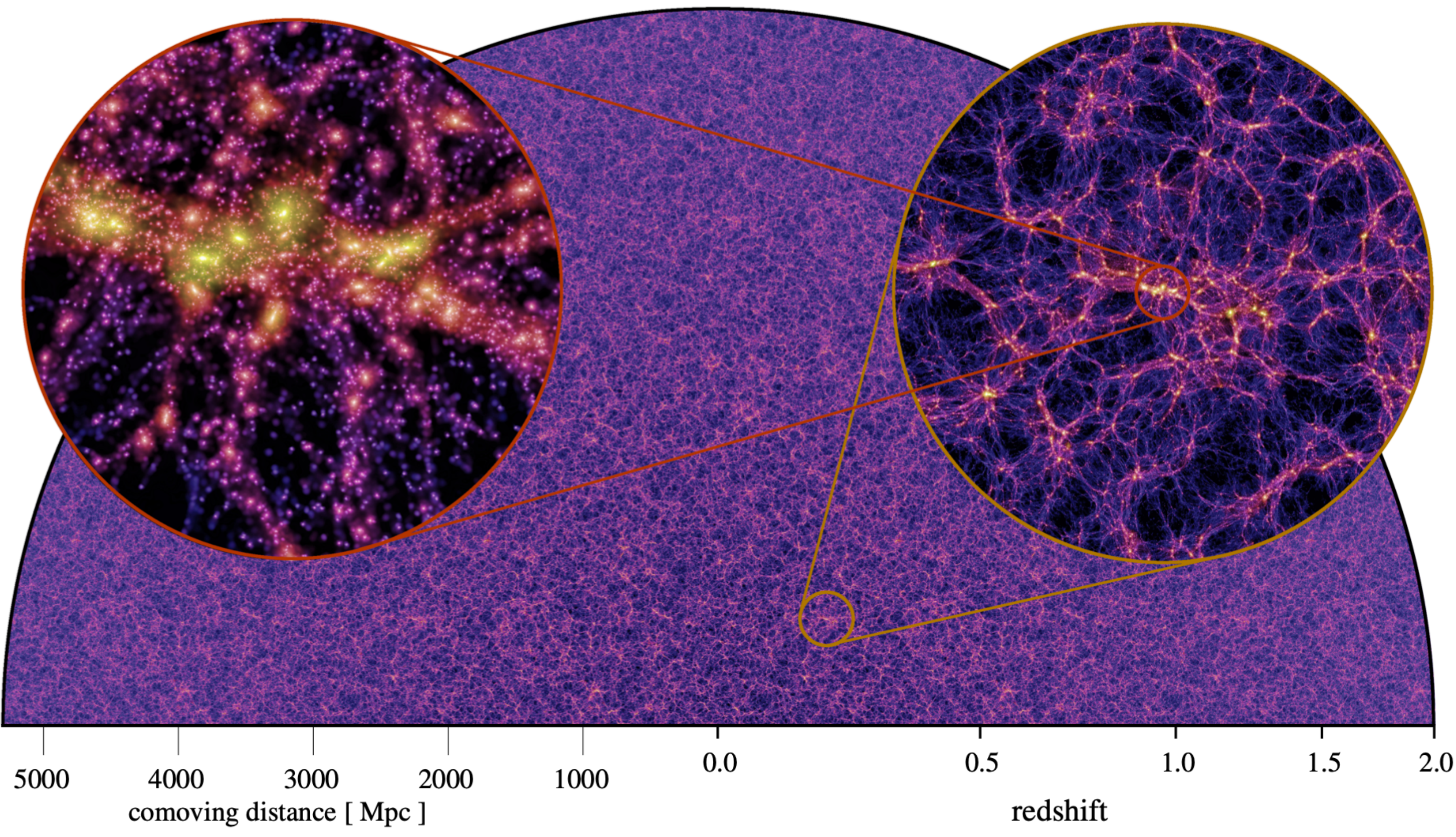}
\caption{Projected dark matter density in one of the lightcones of the MTNG740 dark matter-only simulations (here the MTNG740-DM-2-A simulation was selected, for definiteness). The observer at the present time ($z=0$) is located at the bottom centre of the figure, while the lightcone extents out to $z=2$. The dark matter particles are displayed using comoving coordinates, and the comoving thickness of the projected inclined slice is $15\,h^{-1}{\rm Mpc} = 22.14\,{\rm Mpc}$, independent of distance.  We also show two nested  zoom-in regions (spherical insets) that illustrate the well-known spider web-like large-scale structure and a dense region around a forming galaxy cluster. These zoomed inset images have diameters of
$400\,{\rm Mpc}$ and $40\,{\rm Mpc}$, respectively.}
\label{fig:lightcone}
\end{figure*}

%---------------------------------------------------------------
\section{Data products}
\label{sec:products}
%---------------------------------------------------------------
For the {\sc MillenniumTNG} project, we adopted a new outputting strategy designed to allow the construction of halo merger trees at high time resolution without the need to produce a large number of time\-slices (traditionally called snapshots) on disk. This meant that much of the necessary halo finding and halo linking across time had to move from postprocessing to an on-the-fly treatment. Making this possible has been one of the main new features of the {\sc Gadget-4} code, whose routines we use for this purpose.

Furthermore, we wanted to simplify the comparison with deep wide-angle observational data by making use of a lightcone outputting routine that detects crossings of simulation particles or cell trajectories with the past backwards lightcone of a fiducial observer position. We realise multiple lightcones of various depth and angular extent, and use them, in particular, to also create projected density shells for studies of weak gravitational lensing. Below, we give more details about the various simulation outputs produced by the MTNG simulations.

%---------------------------------------------------------------
\subsection{Output times}
%---------------------------------------------------------------
To produce determinations of group and subhalo catalogues, measurements of matter power spectra, and for (occasionally) storing snapshot particle information, the MTNG simulations define 265 output times between redshift $z=30$ and the present time, $z=0$. The output spacing is constant in the logarithm of the scale factor, with three regions in which the spacing differs by a factor of 2 each, such that the resulting output frequency is finer at low redshift than at very high redshift, as follows:

\begin{itemize}
\item  $\Delta {\rm log}(a)= 0.0325\;\;$ for $\;\;10 \le z < 30\;\;$ (32 times). 
\item  $\Delta {\rm log}(a)= 0.0162\;\;$ for $\;\;\;3 \le z < 10\;\;$ (62 times).
\item  $\Delta {\rm log}(a)= 0.0081\;\;$ for $\;\;\;0 \le z < 3\;\;\;$ (171 times).
\end{itemize}

For the neutrino runs, we have reduced the output frequency by a factor of two to arrive at just 133 output times, omitting every second of the output times defined above. This was done in order to save some computation time, and especially disk storage space, and has also been motivated by our finding that the resulting time resolution for the updating of the group catalogue is sufficient for the semi-analytic galaxy formation code. 

%---------------------------------------------------------------
\subsection{Group and subhalo catalogues}
%---------------------------------------------------------------
At all output times defined above, our simulation codes {\sc Gadget-4} and {\sc Arepo}, respectively, have run the Friends-of-Friends (FoF) group finding algorithm \citep{Davis:1985} combined with the {\sc Subfind-HBT} substructure finder to compute group and subhalo catalogues on-the-fly. The corresponding algorithms are described in the {\sc Gadget-4} paper \citep{Springel:2020plp}. Compared to the traditional {\sc Subfind} algorithm, the `hierachical bound tracing' \citep[][HBT]{Han2018} extension yields better tracking of subhaloes especially close to pericentre. For each FoF group, we compute spherical overdensity mass estimates around the particle with the  minimum gravitational potential  which is  taken as centre of the group. 
Here the gravitational potential is computed just for the set of particles making up the group, i.e.~it is not the global gravitational potential, although the latter could in principle be used with the  `boosted potential' method by \citet{Stuecker2021} to define the tidal boundary of a halo. We refer to \citet{Springel:2020plp} for a description of the technical aspects of the calculation of the gravitational potential in the {\sc Gadget-4} code.
For each gravitationally bound subhalo, a number of further properties are also computed, among them the maximum circular velocity, a measure of the environmental density, shape information, and further properties.

The particle IDs that make up each group or subhalo are not explicitly stored. Instead, the simulation codes remember this membership information until the next group catalogue is computed, and then use it to determine descendant and progenitor pointers that link two subsequent group catalogues. These pointers are stored on disk alongside the group and subhalo catalogues themselves, and constitute the basic information needed to construct detailed merger trees in a postprocessing step detailed below.

%---------------------------------------------------------------
\subsection{Snapshots}
%---------------------------------------------------------------
The above procedure obviates the need to store full snapshot data for all output times desired in a high time-resolution merger tree. In fact, none are in principle needed. However, for certain analysis  (including unforeseen ones) full time slices are still needed, but then only a comparatively small number is usually sufficient, making still an order of magnitude  reduction of the data volume possible. We have decided to store only 10 full particle snapshots at redshifts $z = 0$, 0.25, 0.5, 1.0, 1.5, 2.0, 3.0, 4.0, 5.0, and 7.0\footnote{The corresponding snapshot numbers for the main MTNG runs are 264, 237, 214, 179, 151, 129, 094, 080, 069, and 051.}. For the neutrino simulations, we reduced this even further to just 5 snapshot times.

For output times where a corresponding full snapshot is saved, we note that the particle data is stored in an ordered and nested fashion, with the largest FoF halo coming first (with the others following in descending order), with the particles making up each subhalo within a FoF group being stored in descending order of subhalo size as well. Finally, within each subhalo, the particles are ordered by increasing binding energy, so that the most-bound particle comes first in each substructure. This storage scheme thus makes it possible to selectively load the particles making up each group or subhalo, as they are stored consecutively on disk with a starting offset that is known beforehand. Also, this does not require a separate  storage of the IDs that make up a certain group or subhalo.

%--------- Figure --------------
\begin{figure*}
 \centering
\includegraphics[width=0.47\textwidth]{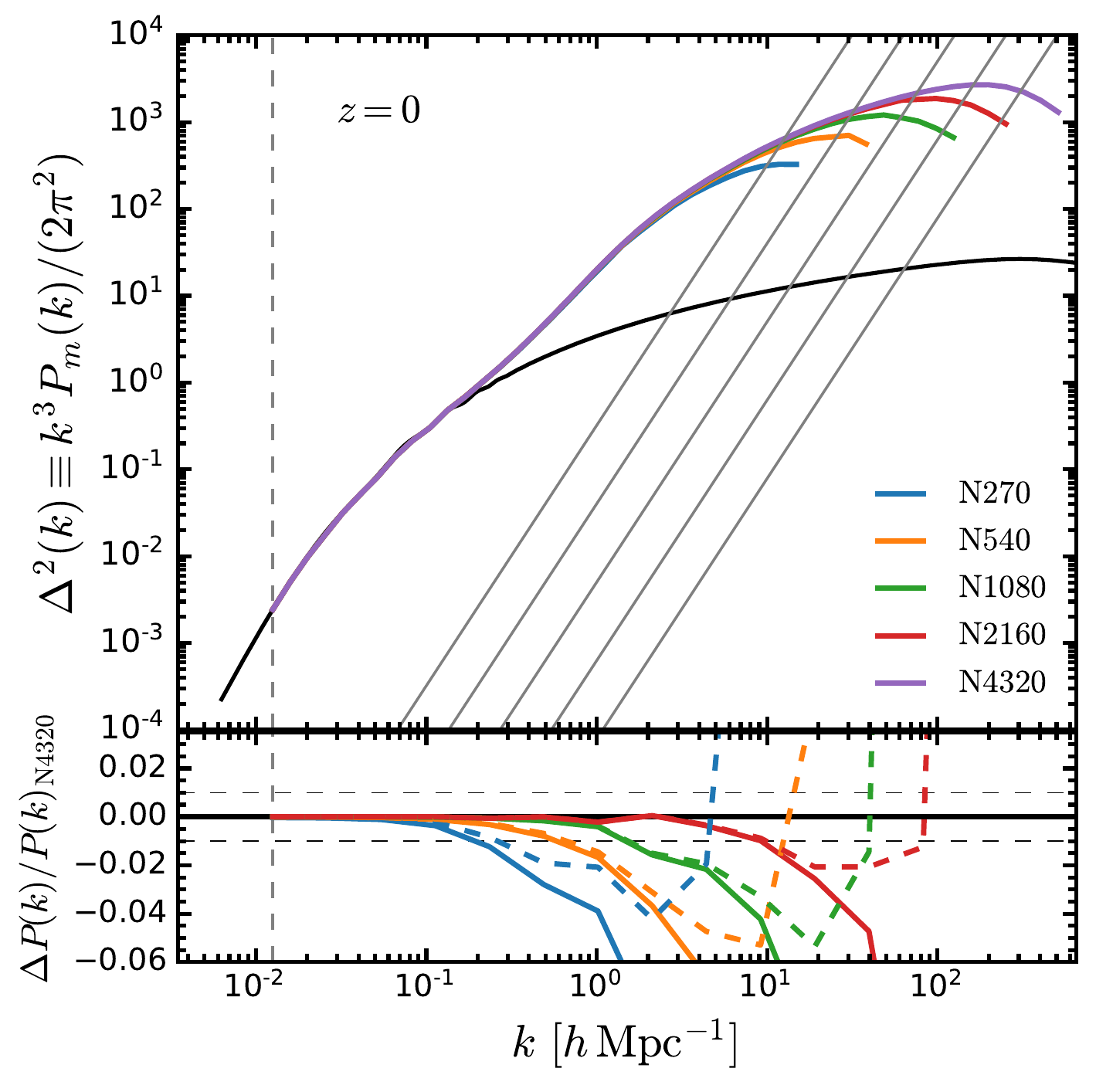}
\includegraphics[width=0.47\textwidth]{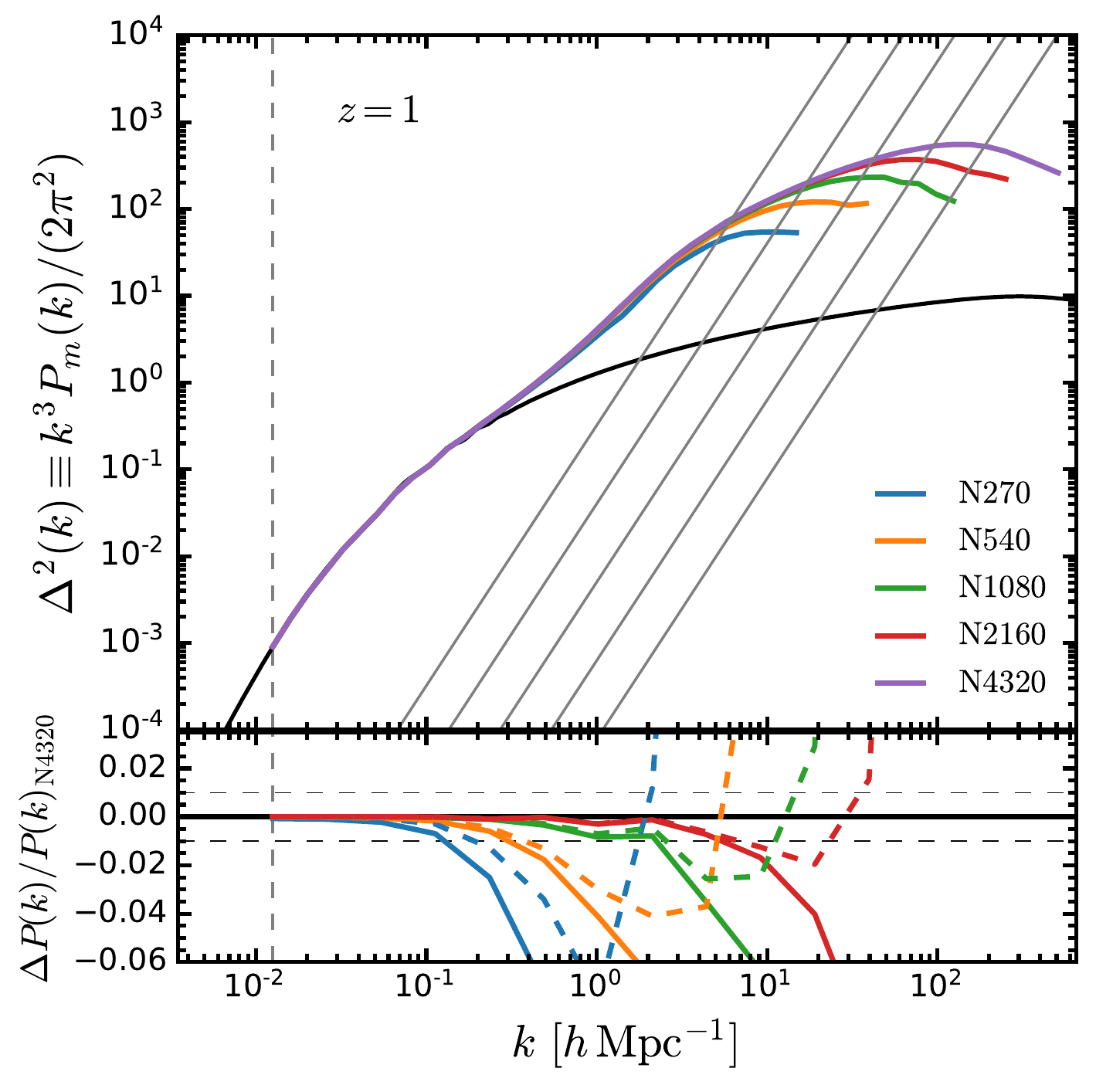}
\caption{Particle mass resolution impact on the non-linear power spectrum of the MTNG740-DM runs at $z=0$ (left panel) and $z=1$ (right panel). In each case we show the average of the A and B realisations that we carried out. Different colours represent the different number of particles, as labelled, corresponding to our resolution levels $1-5$. In both panels, the vertical dashed line indicates the fundamental mode of the box, and the diagonal solid grey lines give the shot-noise contribution for each case. The solid black lines display the linear theory prediction. The lower subpanels show the relative difference with respect to the highest resolution MTNG740-DM runs, and the horizontal dashed lines mark the $1\%$ difference interval. The dashed lines in the lower subpanels correspond to the relative difference of the PS measurements without subtracting the shot-noise contribution.  Note that the shot-noise subtraction is only really adequate for the highly nonlinear regions of the density field (i.e.~inside halos and filaments) whereas for regions that still reflect the sub-Poissonian ‘coldness’ of the initial particle load it is not. Subtracting the shot-noise assuming a Poissonian sampling of the full density field (solid lines) thus represents an over-correction, while not doing a subtraction at all is an under-correction (dashed lines).}
\label{fig:MTNG_Pk_res}
\end{figure*}

For the special application of semi-analytic galaxy formation, we have actually produced an additional set of snapshot files at all output times. These contain only those particles that have been a most-bound particle of a subhalo sometime in the past. These particles can be used to approximately track in semi-analytic models  so-called orphaned galaxies whose dark matter substructures are disrupted by tidal forces before the corresponding galaxies are predicted to have merged with their central galaxies. The fraction of these particles grows monotonically in time and reaches a few percent by $z=0$. This means that even with 265 output times as used here, the cumulative data volume of these special most-bound particle snapshots is still small.

%---------------------------------------------------------------
\subsection{Power spectra measurements}
%---------------------------------------------------------------
Matter power spectra are measured for all defined output times (both for the total matter, and, where available, separately for baryons, dark matter, and neutrinos). Three power spectra measurements are done at each output time, applying the folding technique \citep{Jenkins1998} twice with folding factors of $16$ and $16^2$. This allows the measurements to be combined such that they yield a coverage of the full spatial dynamic range of the simulations, up to $k-$ranges that reach well below the gravitational softening scale. The power spectra are output in a finely binned fashion such that they can be conveniently band-averaged for a new binning, if desired.

%--------- Figure --------------
\begin{figure*}
 \centering
\includegraphics[width=0.47\textwidth]{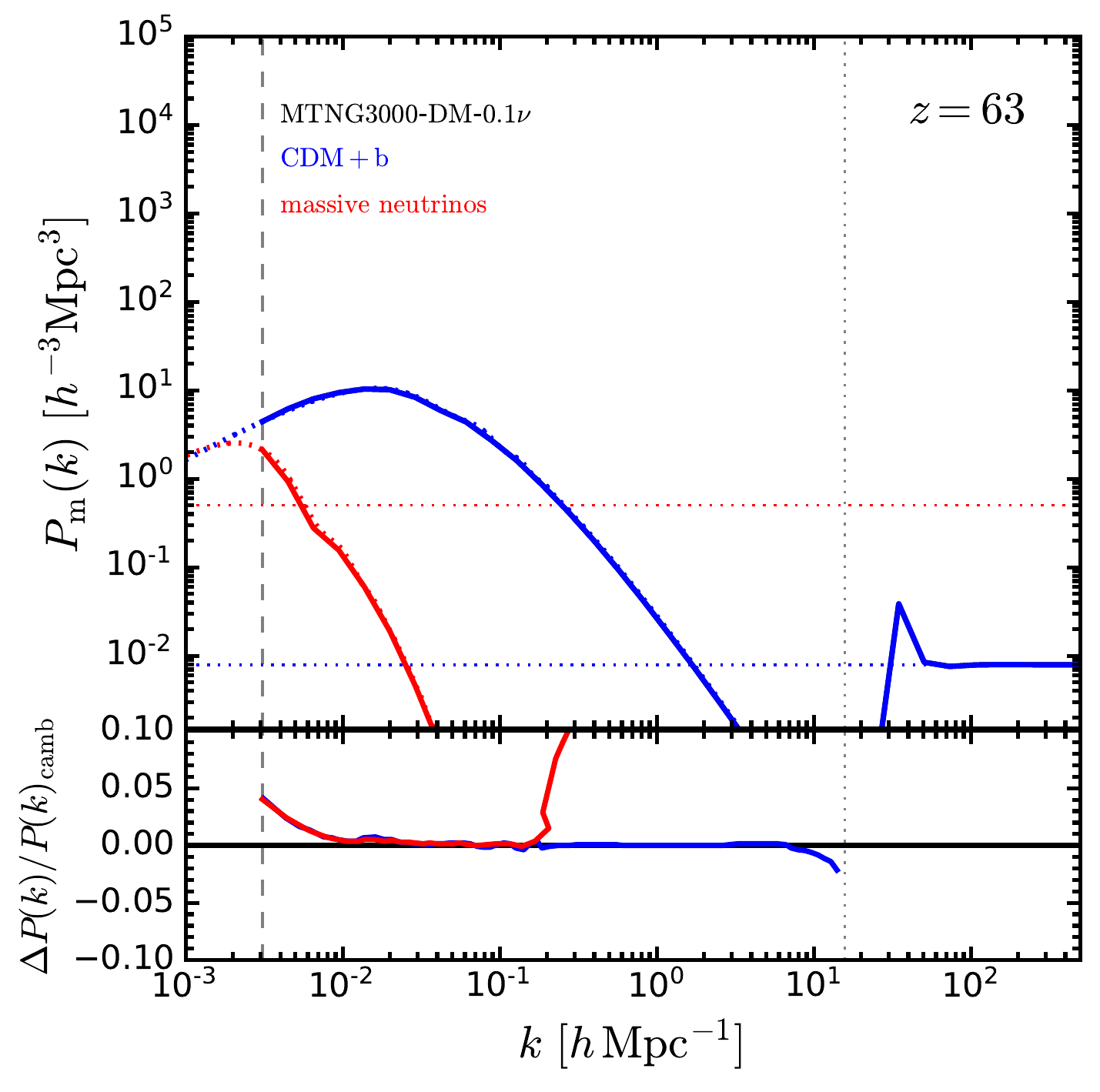}
\includegraphics[width=0.47\textwidth]{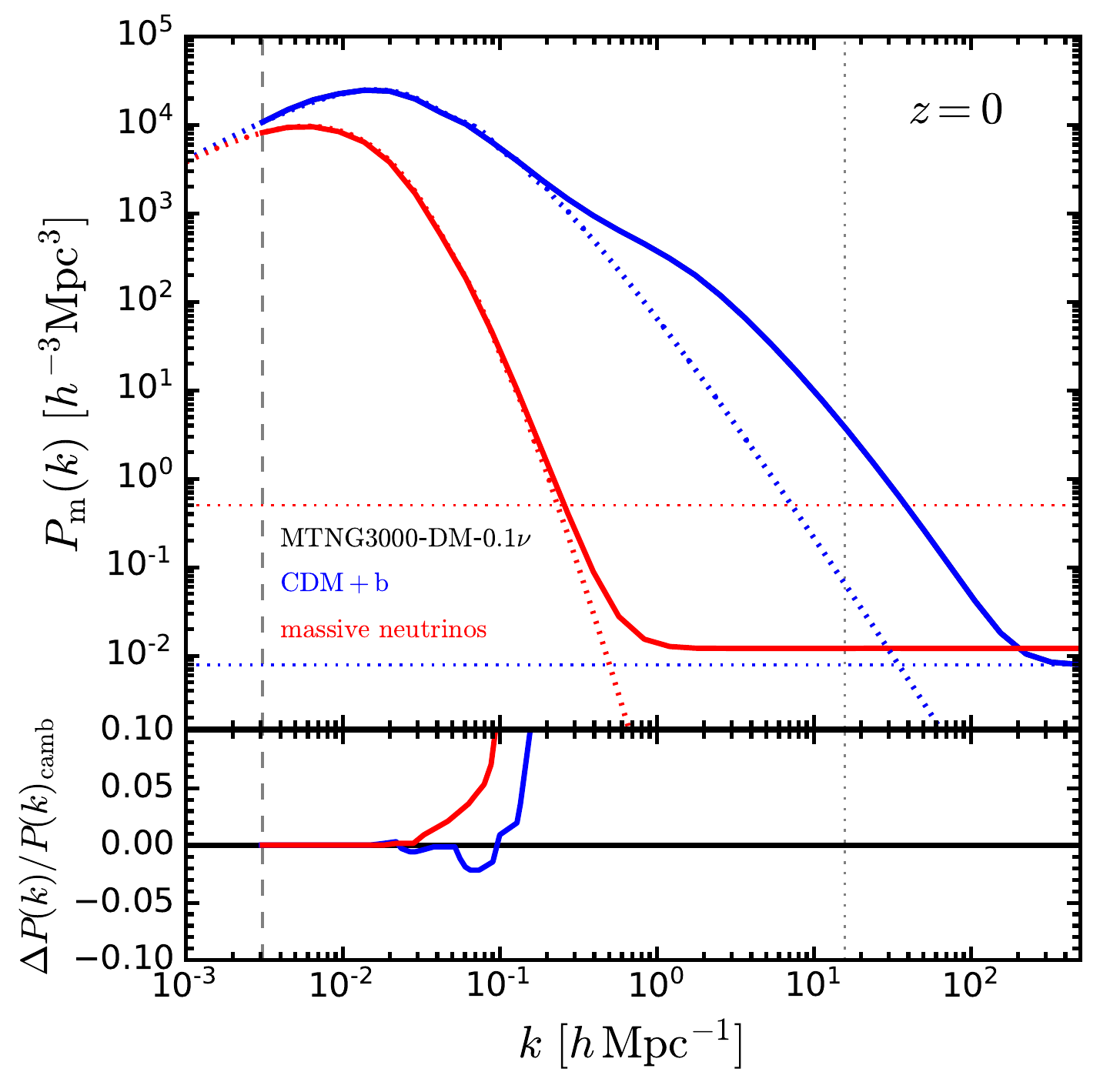}
\caption{Power spectrum measurements for our MTNG3000-DM-0.1$\nu$ run at the initial time (left panel, $z=63$) and the final time (right panel, $z=0$). In both cases we show the measured power spectrum of the cold matter particles (CDM and baryons; solid blue line) and of the neutrino particles (solid red line), and we compare to the expected linear theory power spectrum (thin dotted curves) as computed by {\small CAMB} \citep{Lewis2000} for this cosmology (which includes neutrinos that transition from the relativistic into the non-relativistic regime, as well as a photon background). The dashed vertical line shows the fundamental mode of the box, while the vertical thin dotted line gives the Nyquist frequency of the initial DM particle grid. The dotted horizontal lines gives the nominal shot noise of the two particle sets, $10240^3$ for the DM, and $2560^3$ for the neutrinos. Despite this, the actual shot noise realised by the neutrinos at late times is much lower, thanks to the $\delta f$-simulation technique \citep{Elbers:2020lbn}, and is in fact close to the one of the dark matter at late times. In the bottom panel, we give the relative deviation of the measured power spectrum modes relative to linear theory. Note that in the initial conditions, we deliberately create a small boost in the linear power at the largest scales in order to offset the fact that our code treats super-horizon modes with Newtonian gravity.  With this correction, the simulation-evolved linear power spectrum is fully correct at $z=0$ at large scales by construction, while at higher redshift, there will be a small residual difference. However, at $z=5$, when the outputting of our deepest lightcone starts, the difference to the relativistic result at the fundamental mode of the box has already dropped to just $0.7\%$, and this value shrinks rapidly further towards lower redshift, and in any case also quickly towards higher $k$.
  \label{fig:MTNG3000_Pk}}
\end{figure*}

%---------------------------------------------------------------
\subsection{Particle lightcones}
%---------------------------------------------------------------
For each simulation, we produce a set of different lightcones for a fiducial observer located at the origin of the computational box\footnote{Note that due to the translational symmetry induced by our periodic boundary conditions, this point is not special in any way.}. The periodic simulation box is replicated automatically to the extent necessary to fill the geometry of the specified lightcone. We have created five different lightcones with the following definitions:
\begin{itemize}
    \item {\bf Cone 0:} full-sky particle lightcone between $z = 0 - 0.4$, extending to a comoving distance $\sim 1090\Mpch$.
    \item {\bf Cone 1:} particle lightcone covering one octant on the sky (coordinates $x > 0$, $y > 0$, $z > 0$) for redshifts $z = 0 - 1.5$,  reaching comoving distance $\sim 3050\Mpch$.
    \item {\bf Cone 2:} a pencil beam particle lightcone with a square-shaped footprint of area $(10\deg)^2$ at an oblique angle in the direction $\vec{n}=\{0.26726, 0.53452, 0.80178\}$,  between $z = 0 - 5$, hence extending out to a comoving distance $\sim 5390\Mpch$.
    \item {\bf Cone 3:} a disk-like particle lightcone with a comoving thickness of $15\Mpch$ (tilted against the principal coordinate planes with normal vector $\vec{n}=\{-0.06043, -0.24173, 0.96846 \}$),  over the redshift $z=0-2$, and thus extending to a comoving distance $\sim 3600\Mpch$.
    \item {\bf Cone 4:} a full sky lightcone between $z = 0 - 5$, but only containing `most-bound' particles as described for the partial snapshot, yielding a comoving distance $\sim 5390\Mpch$.    
\end{itemize}

Note that while Cone~1 and 2 redundantly comprise some of the particle data that is contained in Cone~0, they go out to considerably deeper redshift, which is the reason they were added in the first place. The primary purpose of Cone~3 is to allow a two-dimensional visualisation and analysis of structure out to higher redshift than possible by extracting this data from a full-sky lightcone. Finally, Cone~4 can be used in semi-analytic modelling of galaxy formation to improve the orbital treatment of galaxies by allowing a precise determination of when subhaloes or orphans (whose position is both marked by particles which have been, or still are, a most-bound particle of a subhalo) cross the past-backwards lightcone, thus yielding accurate phase-space information for these events. We note in passing that for the neutrino runs, we added a second pencil beam lightcone pointing into a different direction than Cone~2.

As an illustration of the lightcone outputs, in Fig.~\ref{fig:lightcone} we show a cut-out from Cone 3 (the disk-like lightcone). The observer is located at $z=0$ at the bottom centre of the figure, while the lightcone extends out to $z=2$, which is reached at the outer perimeter of the displayed half-sphere. The right circular inset shows a zoomed region of the lightcone, centered on a forming galaxy cluster. This highlights the filaments, knots and voids that comprise the large-scale structure of the Universe. A further zoom by another factor of 10 (left spherical inset with radius $20\,{\rm Mpc}$) displays the distribution of individual dark matter haloes and their embedded substructures in the dense region of the protocluster.

%---------------------------------------------------------------
\subsection{Mass-shell outputs}
%---------------------------------------------------------------
For weak gravitational lensing applications, we additionally create onion-like shells with projections of a fiducial full-sky particle lightcone (which itself is not output to disk due to its prohibitive size) onto a healpix tesselation of the sky. The comoving depth of these shells is $25\Mpch$, and they go out to redshift $z=5$, giving 216 such maps in total. The number of equal-area pixels in each map is $N_{\rm pix} = 12\,N_{\rm side}^2$, where $N_{\rm side}$ is the resolution parameter of the healpix tesselation algorithm. For our highest resolution maps we go up to $N_{\rm side} = 12288$, yielding 1.8 billion pixels and a $0.28$ arcmin angular resolution of these mass maps. Each of the pixels simply contains the total mass of all the particles that fall within the corresponding solid area of the pixel.

%---------------------------------------------------------------
\subsection{Merger trees}
%---------------------------------------------------------------
We build merger trees for all the simulations primarily in order to construct mock galaxy catalogues using semi-analytical models. In principle, the group/subhalo catalogues and the descendant/progenitor information produced on-the-fly during the simulation runs already contain all the data needed for the merger tree. However, to efficiently work with this data (in particular to avoid excessive I/O times to collect it from many different files), it is prudent to rearrange the data such that the subhaloes that are linked together in a single tree are also stored together. This final step in the merger tree construction is carried out in a postprocessing step using the methods implemented in the {\sc Gadget-4} code \citep{Springel:2020plp}.

The result are tree catalogues where each tree is self-contained in the sense that all progenitor and descendant pointers only lead to other subhaloes contained in the same tree. Also, all subhaloes in a common FoF group are always in the same tree. A single tree is thus sufficient for running semi-analytic models of galaxy formation such as {\sc L-Galaxies} \citep{Henriques:2014sga}, which can therefore process the trees of a simulation in an ``embarrasingly parallel'' fashion. As a further convenient data product, {\sc Gadget-4} creates auxiliary files that tell for each subhalo in which tree, and at which place within the tree, the corresponding subhalo can be found. One can then selectively load this tree, if desired, to examine, for example, the history and fate of the chosen subhalo.

%---------------------------------------------------------------
\section{Precision predictions for matter and halo statistics}
\label{sec:matter_halo_stats}
%---------------------------------------------------------------
The {\sc MillenniumTNG} simulations aim to assist ongoing and future galaxy surveys (e.g., DESI, Euclid and PFS) by making accurate predictions of the clustering of matter over cosmic time (and thus the clustering of different types of galaxies). In this section we verify the statistical power of the MTNG simulations by performing a series of convergence tests, by highlighting the advantage of running fixed-and-paired simulations, and by assessing the accuracy of the matter and halo statistics predicted by published emulators. In the following, we will always show the mean of the measurements for the A and B realisations where available, unless stated otherwise.

%---------------------------------------------------------------
\subsection{Numerical convergence assessment of matter and halo statistics}
\label{sec:resolution}
%---------------------------------------------------------------
\subsubsection{Non-linear matter power spectra}
\label{sec:Pkm_res}
%---------------------------------------------------------------
The non-linear matter power spectrum is a key outcome of structure formation and allows us to test the convergence of numerical simulations, and to establish the smallest scales at which accurate results can be obtained \citep[see e.g.][]{coyoteI, Schneider:2015yka, Grove:2021rvc}. Furthermore, large galaxy surveys require accurate predictions of the matter power spectrum in the nonlinear regime, i.e.~up to scales $k \sim 10 \hMpc$ with an accuracy of order 1 per cent \citep{LSST:2009,Laureijs:2011gra,DESI:2016zmz}.

%--------- Figure --------------
\begin{figure*}
 \centering
\includegraphics[width=0.47\textwidth]{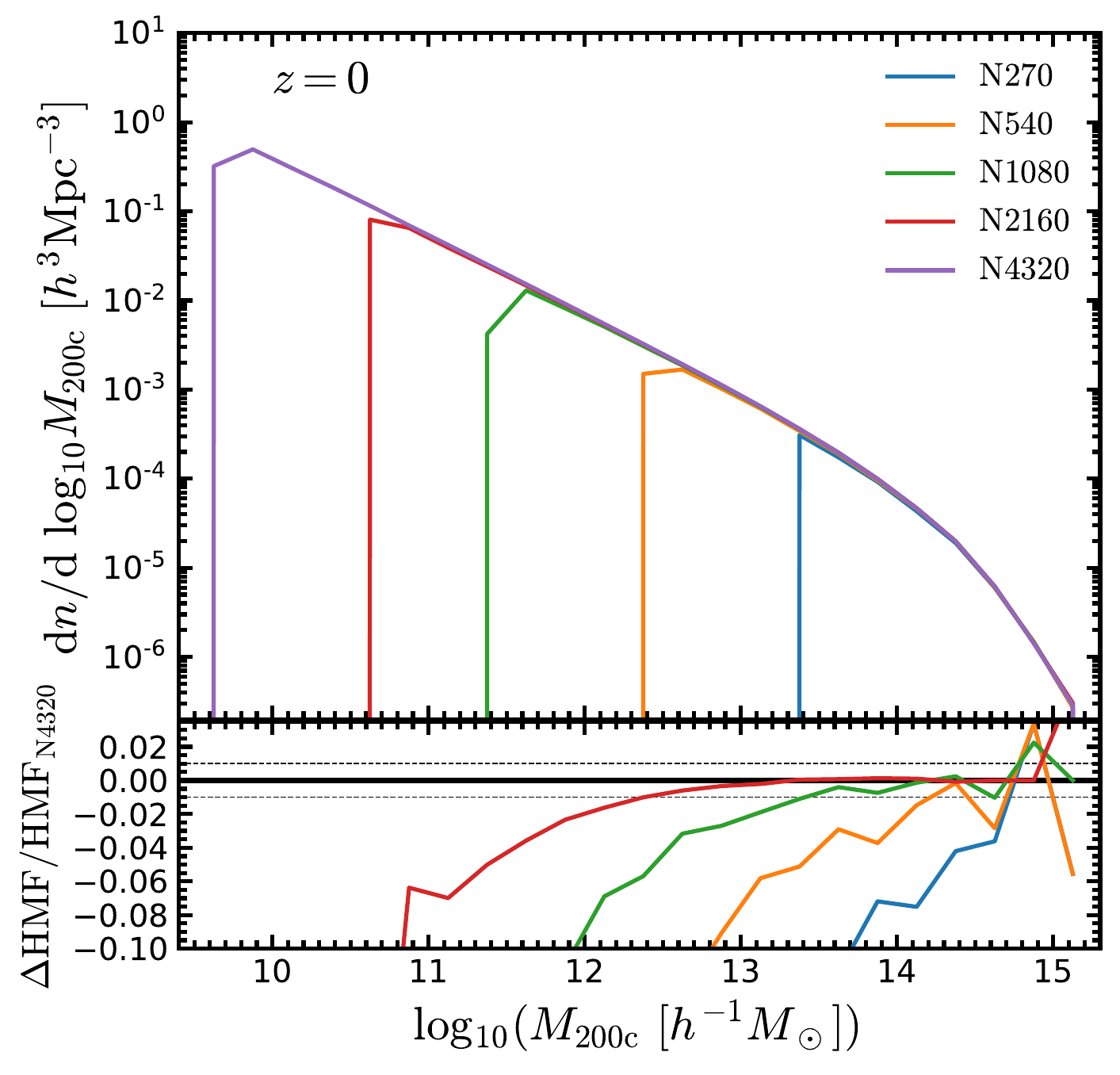}
\includegraphics[width=0.47\textwidth]{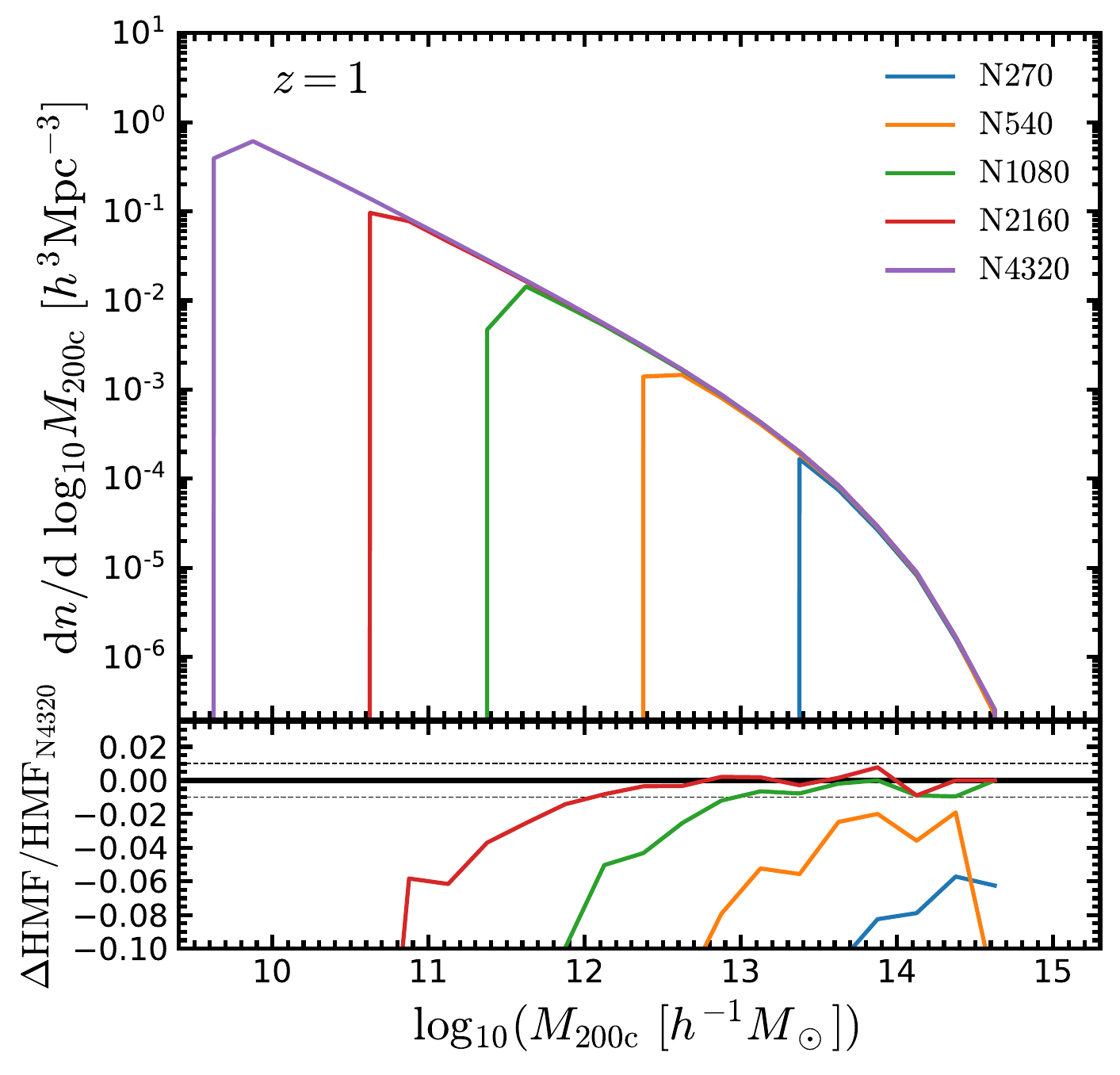}
\caption{Convergence tests of the measured halo mass $(M_{200c})$ function from the MTNG740-DM runs at $z=0$ (left panel) and $z=1$ (right panel). The lower subpanels show the relative difference with respect to the highest resolution MTNG740-DM simulations, with the horizontal dashed lines representing a $1\%$ difference.}
\label{fig:MTNG_HMF_res}
\end{figure*}

%--------- Figure --------------
\begin{figure*}
 \centering
\includegraphics[width=0.47\textwidth]{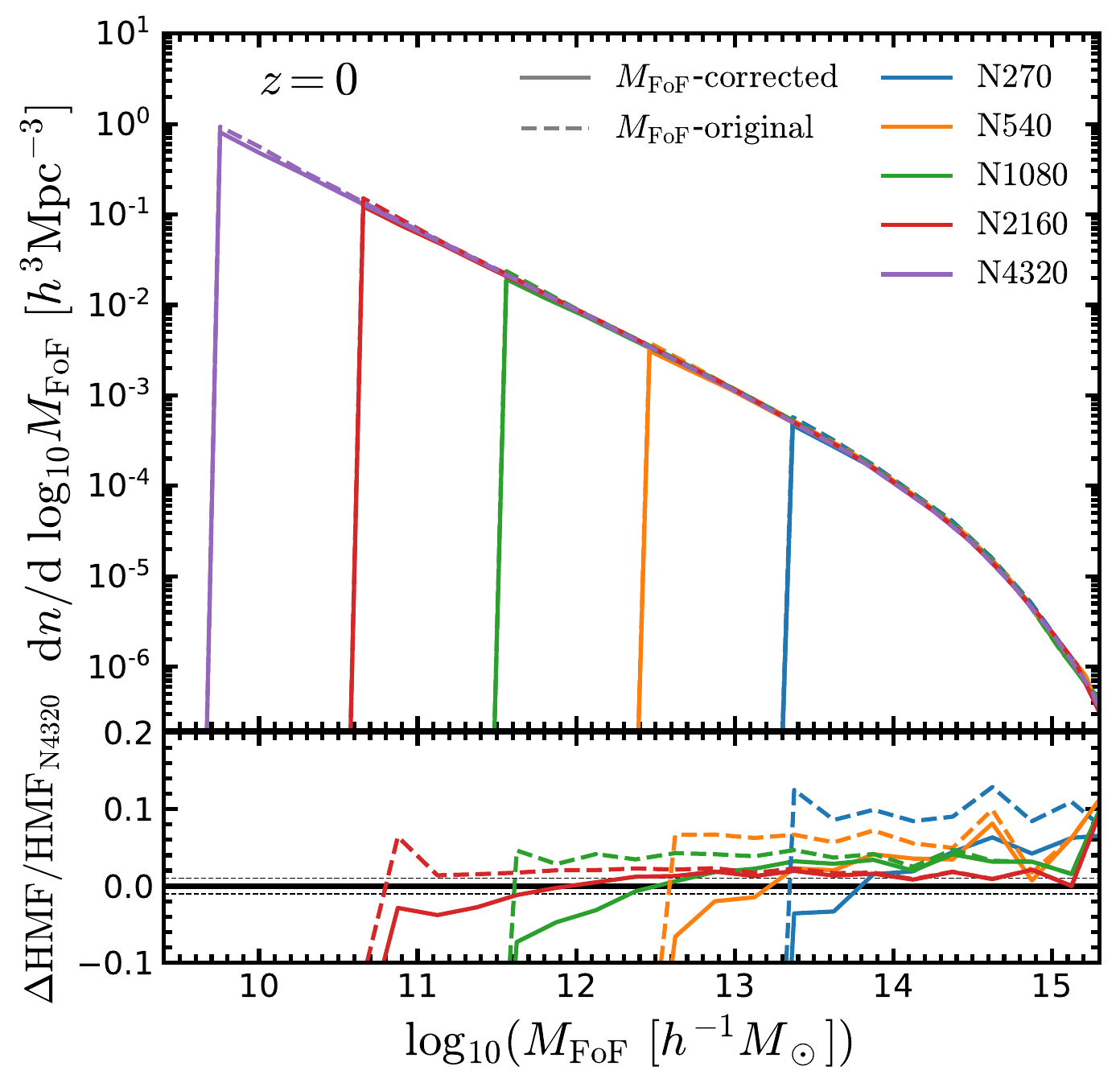}
\includegraphics[width=0.47\textwidth]{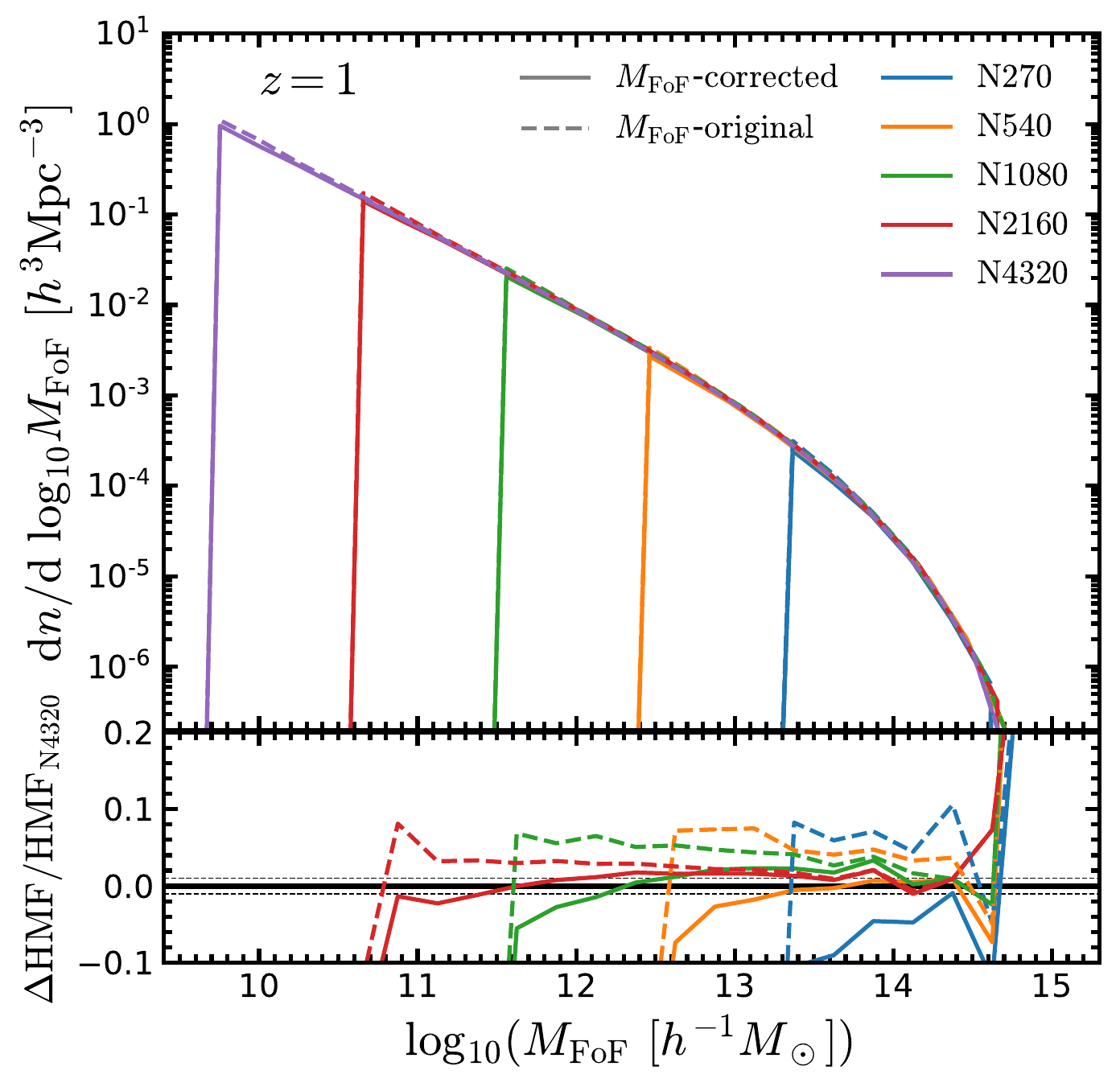}
\caption{Similar to Fig.~\ref{fig:MTNG_HMF_res}, but for the measured halo mass function using the Friends-of-Friends halo mass $(M_{\rm FoF})$ definition. The solid and dashed lines correspond to the measured HMF with and without the $M_{\rm FoF}$ correction given by Eq.~\eqref{eq:Mfof}, respectively.}
\label{fig:MTNG_Mfof_res}
\end{figure*}

As described earlier, we have measured the matter power spectra for our MTNG simulations for all defined output times (both for the total matter, and where available, separately for baryons, dark matter, and neutrinos), and with the folding technique down  to $k-$ranges that reach well below the softening scale, thus covering the full spatial dynamic range of the simulations.

In order to illustrate the convergence of our MTNG simulations, Figure~\ref{fig:MTNG_Pk_res} displays the measured (dimensionless) non-linear matter power spectrum from our MTNG740-DM runs at $z=0$ (left panel) and $z=1$ (right panel). Each coloured curve shows the average of the A and B realisations for each resolution, while the black solid line displays the linear theory prediction. The dimensionless power spectrum is expressed as,
\begin{equation}
    \Delta^2(k) = \frac{k^3}{2\pi^2}P_m(k)\,,
\end{equation}
where $P_m(k)$ is the non-linear matter power spectrum measured from our MTNG simulations.
The vertical dashed line represents the fundamental mode of the box, given by
\begin{equation}\label{eq:kbox}
    k_{\rm box} = \frac{2\pi}{L}\,,
\end{equation}
where $L=500\Mpch$ is the box length of the MTNG740-DM simulations. In addition, the diagonal lines display the Poisson {\it shot-noise} contribution,
\begin{equation}
    \Delta^2(k)_{\rm shot} = \frac{k^3}{2\pi^2}P_{\rm shot}\,,
\end{equation}
with $P_{\rm shot} = L^3/N_{\rm p}$ and $N_{\rm p}$ being the total number of particles. We have here subtracted the shot-noise from our power spectra measurements \citep[but see][]{Maleubre2022}.

From the upper panels of Fig.~\ref{fig:MTNG_Pk_res} we can clearly see that the size of the MTNG740 simulations is enough to measure the clustering of matter up to the BAO scale, as explored further below. Also, we have a first impression of the mass resolution impact on the power spectra measurements, low-resolution simulations predict a lack of power on small-scales compared with higher resolution runs. We quantify the dependence of mass resolution on the power spectrum through the relative difference between the lower resolution runs of level-5 to 2 (denoted based on their particle numbers as N270, N540, N1080 and N2160) and the highest resolution level-1 simulation (N4320, see the lower subpanels of Fig.~\ref{fig:MTNG_Pk_res}). We find sub percent $(\lesssim 1\%)$ agreement at $z=0$ on scales $k < 0.2\hMpc$, $0.6\hMpc$, $2\hMpc$ and $10\hMpc$, for the N270, N540, N1080 and N2160 simulations, respectively. In addition, the dashed lines in the lower subpanels of Fig.~\ref{fig:MTNG_Pk_res} indicate the relative difference between the PS measurements without subtracting the shot-noise contribution. We find a slightly improved agreement between the lower resolution and the level-1 runs on small-scales. Similar values are obtained at $z=1$, as seen in the right panel of Fig.~\ref{fig:MTNG_Pk_res}.  With this exercise we have obtained a sense of the numerical precision of our simulations.  Based on these results, we conclude that cosmological simulations with mass resolution better than $m_{\rm p} = 1.06\times 10^9 \Msh$ are needed to achieve the precision of theoretical predictions required by the current and upcoming galaxy surveys.  Due to the large number of particles, the convergence of the measured non-linear $P(k)$ of the MTNG740-DM-1 simulation extends slightly beyond its Nyquist frequency, to $k \sim 30\hMpc$.

To illustrate the accuracy of our MTNG3000-DM-0.1$\nu$ simulation, we show in Figure~\ref{fig:MTNG3000_Pk} the power spectrum measurements for the cold (dark) matter (CDM and baryons) and neutrino components both at the initial time, $z=63$, and at the final time $z=0$. We compare directly to the linear theory prediction of {\small CAMB} \citep{Lewis2000} for the corresponding cosmology, assuming that baryons are present but represented by the dark matter as well. The neutrinos consist of two degenerate $50\,{\rm meV}$ species, and a third massless neutrino. We also include the photon background, and have modified {\small GADGET-4} such that it correctly accounts for relativistic backgrounds in the clustering evolution. At the initial time, we however deliberately put in a small correction in the form of a power increase of up to $\sim 4\%$ at the fundamental mode, because at this time these modes are still outside of the horizon, meaning that our Newtonian code would not capture their faster growth accurately. The correction is computed such that at low redshift our results match linear theory precisely. The bottom panel of the $z=0$ result shows that this succeeds with impressive precision. Our simulation reproduces the expected linear theory results on the largest scales to fractions of a percent, both in the dark matter and in the neutrino components.

Note that the power in the cold dark matter necessarily cuts off at the Nyquist frequency in the initial conditions, at which point the cold dark matter power actually lies far below the shot noise limit. The `coldness' of the dark matter allows this sub-Poissonian distribution to be preserved during the subsequent evolution.
As our power spectrum measurement extends to much smaller scales, we find, however, for scales below the Nyquist frequency in the initial conditions the shot noise power, simply because these scales are undetermined in the initial conditions. They are however filled in during non-linear evolution by power transfer from larger scales, such that the non-linear power at $z=0$ can be predicted accurately well below the initial Nyquist frequency. Eventually, however, the dark matter power spectrum hits the shot noise limit in the deeply non-linear regime.

For the neutrinos, the initially imprinted power spectrum would normally not be preserved below the neutrino shot noise, due to them being extremely relativistic particles,  letting them move with velocities close to the speed of light. Interestingly, this unfavourable outcome does not occur for the neutrinos, and their effective shot noise (horizontal plateau of the measurement) stays at a level far below the nominally expected shot-noise for the $2560^3$ neutrino particles we used. This is due to the $\delta f$-method \citep{Elbers:2020lbn} we employed  to simulate the neutrino component, which allows us to keep the shot-noise level at late times nearly a factor of 100 lower than naively expected, making it in fact very close to that of the dark matter particles, a situation we intended to achieve with our choice of the particle numbers.

In the remainder of this paper, we shall focus on an analysis of the $500\,h^{-1}{\rm Mpc}$ simulations (MTNG740),
deferring a further analysis of the neutrino predictions to forthcoming work \citep{paperII}. We proceed in this way because we consider it prudent to first validate the foundations of the MTNG simulation project and test for numerical convergence and the impact of baryons on basic clustering statistics, before exploring the additional effects of neutrinos in more quantitive detail.

%--------- Figure --------------
\begin{figure*}
 \centering
\includegraphics[width=0.47\textwidth]{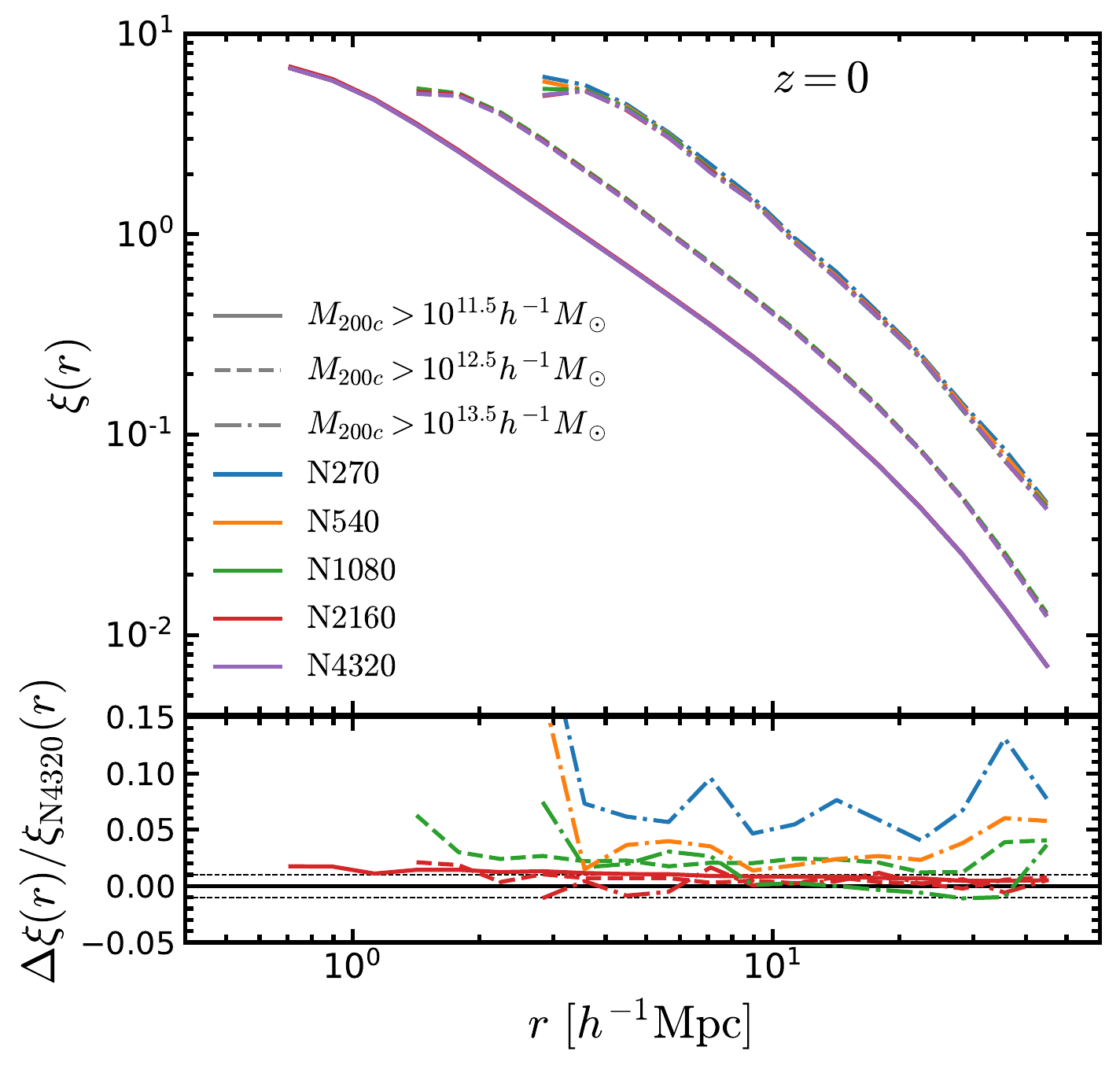}
\includegraphics[width=0.47\textwidth]{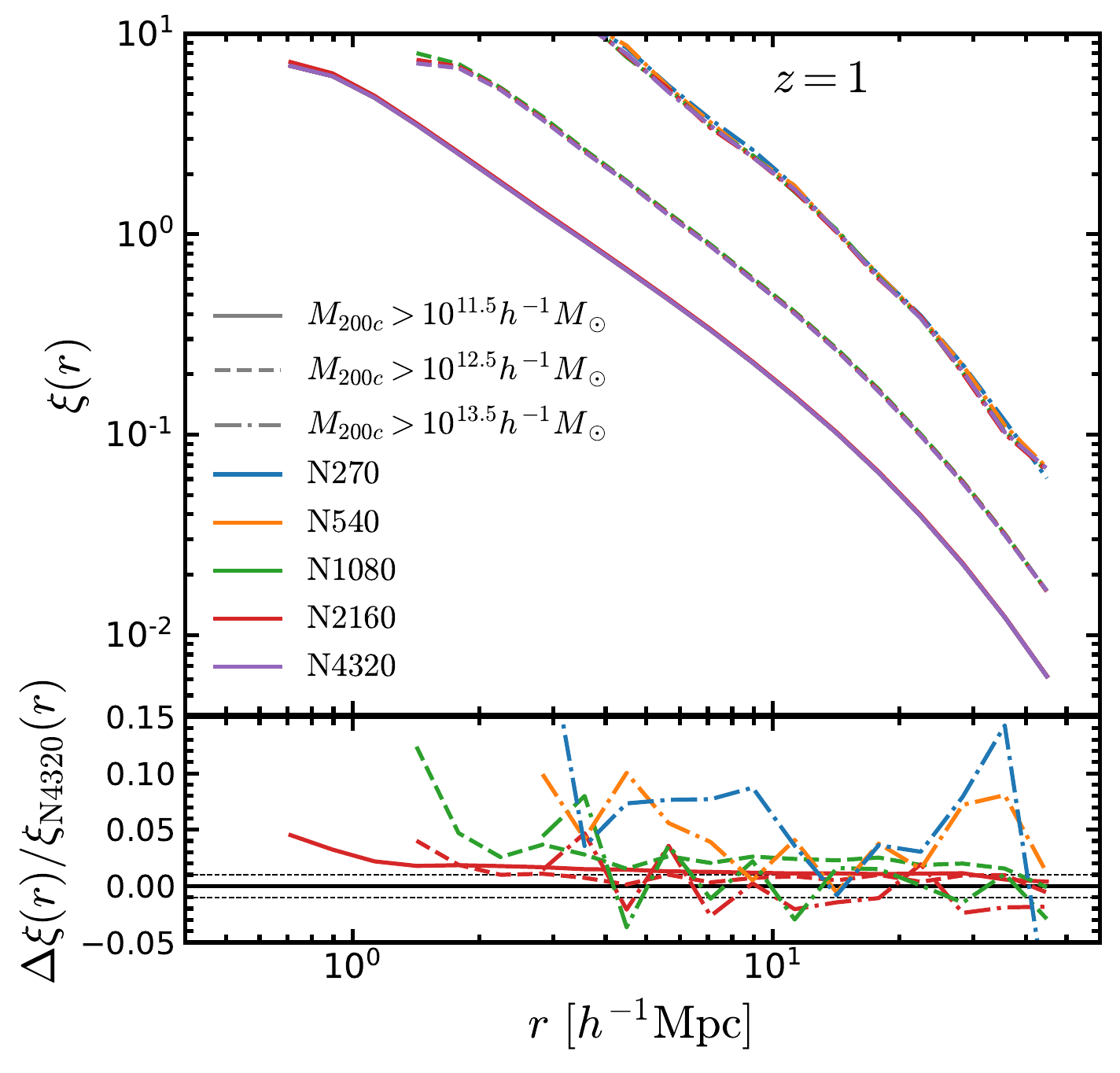}
\caption{Convergence test of the real-space halo two-point correlation functions at $z=0$ (left panel) and $z=1$ (right panel), for three different halo samples with masses $M_{200c}>10^{11.5}\Msh$ (solid lines), $M_{200c}>10^{12.5}\Msh$ (dashed lines) and $M_{200c}>10^{13.5}\Msh$ (dash-dotted lines), based on the MTNG740-DM simulations. The relative difference between the low and the highest resolution simulations is shown in the lower subpanels. The different resolution levels are distinguished by their particle number per dimension, as labelled.}
\label{fig:xih_res}
\end{figure*}

%---------------------------------------------------------------
\subsubsection{Halo mass function and halo clustering}
\label{sec:halo_res}
%---------------------------------------------------------------
%--------- Figure --------------
\begin{figure*}
 \centering
\includegraphics[width=0.45\textwidth]{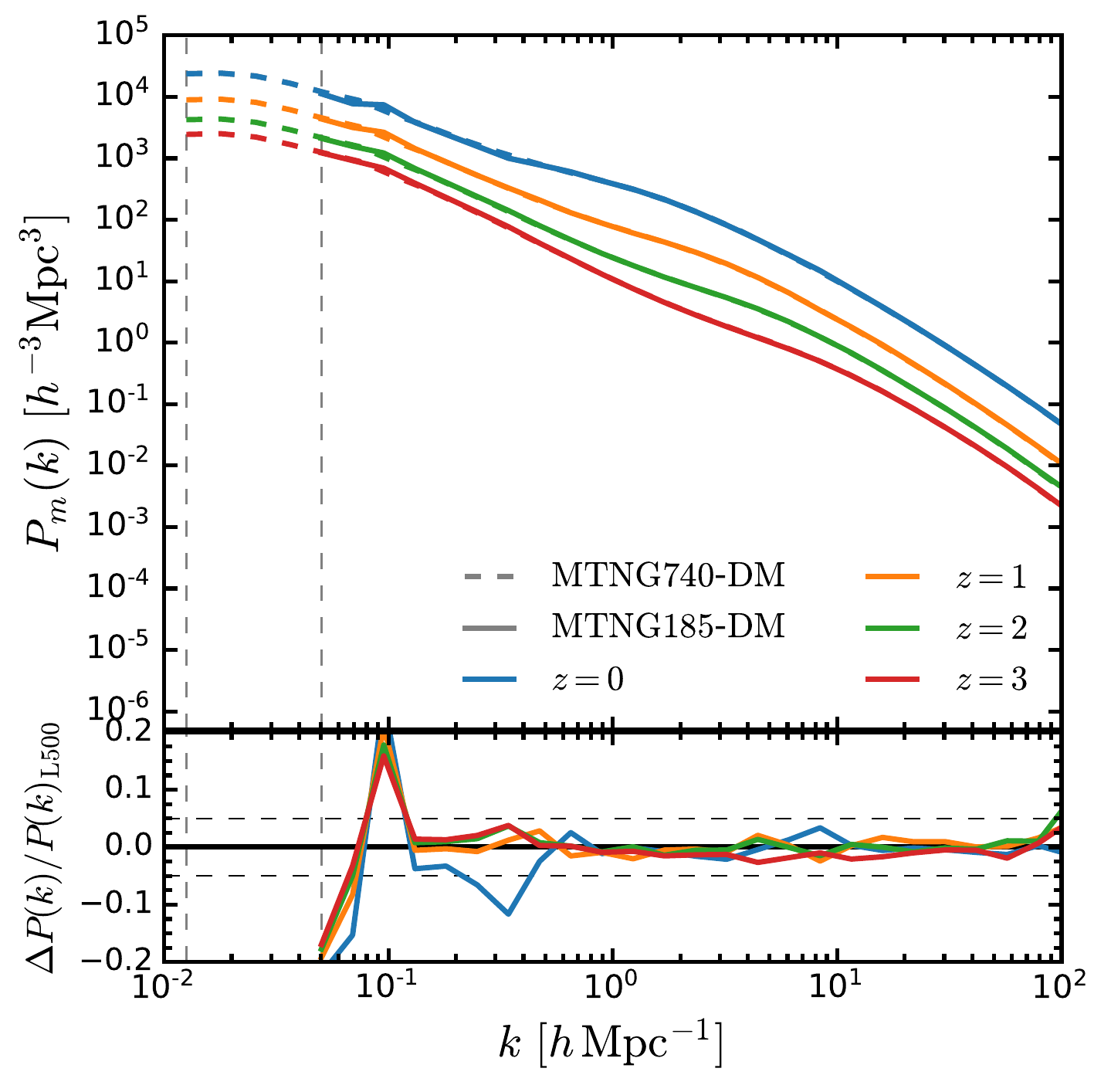}
\includegraphics[width=0.45\textwidth]{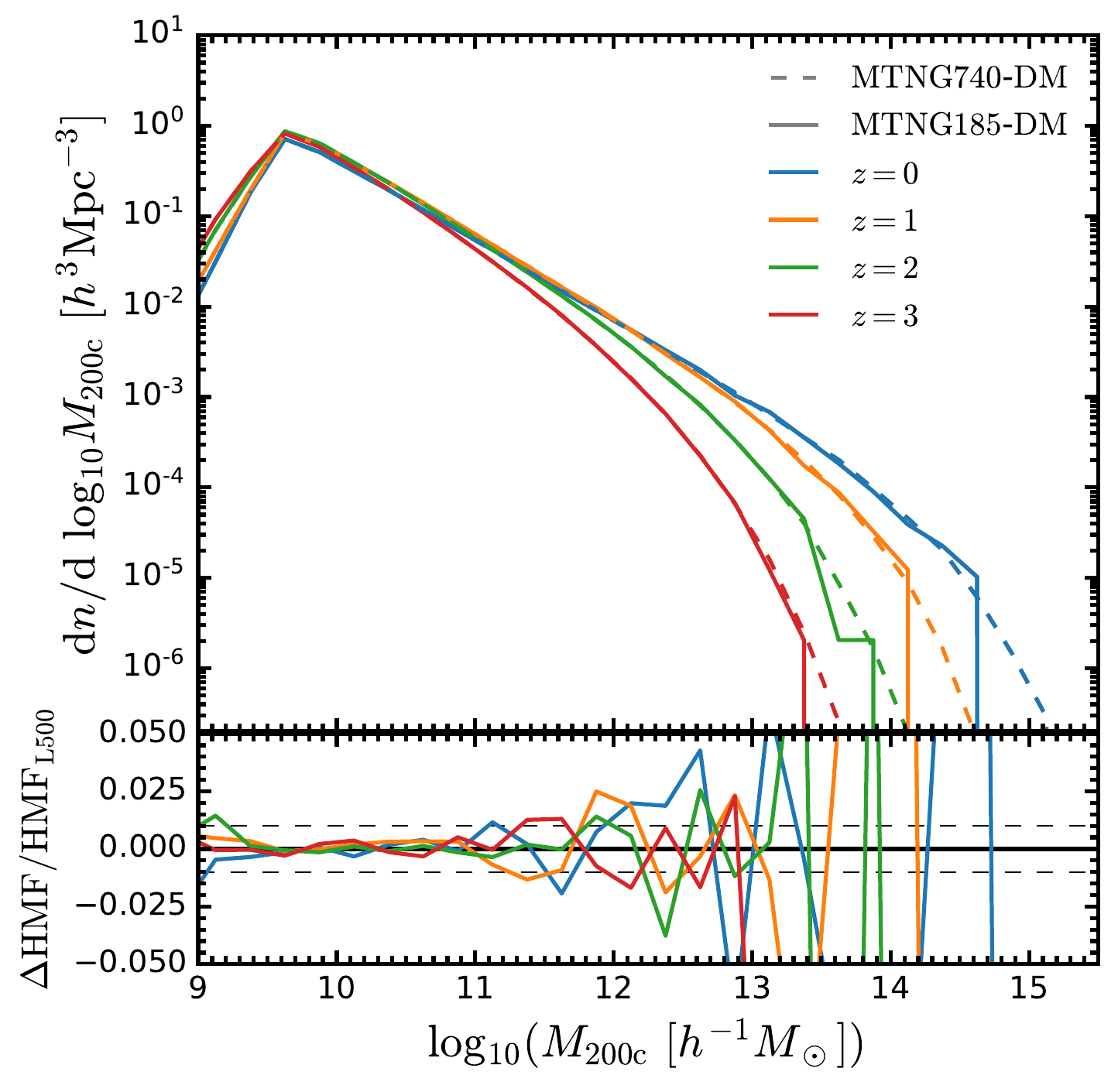}
\caption{Box size impact on the measured non-linear matter power spectrum (left panel) and the halo mass function (right panel) at $z=0$ (blue lines), $z=1$ (orange lines), $z=2$ (green lines), and $z=3$ (red lines) from the MTNG740-DM-1 (dashed lines) and MTNG185-DM (solid lines) simulations, the dark matter-only versions of the MTNG hydro simulations. The lower subpanels show the relative difference with respect to the larger MTNG740-DM-1 simulation, where the horizontal dashed lines represent differences of $5\%$ (left panel) and $1\%$ (right panel), respectively. Note that for MTNG740-DM we use the average result of two paired realisations, whereas for MTNG185-DM only one (but still `fixed') realisation is available.}
\label{fig:MTNG_box}
\end{figure*}

Dark matter haloes are the building blocks of the cosmic structures since they host the different types of galaxies in the Universe. Therefore, the study of their abundance and clustering is important for modern cosmological analysis. The abundance of dark matter haloes can be quantified by the halo mass function (HMF), which gives the number of dark matter haloes as a function of halo mass, cosmic time and cosmological parameters, in different mass intervals. Since cosmic structure formation is a hierarchical process, small haloes form first and their subsequent mergers build ever larger groups, up to haloes with the mass of galaxy clusters, which are the most massive objects in the Universe. We expect to find a higher abundance of less massive objects at early-times (or higher redshifts), whereas the most massive haloes are rare (and hence require large simulation volumes for good statistics) and form comparatively late. In any case, the HMF can be considered one of the most important quantities predicted by cosmological simulations. Exploring the completeness and convergence of the HMF, which is strongly related to the mass resolution and the simulated volume, is thus of great importance to accurately predict galaxy statistics.

In Figure~\ref{fig:MTNG_HMF_res}, we show the measured halo mass functions at $z=0$ (left panel) and $z=1$ (right panel) from the MTNG740-DM dark matter-only simulations. We use here the $M_{200c}$ spherical overdensity halo mass definition which corresponds to the mass enclosed within a radius in which the density is 200 times the critical density of the Universe,  and we only consider haloes with masses $M_{200c} > 32m_{\rm p}$, where $m_{\rm p}$ is the mass of the dark matter particle (see Table~\ref{tab:sims}). The halo centres adopted for the spherical overdensity calculation are taken as the positions of the particles with the minimum gravitational potential of the bound part of each FOF group (provided this has at least 20 particles, otherwise the group is dropped). Note also that not all groups with low particle number necessarily reach the 200c overdensity value and thus have a $M_{200c}$ value assigned to them. 

From the upper panels of Fig.~\ref{fig:MTNG_HMF_res}, we see that for different resolutions we can find very similar HMFs at the high-mass end. In addition, we note a cutoff at the low-mass end due to the limited number of particles in low-mass haloes. This cutoff moves systematically to higher-mass haloes when we reduce the resolution of the simulations. The cutoff scale can be used to assess the completness of the HMFs; i.e., the point where the HMFs show the largest differences due to the resolution of the simulations. The level-2 (N2160) and level-1 (N4320) simulations predict almost the same number of haloes for masses $M_{200c} > 10^{12}\Msh$ at both $z=0$ and $z=1$. Sub percent agreement between the N1080 and N4320 runs is found for haloes more massive than $M_{200c} = 10^{13}\Msh$. The very low resolution of the N540 and N270 runs have, however, a greater impact on their measured HMFs, as we can see through the large differences when compared with the HMF from the N4320 run. Furthermore, it appears that the HMFs for the N540 and N270 simulations is not converged at $z=1$ for any halo mass, since they fall short of producing the same number of massive haloes as the N4320 run (see right panel of Fig.~\ref{fig:MTNG_HMF_res}). Nevertheless, these low-resolution simulations are still useful to test and debug analysis pipelines.    

As an additional convergence test of the halo population, Fig.~\ref{fig:MTNG_Mfof_res} displays the measured HMFs based on the Friends-of-Friends $(M_{\rm FoF})$ halo mass definition at redshifts $z=0$ and $z=1$. However, it has been shown that the FoF definition has discreteness limitations when estimating the halo mass from low-resolution simulations, producing on average an overestimation of the HMF measured from low-resolution simulations \citep{Warren:2005ey}. We observe this trend when measuring the HMF directly from our halo (FoF group) catalogues as shown in the lower subpanels of Fig.~\ref{fig:MTNG_Mfof_res} (dashed lines). We find an excess of haloes between $2-10$ percent when comparing the level $2-5$ runs with our flagship MTNG740-DM-1 simulations. 
For this reason we apply a correction suggested by \citet{Warren:2005ey} to each FoF halo mass, given by
\begin{equation}\label{eq:Mfof}
    M_{\rm FoF}= (1 - N^{-0.6}_p)M^{\rm sim}_{\rm FoF}\,,
\end{equation}
where $N_p$ is the number of member particles of a halo and $M^{\rm sim}_{\rm FoF}$ is the FoF mass of each halo in the original catalogue.

%--------- Figure --------------
\begin{figure*}
 \centering
\includegraphics[width=0.45\textwidth]{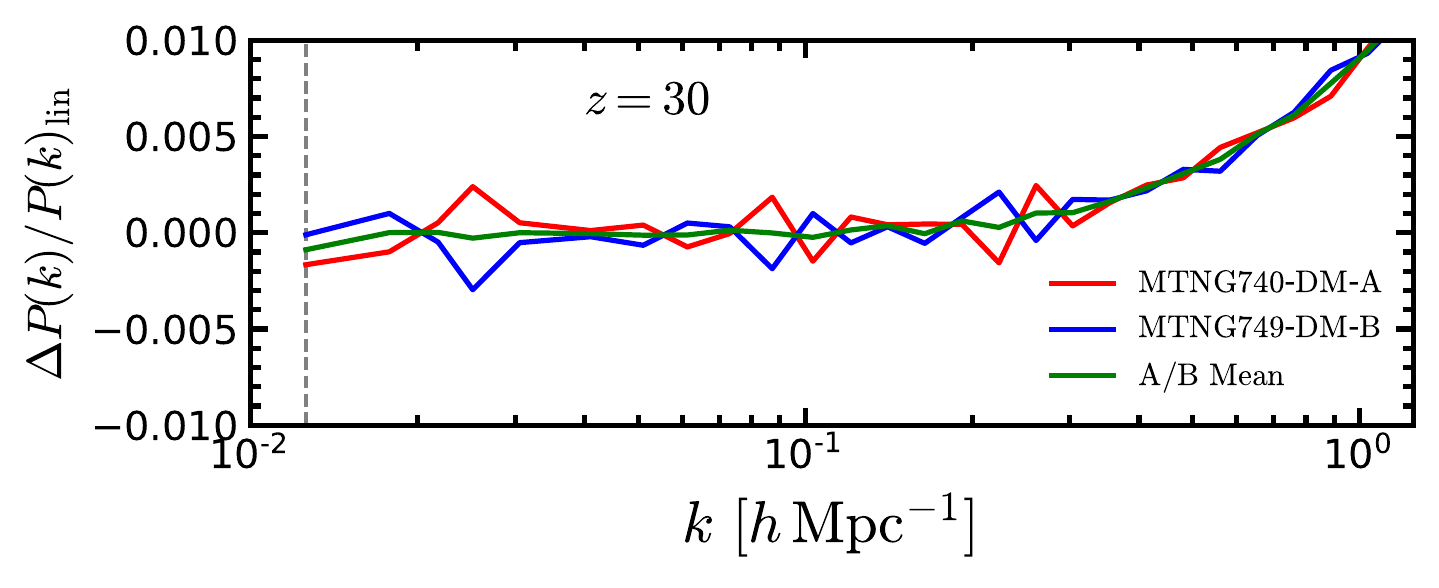}
\includegraphics[width=0.45\textwidth]{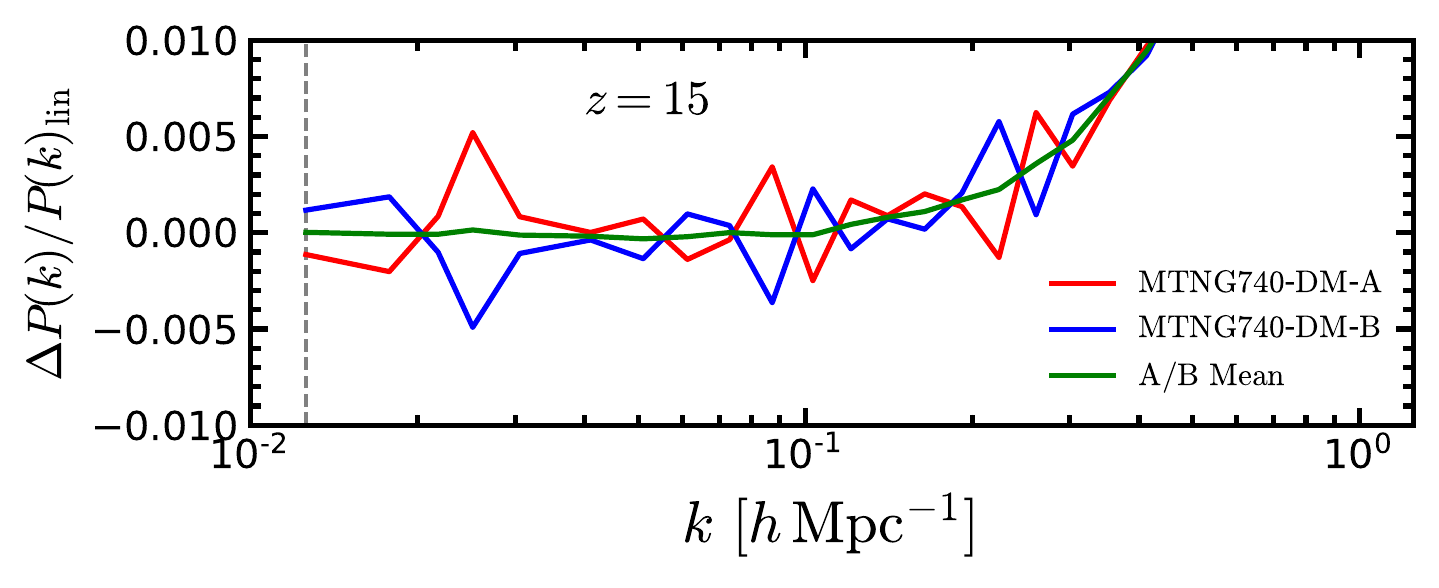}
\includegraphics[width=0.45\textwidth]{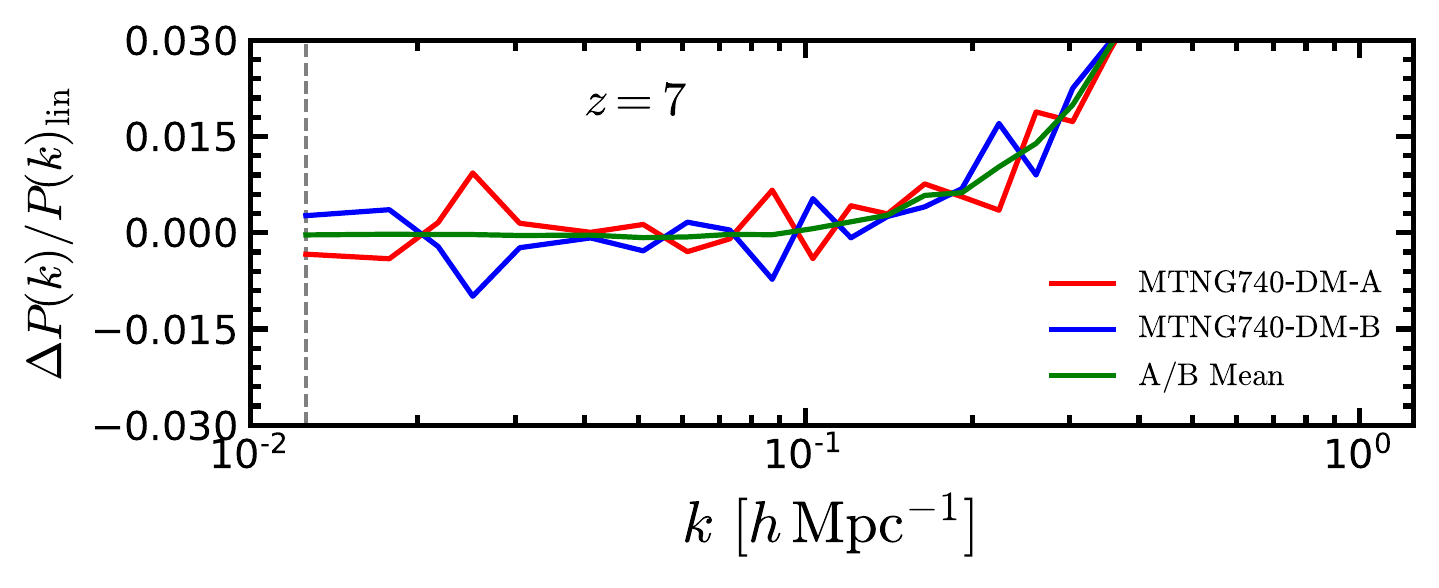}
\includegraphics[width=0.45\textwidth]{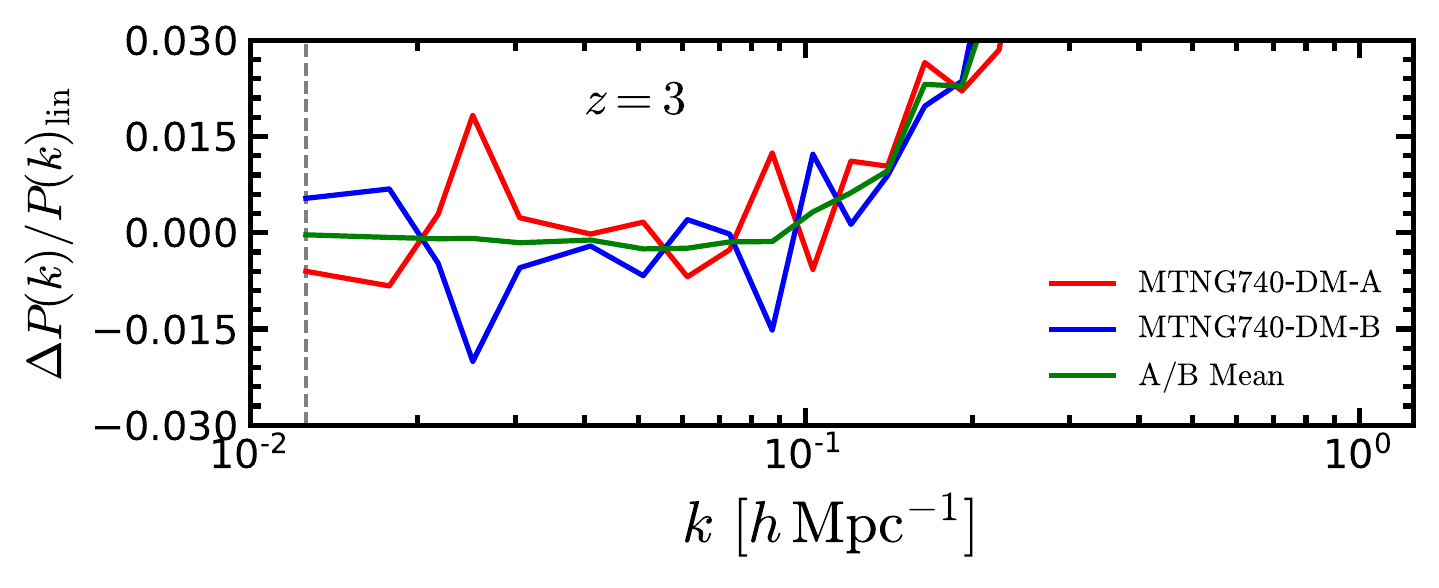}
\includegraphics[width=0.45\textwidth]{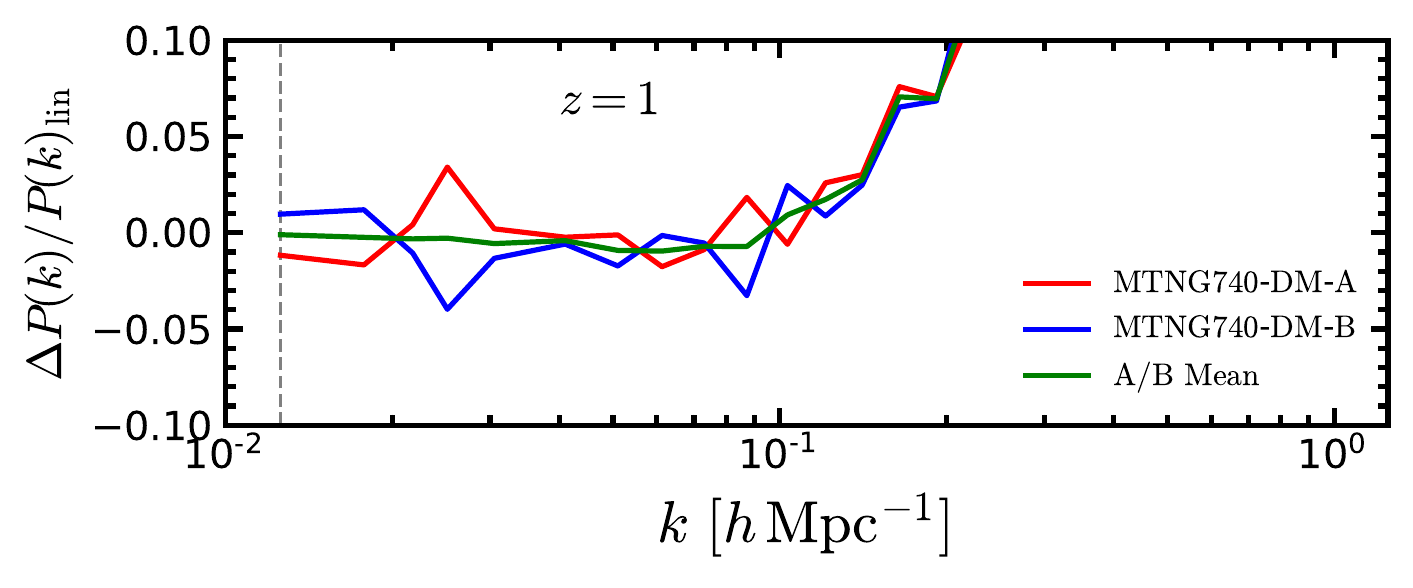}
\includegraphics[width=0.45\textwidth]{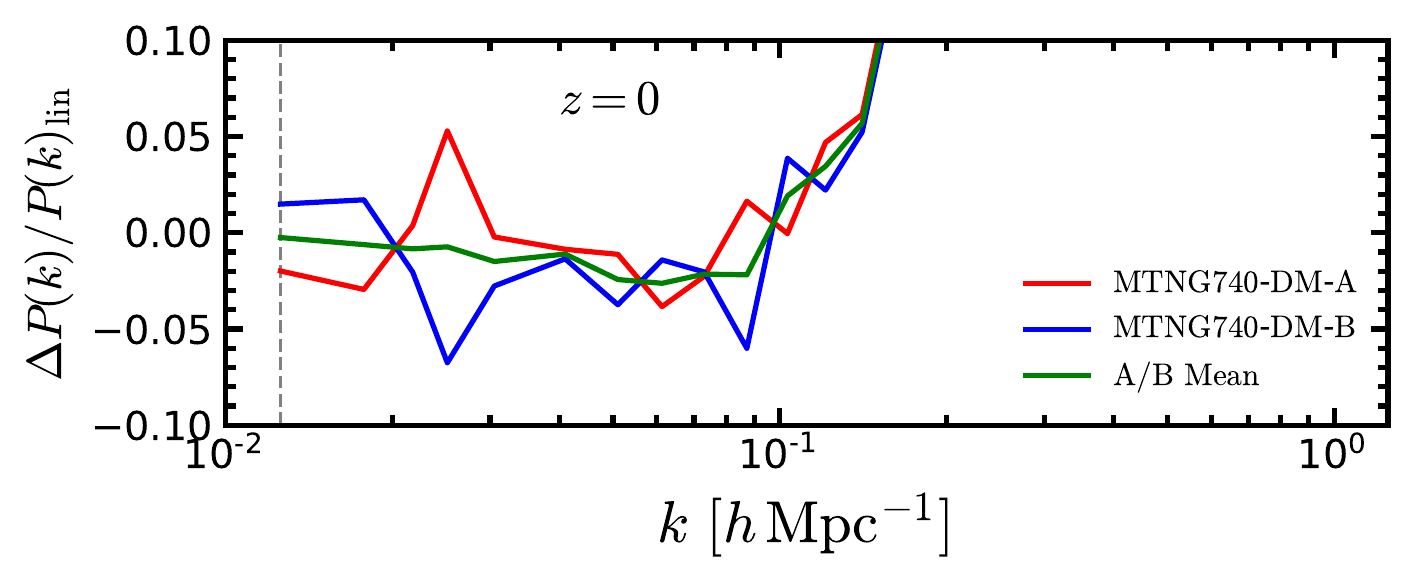}
\caption{The ratio between the measured non-linear matter power spectrum of the MTNG740-DM-1-A (red lines) and -B (blue lines) simulations relative to the linear theory power spectrum prediction, at different redshifts, as labelled. The green line represents the mean of the A and B realisations.}
\label{fig:diff_Pkm}
\end{figure*}

We then find a $\sim 1$ per cent agreement between the FoF halo mass functions of the N2160 and N4320 runs for haloes with masses $M_{\rm FoF} > 10^{11.5} \Msh$, at both redshifts. We still see a slight enhancement of the FoF mass for the N1080, N540 and N270 runs compared to the higher resolution simulations, amounting to $3-5$ per cent at the present time, but these differences are smaller at early times, e.g., at $z=1$, where the differences are $~2\%$ (see the right lower-subpanel of Fig.~\ref{fig:MTNG_Mfof_res}).

We next explore the real-space halo two-point correlation function and consider its numerical convergence for our simulations. The correlation function gives the excess probability to find a pair of tracers (e.g.~dark matter haloes, or galaxies) separated by a distance $r$ compared to a random, uniform distribution. Since haloes are biased tracers of the underlying dark matter field, their clustering amplitude can be different for different halo populations, e.g., massive haloes are more strongly clustered than low-mass haloes (the linear halo bias will be discussed in Sec.~\ref{sec:halo_bias}).

We therefore measure the real-space clustering for three different halo samples with fixed $M_{200c}$ halo mass, and using 20 logarithmically-spaced radial bins between $0.5 < r/[\Mpch] < 50$ using the {\sc Nbodykit} toolkit \citep{nbodykit}. We select haloes with masses $M_{200c} > 10^{11.5}\Msh$ (sample 1), $M_{200c} > 10^{12.5}\Msh$ (sample 2), and $M_{200c} > 10^{13.5}\Msh$ (sample 3). Note that not all simulations are able to resolve haloes for each halo sample (see the completeness of the $M_{200c}$ HMF in Fig.~\ref{fig:MTNG_HMF_res}), for this reason sample~1 contains haloes from the N4320 and N2160 runs only, sample~2 covers the N4320, N2160 and N1080 runs, and haloes of sample~3 can be found in all of the MTNG runs (N4320, N2160, N1080, N540 and N270).

The measured halo clustering is shown in the upper subpanels of Fig.~\ref{fig:xih_res} at redshifts $z=0$ and $1$. We can clearly see that the most massive haloes (sample 3; dash-dotted lines) have a higher clustering amplitude, which means that they are more biased than low-mass haloes (sample 1 and 2; solid and dashed lines, respectively). In the lower subpanels of Fig.~\ref{fig:xih_res}, we compare the halo clustering of the different resolution runs with respect to N4320. The halo correlation functions of the N4320 and N2160 runs agree within 1 per cent for all the samples at $z=0$ and at all scales, however, the same agreement is found at $z=1$ for all length-scales $r > 1\Mpch$. The clustering of samples 2 and 3 of the N1080 case show a slightly larger difference of $\sim 3\%$ in comparison with the N4320 runs for scales greater than $1\Mpch$; similar differences are found for sample 3 of the N540 simulation (see orange dash-dotted lines). The lowest resolution runs (N270) produce a larger difference of $\sim 6$ per cent, this is due to the incompletness of the halo mass function displayed in Fig.~\ref{fig:MTNG_HMF_res}, where we found a deficit of more than 10 per cent for haloes with masses $M_{200c} > 10^{13.5}\Msh$.

%---------------------------------------------------------------
\subsubsection{Box size effects}
\label{sec:box_test}
%---------------------------------------------------------------
Finally, we show the impact of the box-size on the measured matter power spectrum and the halo mass function of the MTNG simulations in the left and right panel of Figure~\ref{fig:MTNG_box}, respectively. It is important to perform convergence tests by varying the size of the cosmological box and keeping the same particle resolution, to understand the limitations in cosmological calculations due to the finite volume of numerical simulations. To this end we compare the measurements for the MTNG740-DM and MTNG185-DM simulations, which are the dark matter-only counterparts of the corresponding full-physics simulations. The  MTNG185 run has a volume 64 times smaller than our flagship MTNG740 run, while keeping the same mass resolution (see Table~\ref{tab:sims}).

The box-size effects on the non-linear power spectrum at $z=0$, $1$, $2$ and $3$ are displayed in the left panel of Fig.~\ref{fig:MTNG_box}. The vertical dashed lines show the fundamental mode of each box given by Eq.~\eqref{eq:kbox}, which delimits the minimum scale we can measure the power spectra from the corresponding simulations, i.e., $k_{\rm L500} \sim 0.0125\hMpc$ and $k_{\rm L125} \sim 0.05 \hMpc$, respectively. The clustering of matter agrees very well between the two boxes at all redshifts; we find only a $\sim 2\%$ difference on scales $k\gtrsim 0.13\hMpc$ in all cases (see the left lower-subpanel of Fig.~\ref{fig:MTNG_box}). We observe larger deviations of up to $20\%$ for the large-scale modes close to the $k_{\rm L125}$ value. This is due to the limited box size of the MTNG185 simulation, which does not allow us to measure the clustering on these large scales with sufficient precision, both because they are already affected by mild non-linear evolution (which is distorted by missing, still larger scale modes) and because we have run only one MTNG185-DM realisation, which is thus still dominated more strongly by {\it sample variance} as we can not eliminate in this case the leading high-order perturbations around linear evolution by means of the pairing technique.

The evolution of the HMF from $z=3$ to $z=0$ of the MTNG740-DM and MTNG185-DM boxes is shown in the right panel of Fig.~\ref{fig:MTNG_box}. At first glance, we can see that the HMFs of the MTNG185 run follow the shapes of their MTNG740 counterparts very well, showing completness down to the same value of $M_{200c} = 10^{10}\Msh$, which corresponds to well-resolved dark matter haloes with 75 particles. The agreement between the HMFs is much better than $1\%$ for haloes with masses $M_{200c} < 10^{13}\Msh$ at all redshifts. We find some small fluctuations at the high-mass end ($M_{200c} > 10^{13}\Msh$) due to the limited volume of the  MTNG185 simulation, which does not contain enough massive haloes, such as groups and clusters-like objects, to avoid being severely affected by small number statistics. 

%--------- Figure --------------
\begin{figure*}
 \centering
\includegraphics[width=0.45\textwidth]{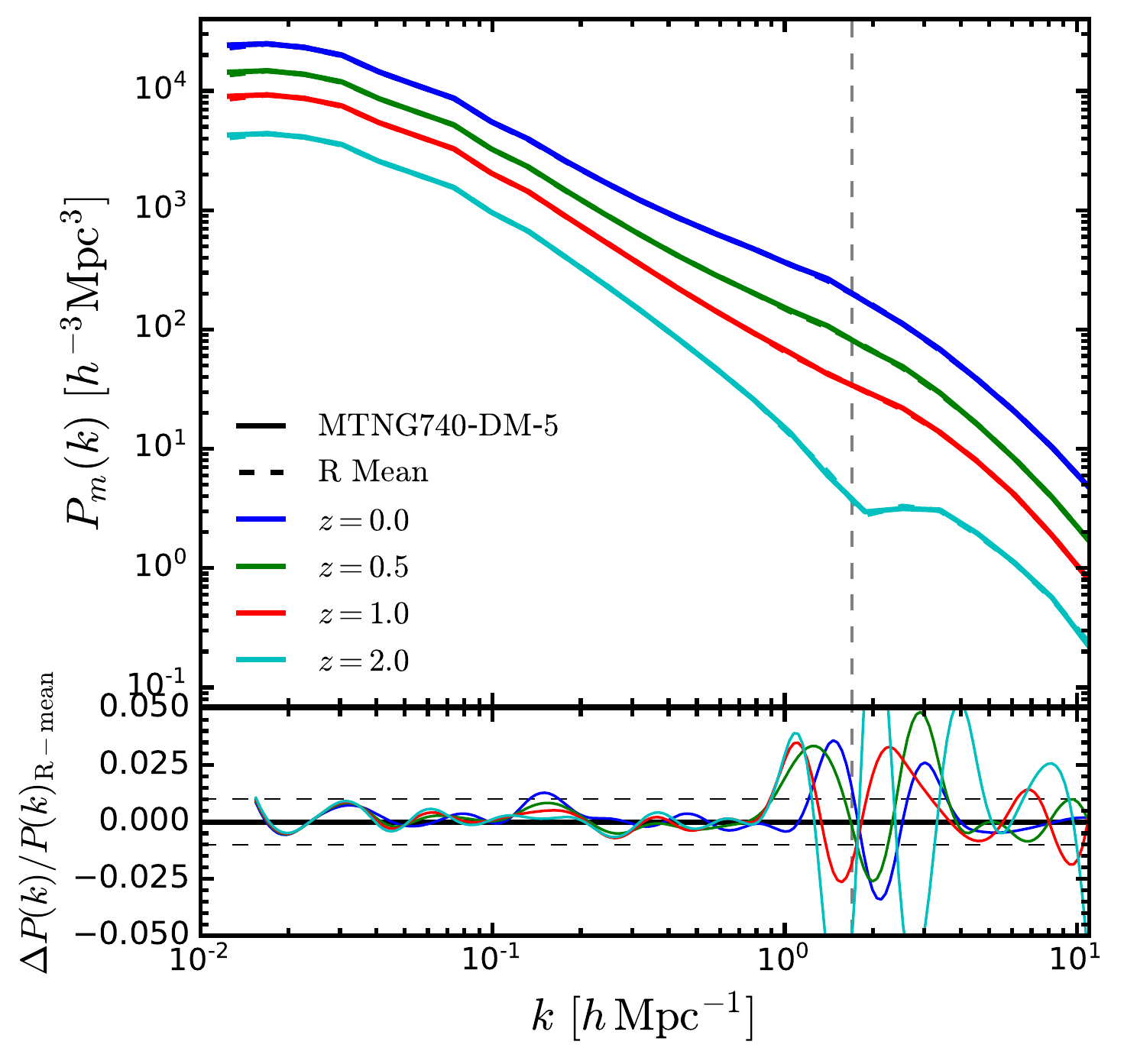}
\includegraphics[width=0.45\textwidth]{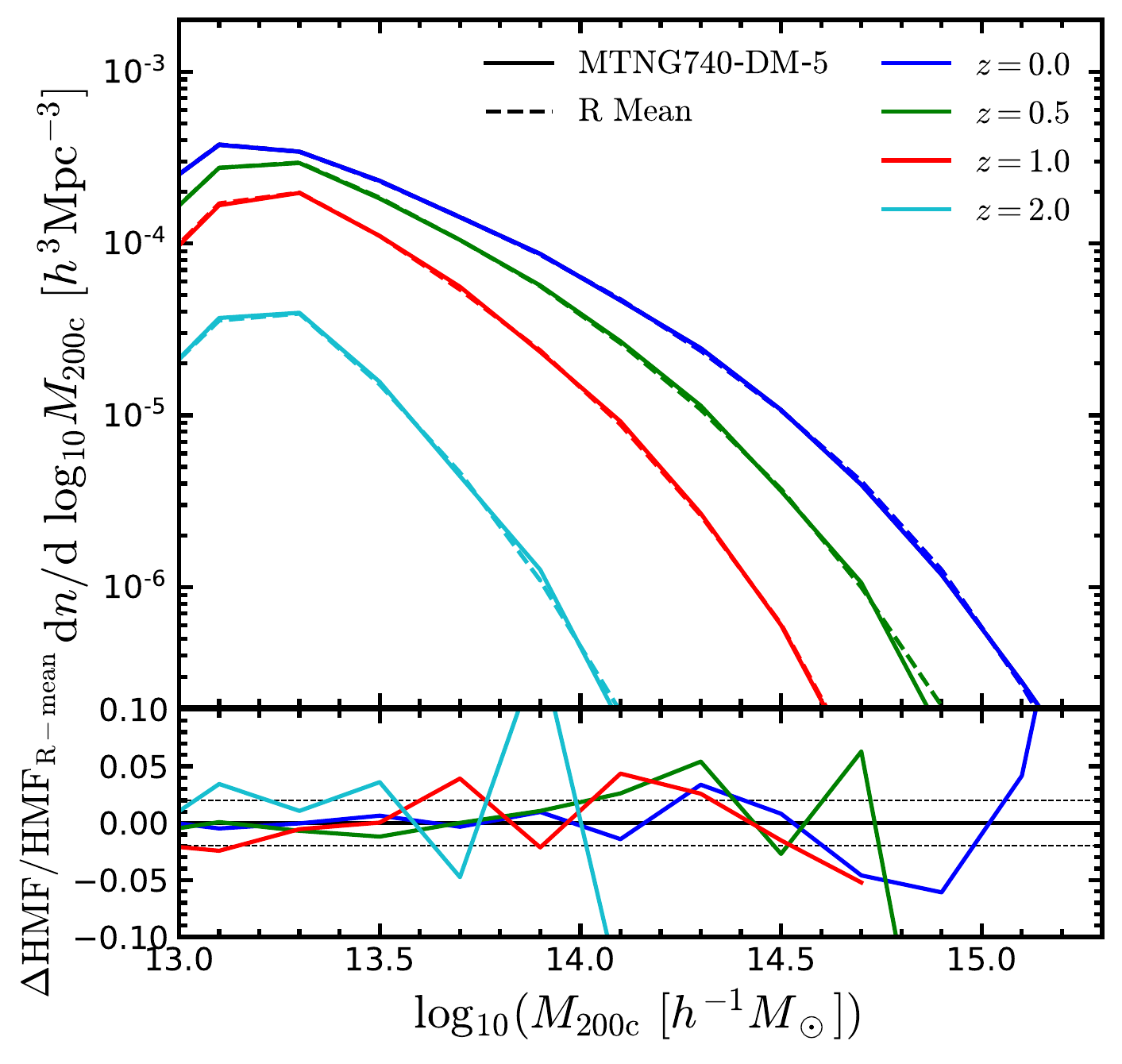}
\caption{Comparison of the measured power spectrum (left panel) and halo mass function (right panel) between the mean over 100 independent Gaussian realisations (dashed lines) and the A$-$ and B$-$ fixed-and-paired (solid lines) MTNG740-DM-5 runs at $z=0$ (blue lines), $z=0.5$ (green lines), $z=1$ (red lines), and $z=2$ (cyan lines). The lower subpanels show the relative difference between the results from the 100 Gaussian realisation (R-mean)s and the A/B pair, while the dashed lines indicate a $1\%$ (left panel) and $2\%$ (right panel) difference.}
\label{fig:R_test}
\end{figure*}

Thus far, we have demonstrated that our MTNG740-DM runs reach enough cosmological volume and mass resolution to show good convergence in the non-linear matter power spectrum and halo mass functions. This should make especially the level-1 (N4320) and level-2 (N2160) runs very helpful for the generation of mock catalogues for large-scale galaxy surveys such as DESI, Euclid or PFS.

%---------------------------------------------------------------
\subsection{Reducing cosmic variance with fixed-and-paired realisations}
\label{sec:realisations}
%---------------------------------------------------------------
As mentioned above, we have used the fixed-and-paired technique by \cite{Angulo:2016hjd} to produce two DM-only cosmological simulations for a given model. The method consists of running two realisations of the initial density perturbation fields, both set-up with Fourier modes amplitudes {\it fixed} to the ensemble average power spectrum, and in addition, both are {\it paired} through the use of opposite phases for each mode (i.e., with a phase difference of $\pi$, or equivalently, with a sign reversal of the density perturbation field). For pure linear theory, the pegging of the mode amplitudes to $\propto \sqrt{P(k)}$ reproduces the linear theory input power spectrum by construction at all times, even when considering each of these realisations individually. However, mild non-linear evolution will still introduce deviations of the mean power in the large-scale modes. But, as \cite{Angulo:2016hjd} show, these can be cancelled to leading order when the results of the two realisations are averaged. In this way, such a pair of simulations can yield equivalent results as obtained from the mean of many independent realisations. 

Figure~\ref{fig:diff_Pkm} shows the evolution of the ratio between the measured non-linear dark matter power spectrum from the MTNG740-DM-1-A and -B realisations and the linear theory prediction, from $z=30$ to $z=0$. We can see that taking the average of the two fixed-and-paired realisations (green lines) eliminates the higher order fluctuations (red and blue lines) around linear theory, allowing us to obtain much more accurate cosmological predictions at large scales for these still comparatively `limited-volume' simulations. Importantly, this method works well down to the present time on scales not yet strongly affected by non-linear evolution, even though individual modes already differ on these scales by 10s of percent from pure linear theory due to higher order terms affecting the evolution.

%--------- Figure --------------
\begin{figure*}
 \centering
\includegraphics[width=0.45\textwidth]{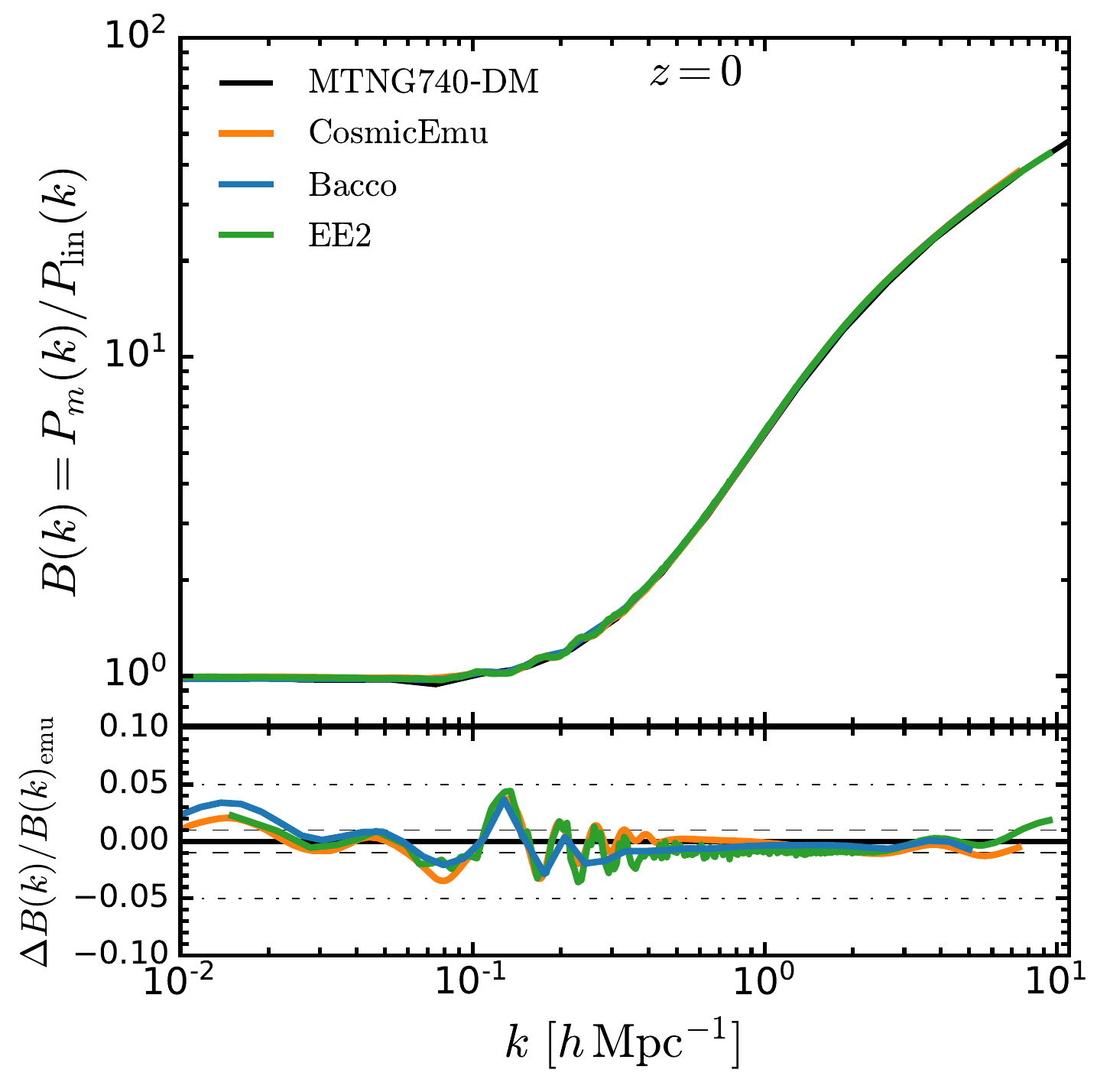}
\includegraphics[width=0.45\textwidth]{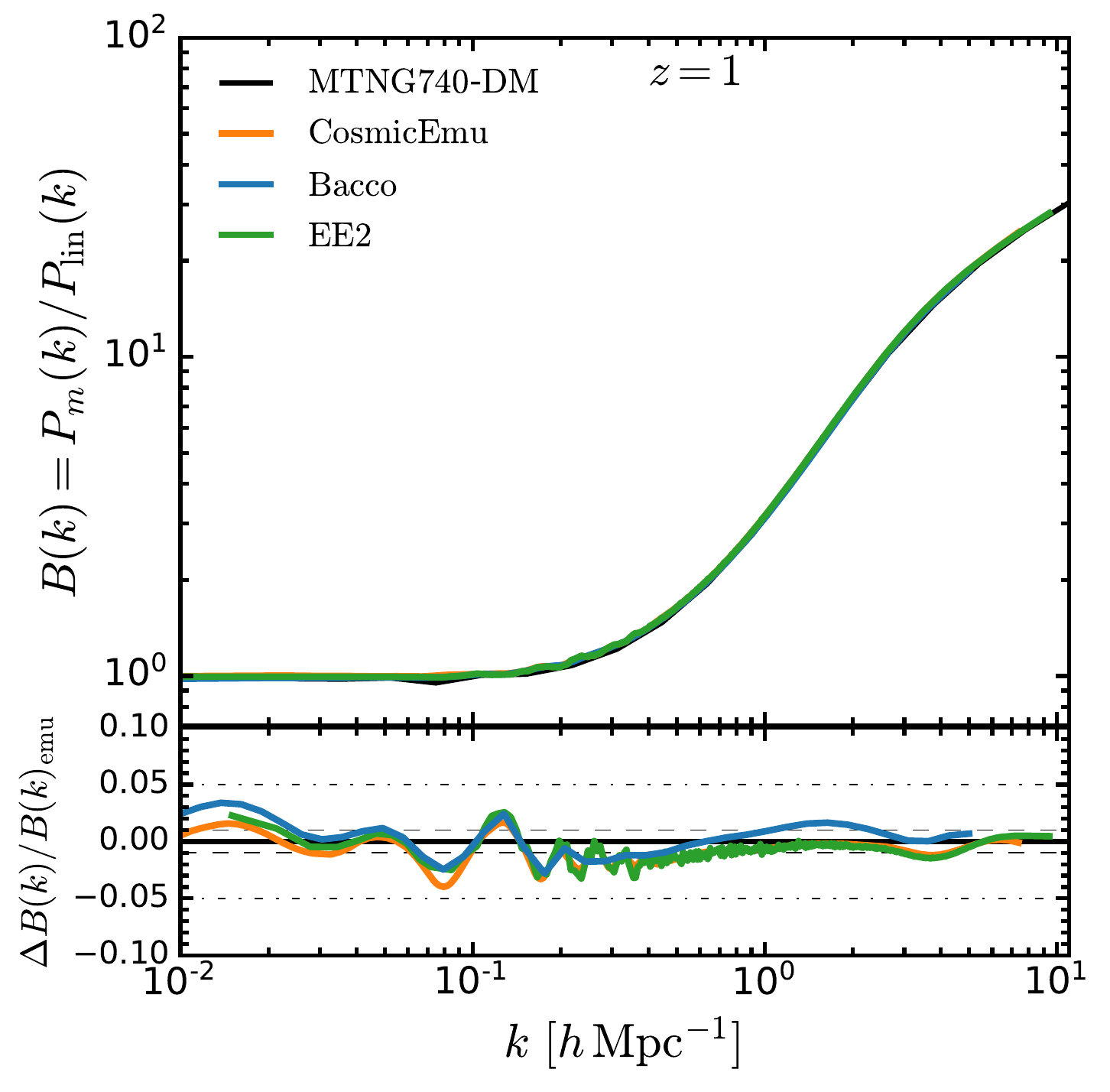}
\caption{Boost factor, $B(k)$, comparison between the measurements from the mean of the MTNG740-DM-1-A/B pair of simulations (black lines) and the predictions from the {\sc CosmicEmu} (orange lines), {\sc Bacco} (blue lines) and {\sc EuclidEmulator2} (green lines) emulators at redshifts $z=0$ (left panel) and $z=1$ (right panel). The lower subpanels show the relative difference between our simulations and the emulator predictions, while the dashed and dotted horizontal lines indicate a $1\%$ and $5\%$ difference, respectively.}
\label{fig:Pkm_emu}
\end{figure*}

As an additional consistency check of the advantage of running fixed-and-paired simulations to suppress the cosmic variance, we have run 100 independent, fully Gaussian realisations with the same specifications as the MTNG740-DM-5 simulations (i.e.~box size, number of particles, and cosmological parameters kept the same). We then compare the matter clustering and halo mass functions of our fixed-and-paired simulations with those from the Gaussian realisations.

The matter clustering results at $z=0$, $0.5$, $1$ and $2$ are shown in the left panel of Fig.~\ref{fig:R_test}. From the lower subpanel we can see that the average of the A and B realisations of MTNG740-DM-5, and the mean over the 100 independent realisations agree to $1\%$ at scales $k < k_{\rm Ny}$, where $k_{\rm Ny}$ is the Nyquist frequency (indicated by the vertical dashed line) given by,
\begin{equation}\label{eq:k_ny}
k_{\rm Ny} = \frac{\pi N_{\rm grid}}{L}\,,  
\end{equation}
where $N_{\rm grid}$ is the number of particles per dimension, in this case $N_{\rm grid}=270$. Even on scales smaller than the Nyquist frequency, where power is created due to non-linear clustering, there is no systematic difference, only the random fluctuations become larger and reach levels of several percent. 

The halo population is also affected by cosmic variance, since haloes form and evolve from peaks in the dark matter density fluctuations. In order to quantify the impact of the variance suppression technique on this statistic, we investigate the differential halo mass functions at the same output times used above, at $z=0$, $0.5$, $1$ and $2$, and give the result in the right panel of Fig.~\ref{fig:R_test}. The HMF shows qualitatively the same level of agreement as the power spectrum. The predicted halo population of one set of paired-and-fixed simulations is almost identical as the one obtained by averaging 100 independent realisations. The somewhat larger differences at the high-mass end are here due to the poor counting statistics in the exponential tail of the mass function that is invariably still present as a result of the still limited volume of the MTNG740 runs.

%--------- Figure --------------
\begin{figure*}
 \centering
\includegraphics[width=0.45\textwidth]{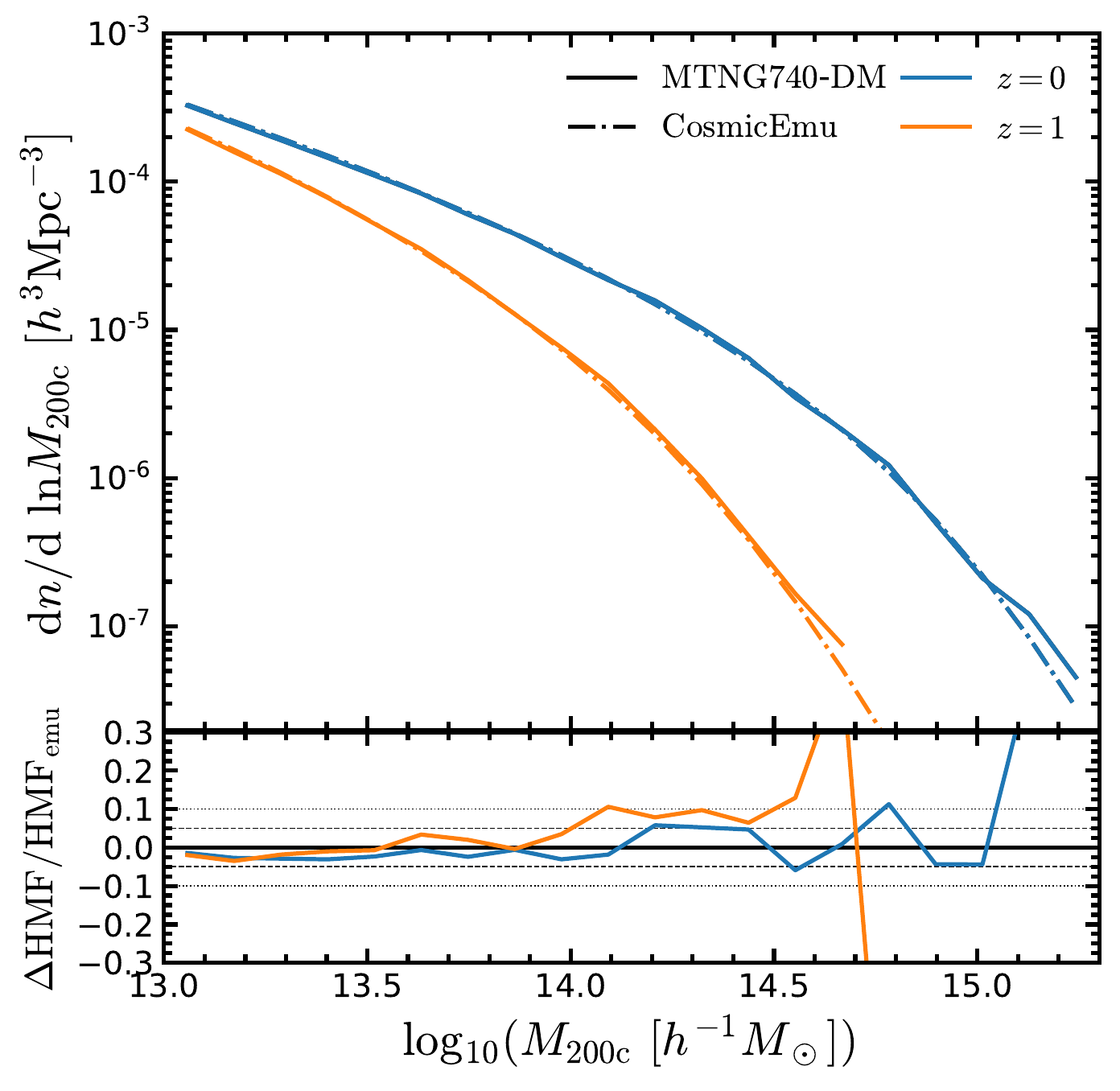}
\includegraphics[width=0.45\textwidth]{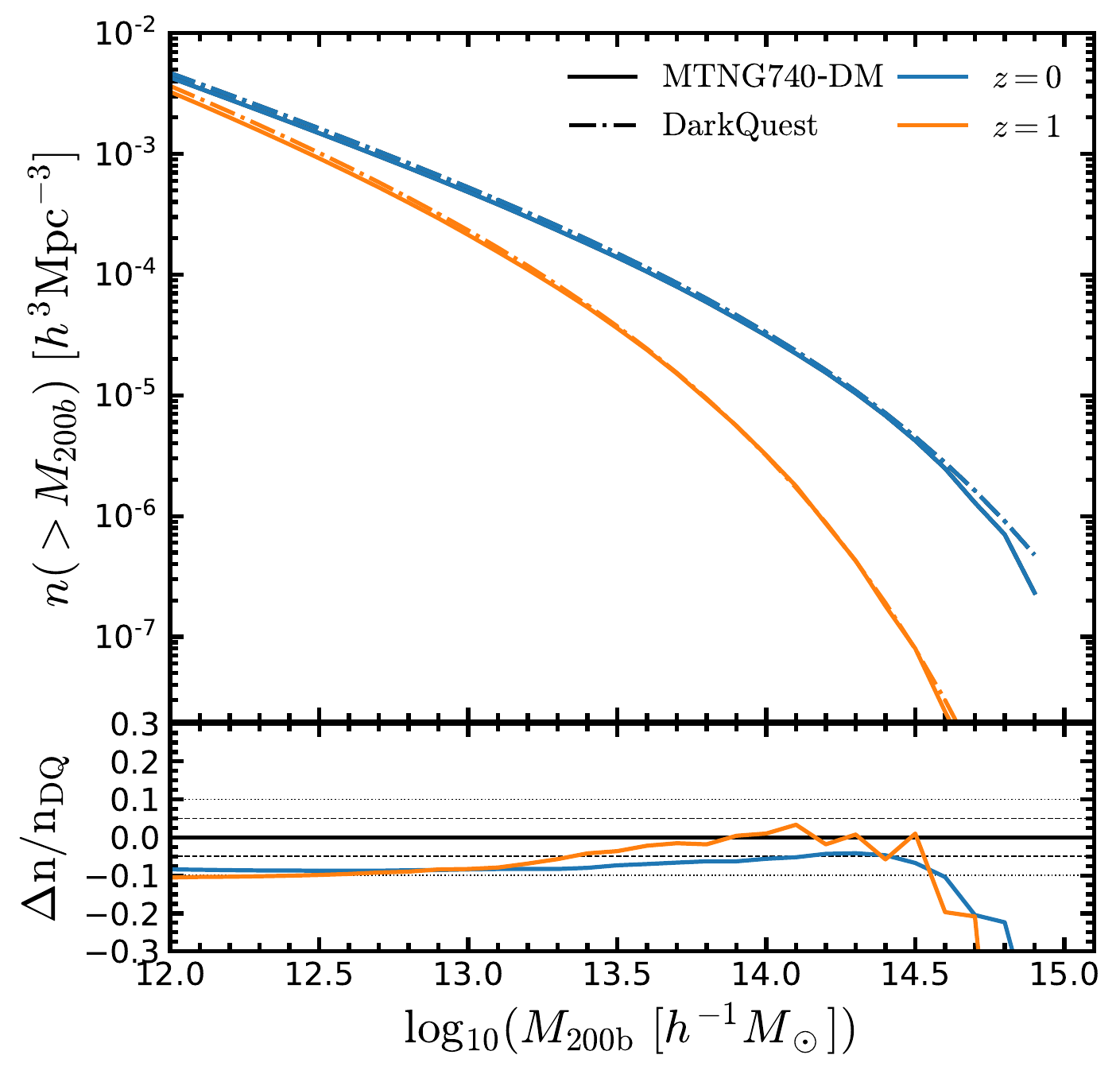}
\caption{Halo mass function comparison between the MTNG740-DM simulations and the {\sc CosmicEmu} (left panel) and the {\sc DarkQuest} (right panel) emulators at $z=0$ and $z=1$, as labelled. The lower subpanels show the relative difference between our simulations and the emulator predictions, the dashed and dotted horizontal lines indicate a $5\%$ and $10\%$ difference, respectively.}
\label{fig:hmf_emu}
\end{figure*}

%---------------------------------------------------------------
\subsection{Comparison with emulators}
\label{sec:emu}
%---------------------------------------------------------------
It is well known that cosmological simulations are computationally quite expensive, especially if one wants to push towards the large volumes and the high-resolution required by modern galaxy surveys, as we have done here with MTNG. In order to alleviate the computational cost of running such numerical calculations, researchers have developed cosmological emulators based on fits to simulation results, with the goal to use them as interpolation (or even extrapolation) tools to predict the outcome of nonlinear structure formation for a variety of cosmological models in a very rapid, but ultimately still approximative fashion \citep[see e.g.,][]{Heitmann:2006hr,Habib:2007ca}.

%--------- Figure --------------
\begin{figure*}
 \centering
\includegraphics[width=0.45\textwidth]{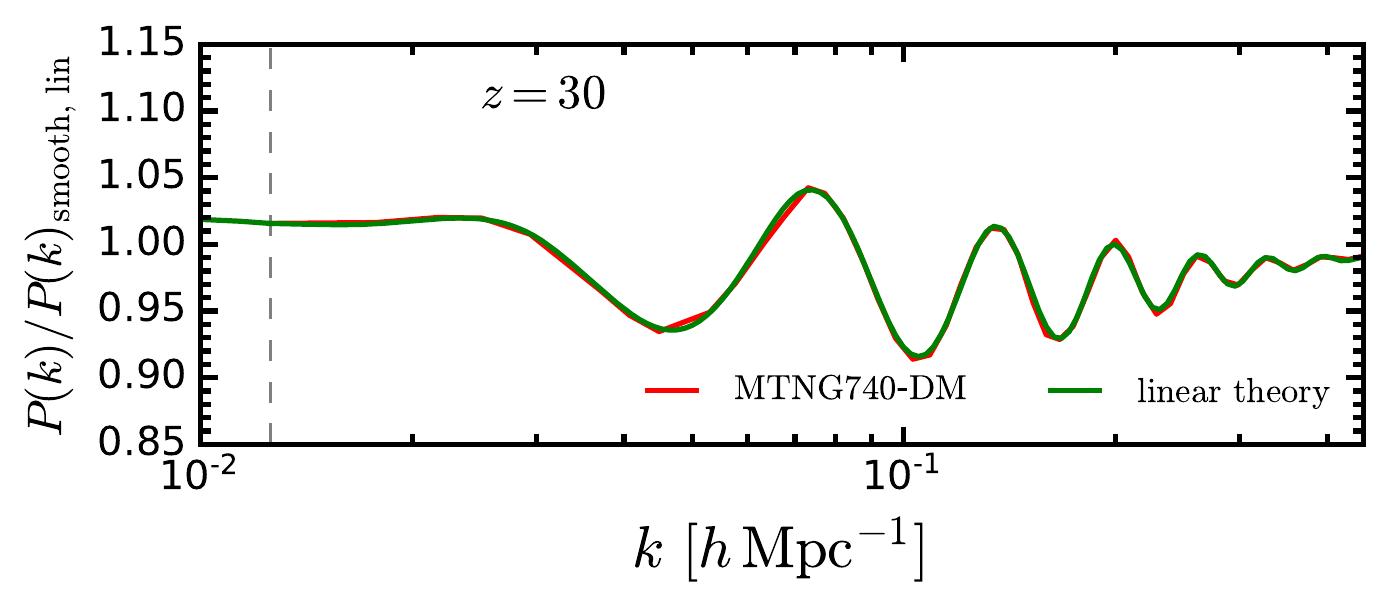}
\includegraphics[width=0.45\textwidth]{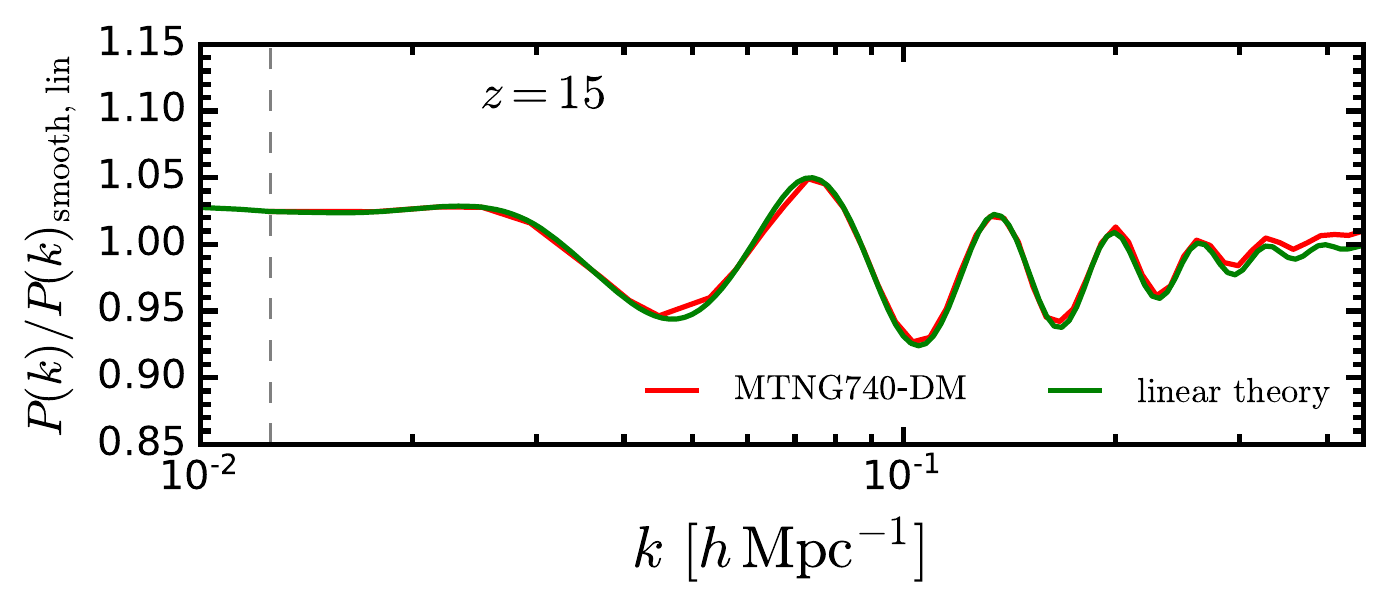}
\includegraphics[width=0.45\textwidth]{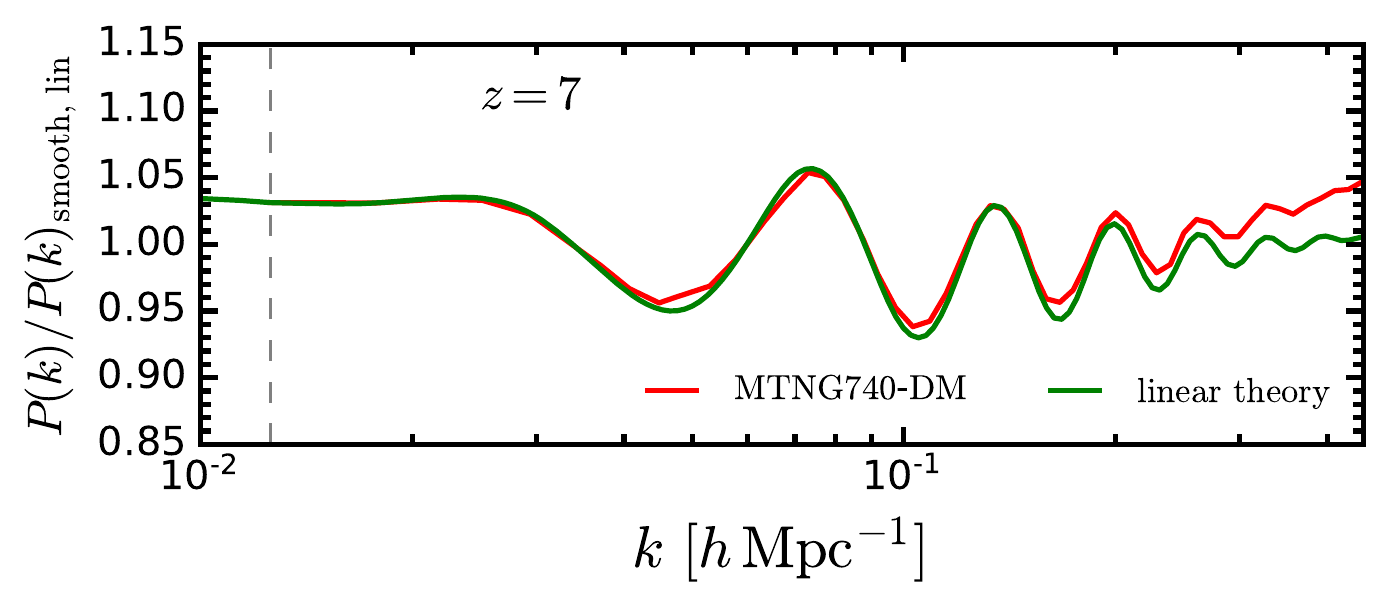}
\includegraphics[width=0.45\textwidth]{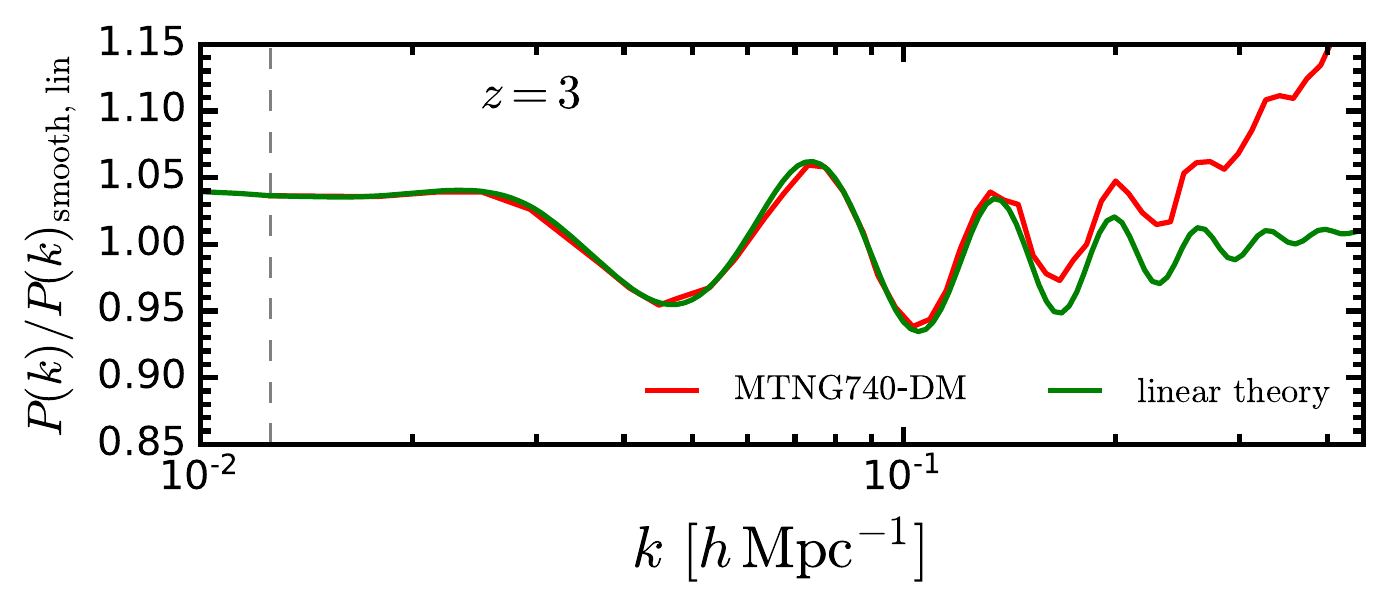}
\includegraphics[width=0.45\textwidth]{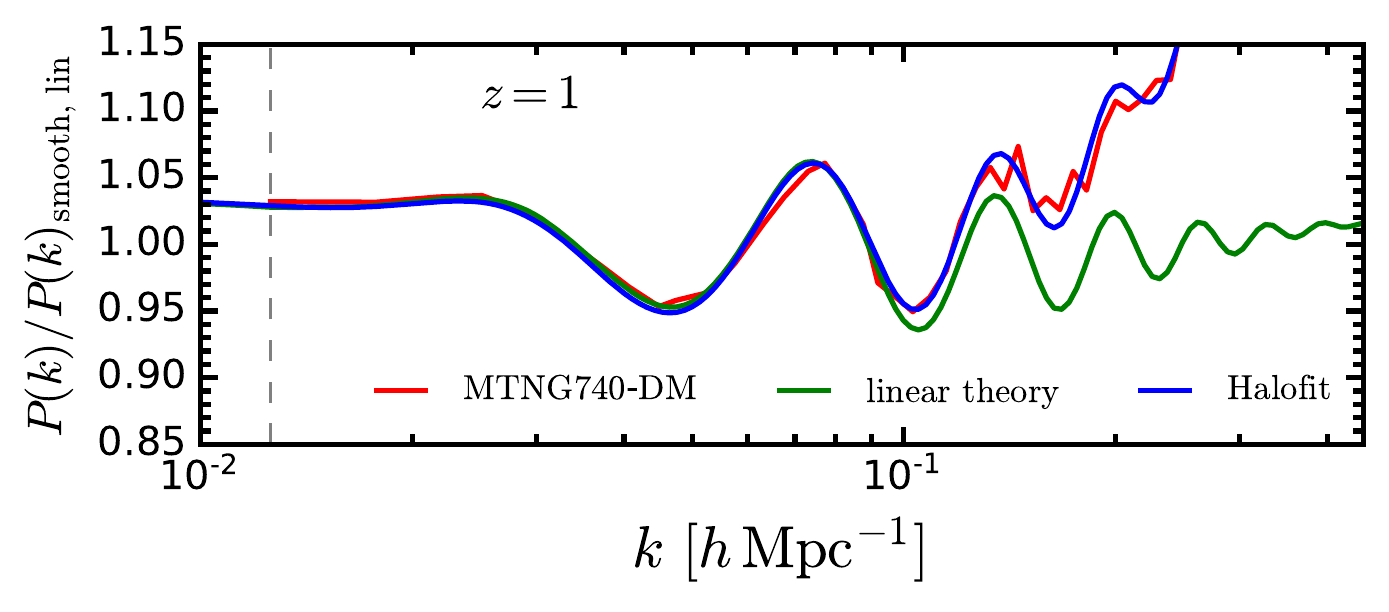}
\includegraphics[width=0.45\textwidth]{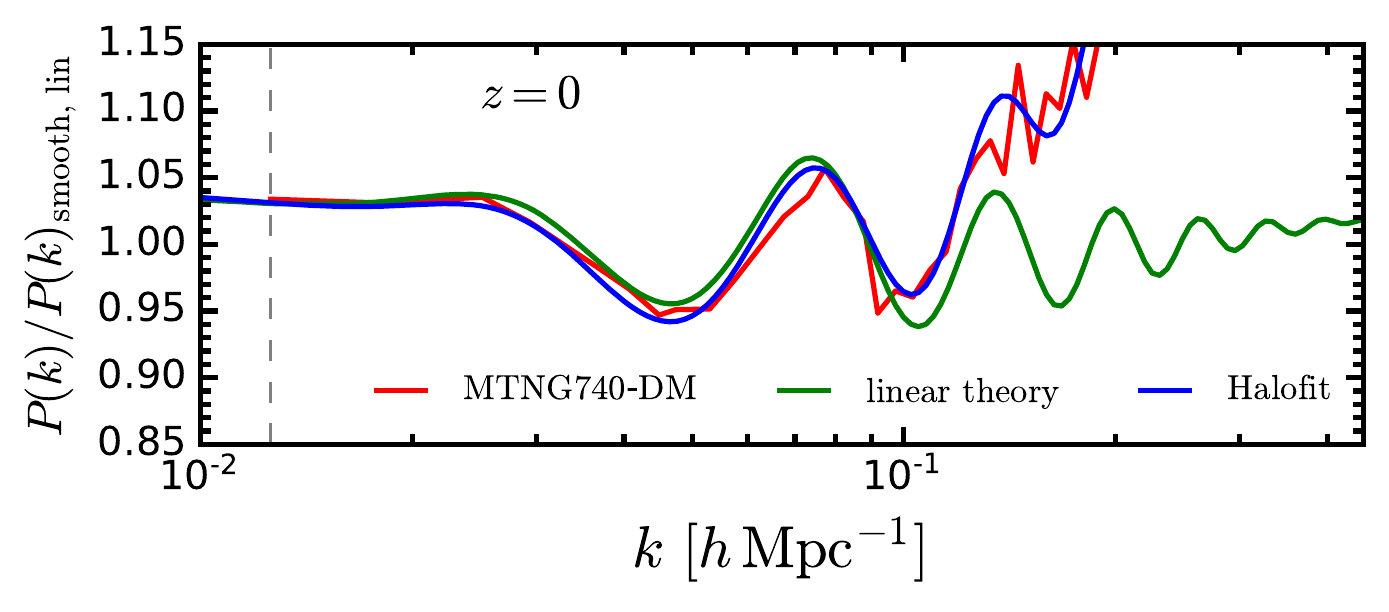}
\caption{The ratio between the measured non-linear power spectrum of the MTNG740-DM simulations and the smooth (no-wiggle) linear theory power spectrum from $z=30$ (upper left panel) to $z=0$ (lower right panel). Note that we are showing the mean over the A and B runs (red line), while the green line in each panel shows the ratio between the linear theory $P(k)$ with respect to its smooth counterpart for each selected redshift. The vertical dashed line indicates the fundamental mode of the box, $k_{\rm box}$, given by Eq.~\eqref{eq:kbox}. In addition, we show the non-linear prediction from {\sc Halofit} at redshifts $z=1$ and $z=0$.}
\label{fig:Pkm_BAO}
\end{figure*}

%--------- Figure --------------
\begin{figure*}
 \centering
\includegraphics[width=0.46\textwidth]{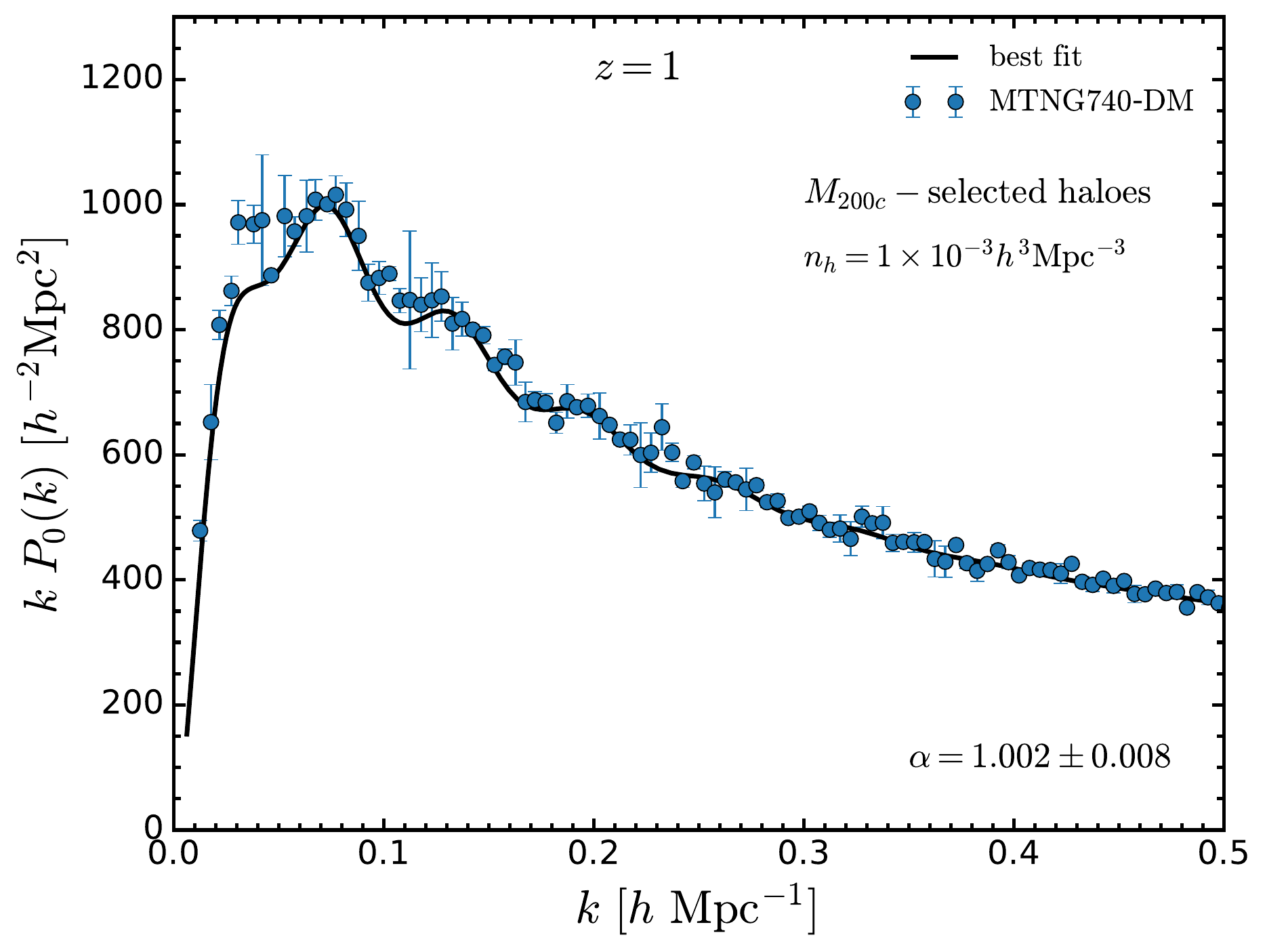}
\includegraphics[width=0.46\textwidth]{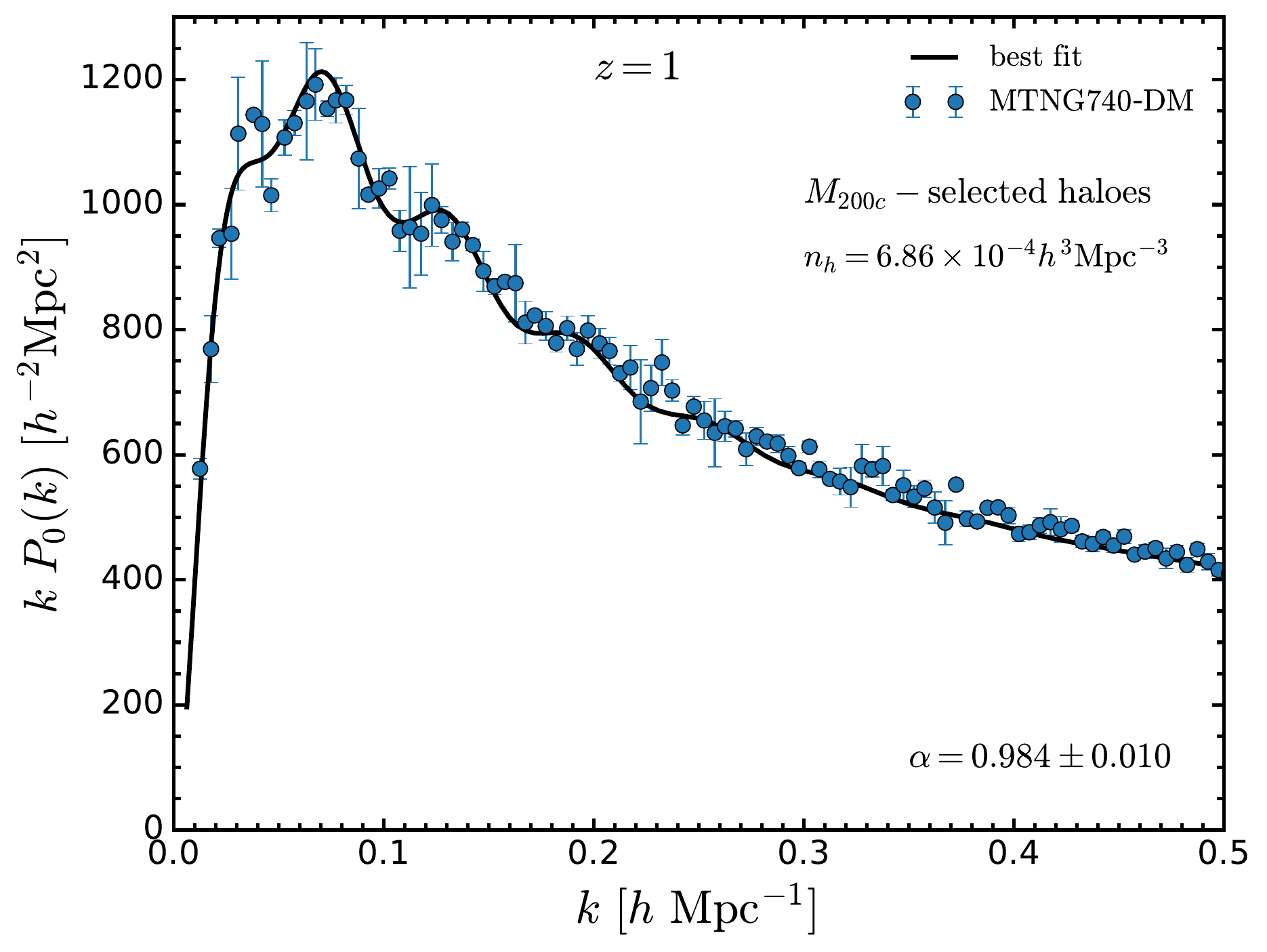}
\includegraphics[width=0.46\textwidth]{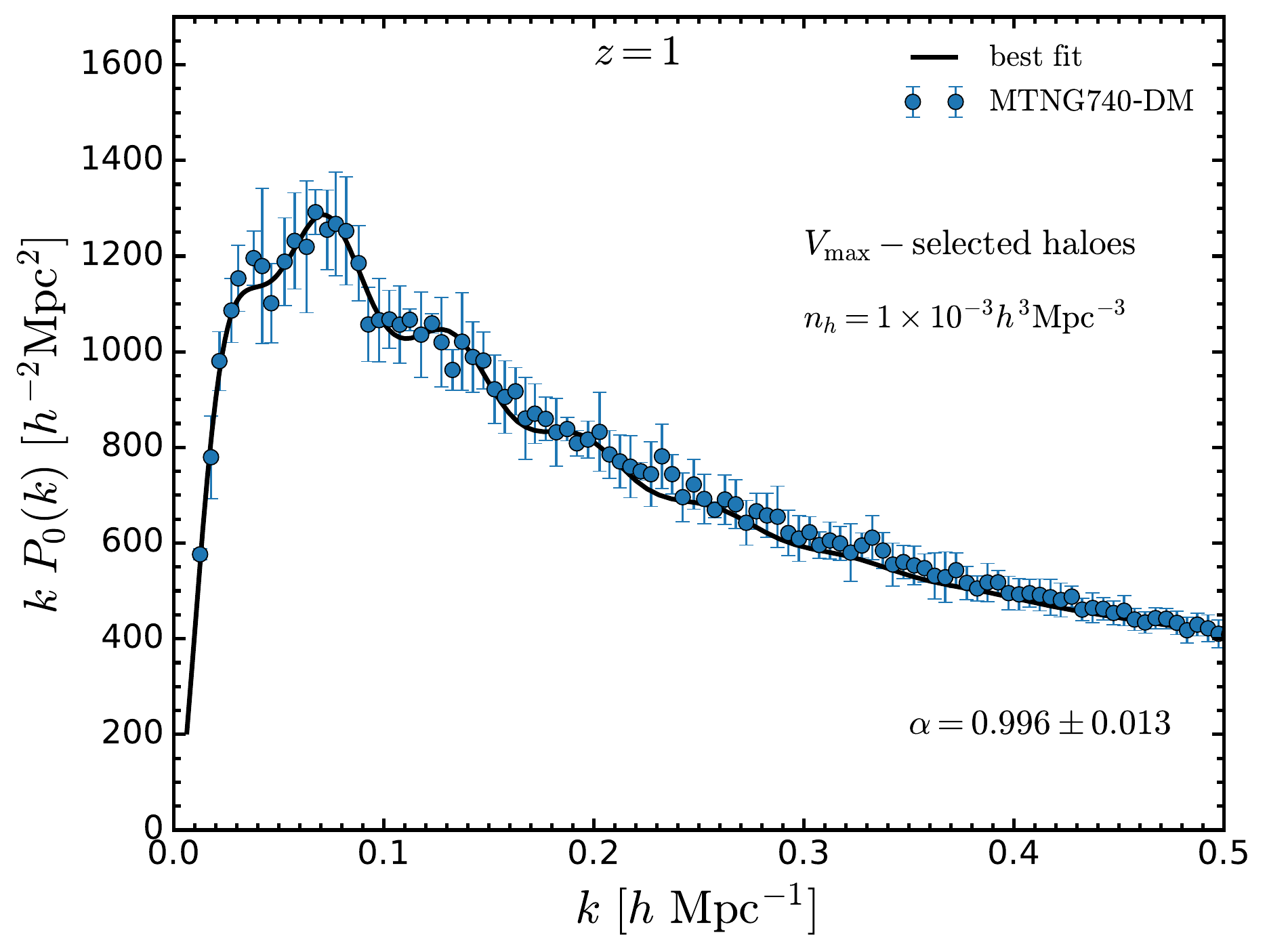}
\includegraphics[width=0.46\textwidth]{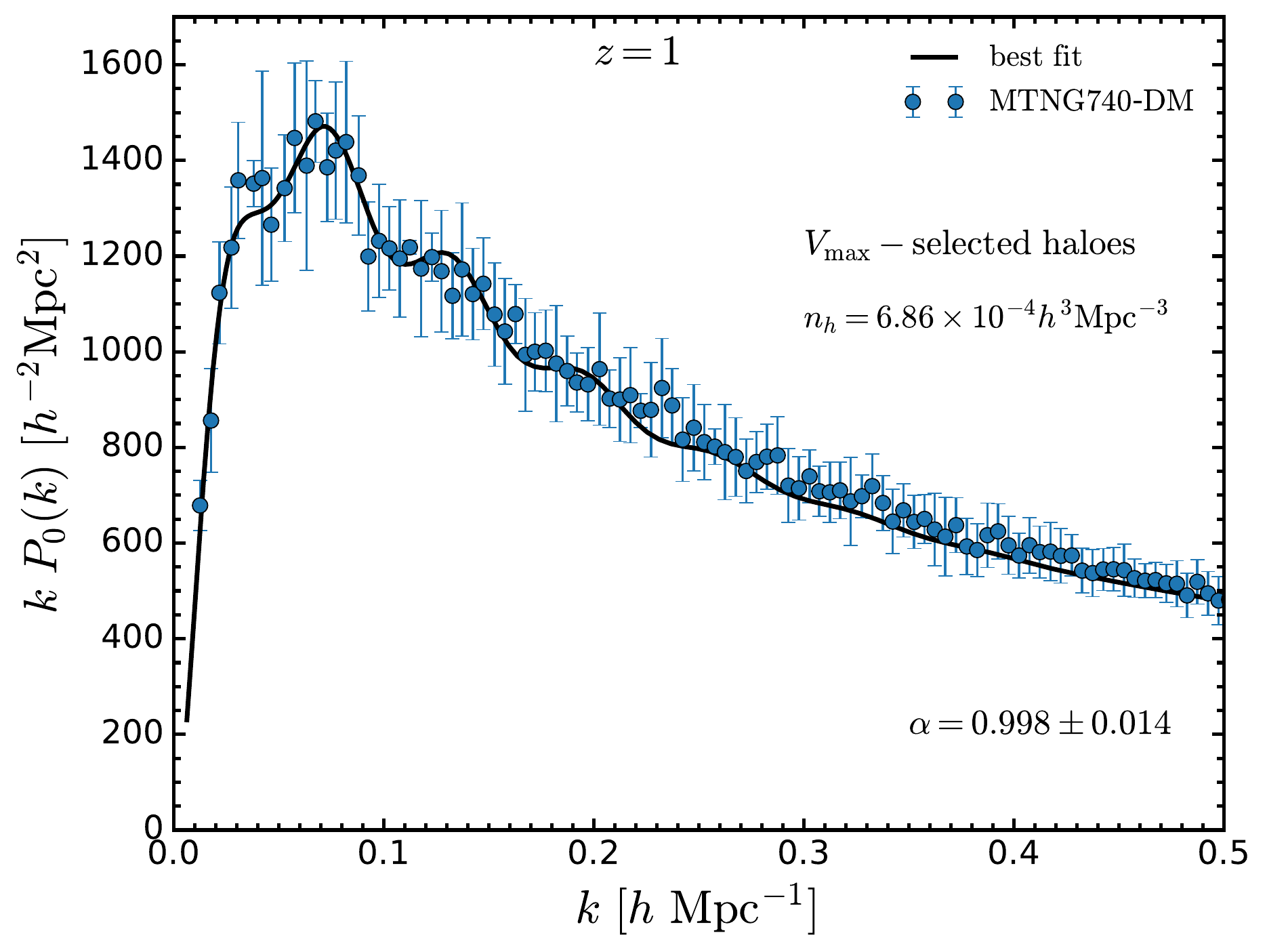}
\caption{The measured redshift-space dark matter halo power spectrum at $z=1$ (blue dots with error bars) for two different $M_{200c}-$ (upper panels) and $V_{\rm max}-$selected (lower panels) halo samples with number densities $n_h = 1\times 10^{-3}\hMpcc$ (left panels) and $n_h = 6.86\times 10^{-4}\hMpcc$ (right panels), as well as the best-fitting model (black solid line). The error bars correspond to the standard deviation over six $P_0(k)$ measurements.}
\label{fig:BAO_halo}
\end{figure*}

%--------- Figure --------------
\begin{figure*}
 \centering
\includegraphics[width=0.45\textwidth]{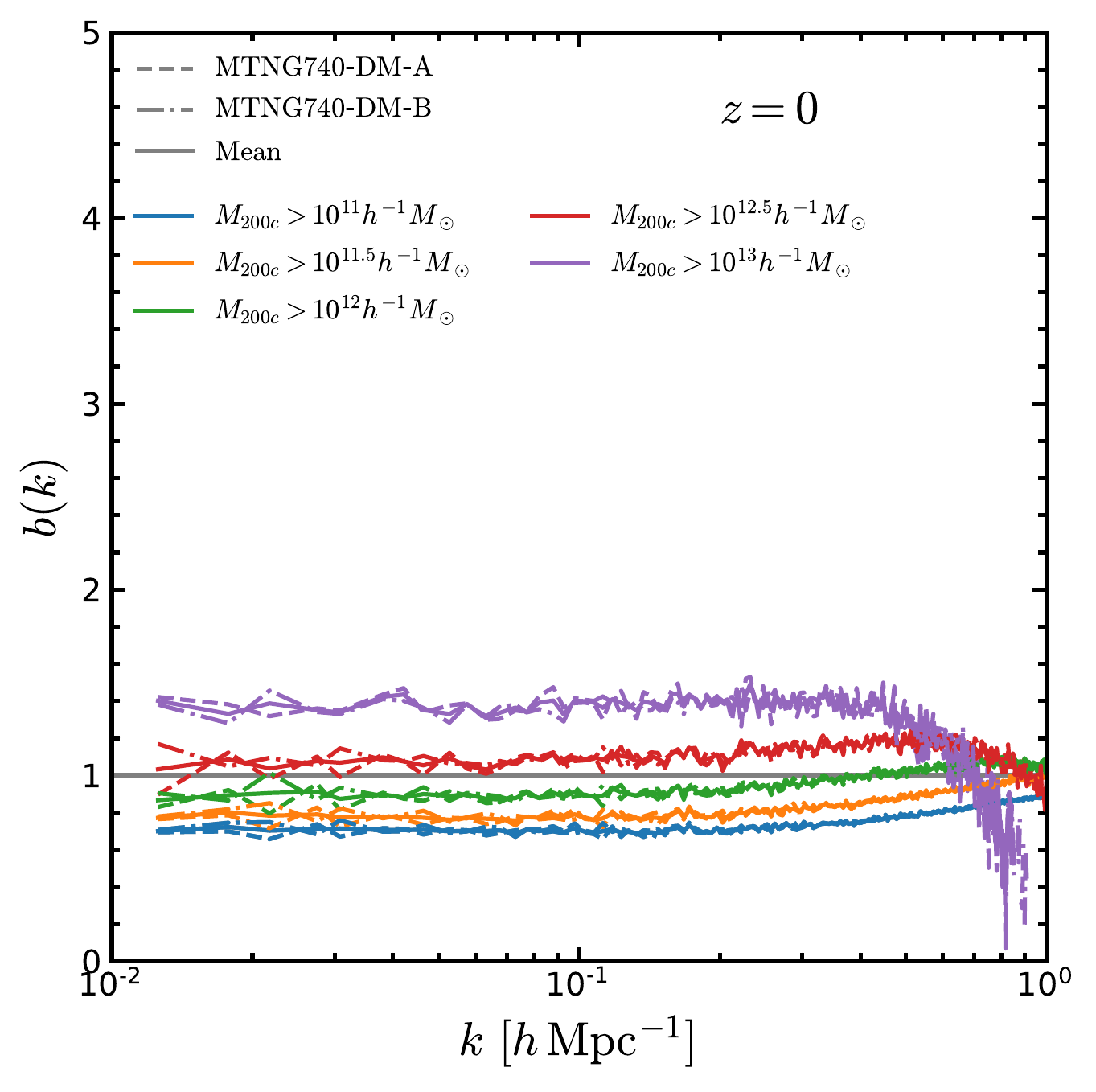}
\includegraphics[width=0.45\textwidth]{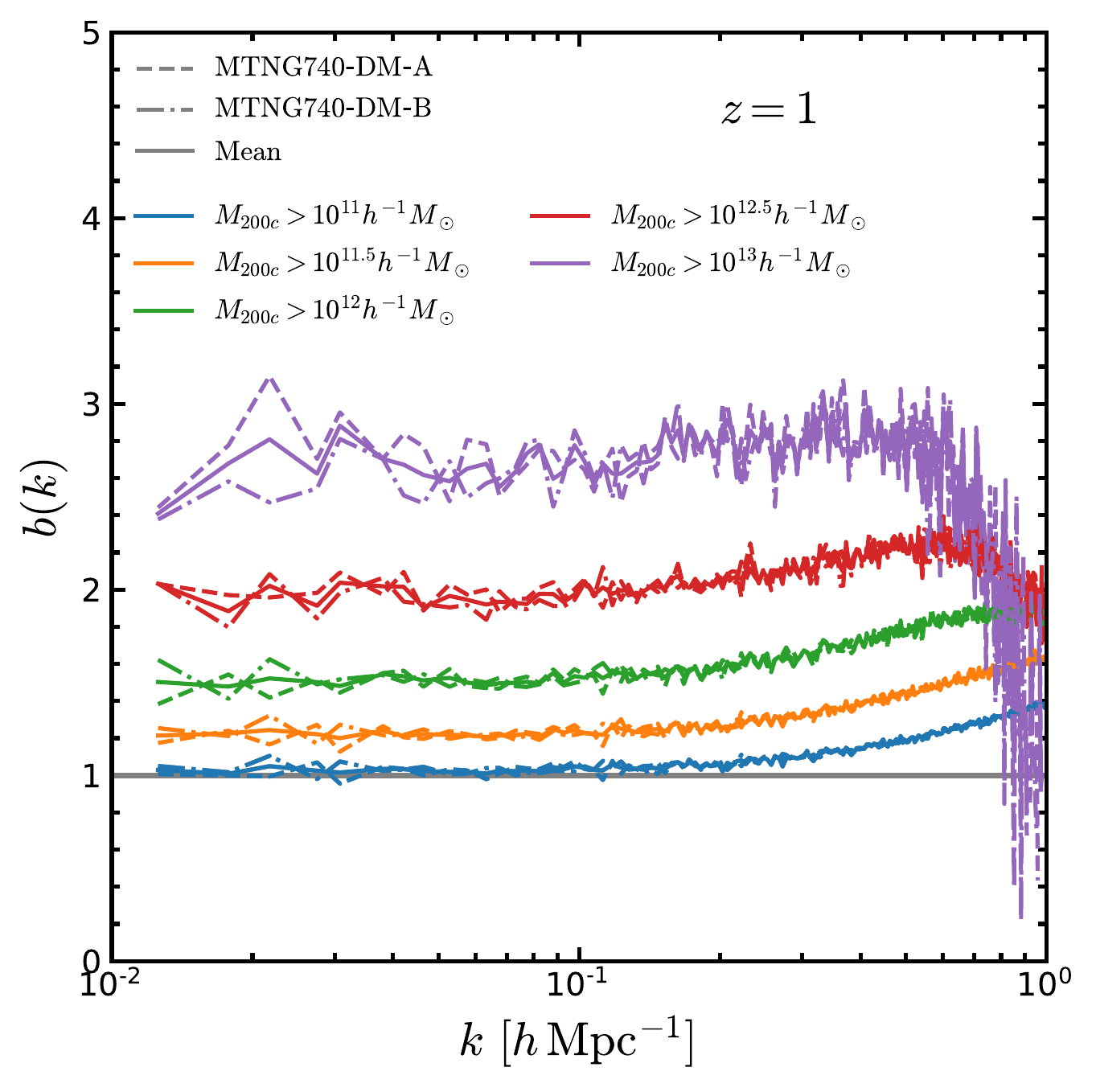}
\includegraphics[width=0.45\textwidth]{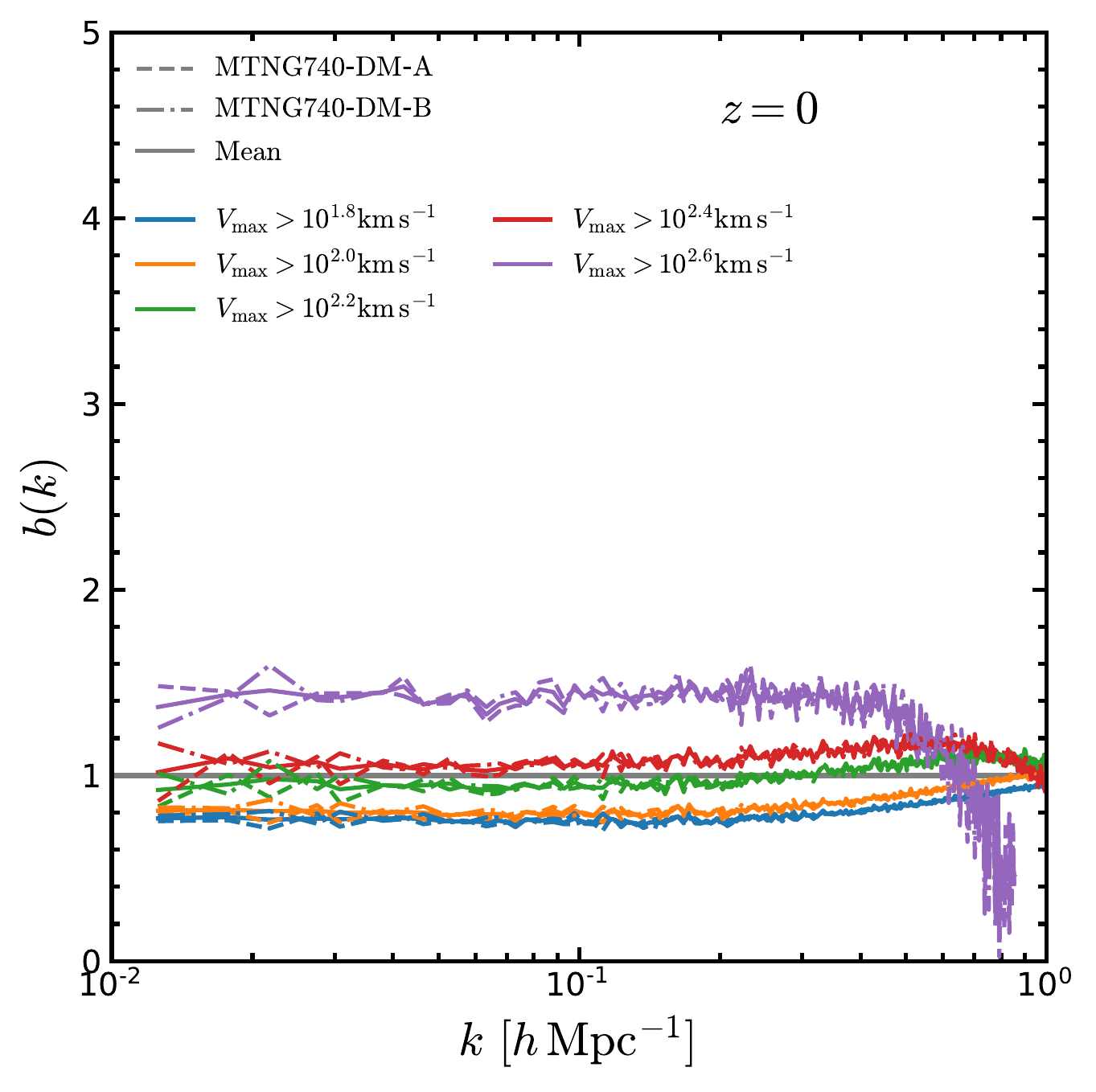}
\includegraphics[width=0.45\textwidth]{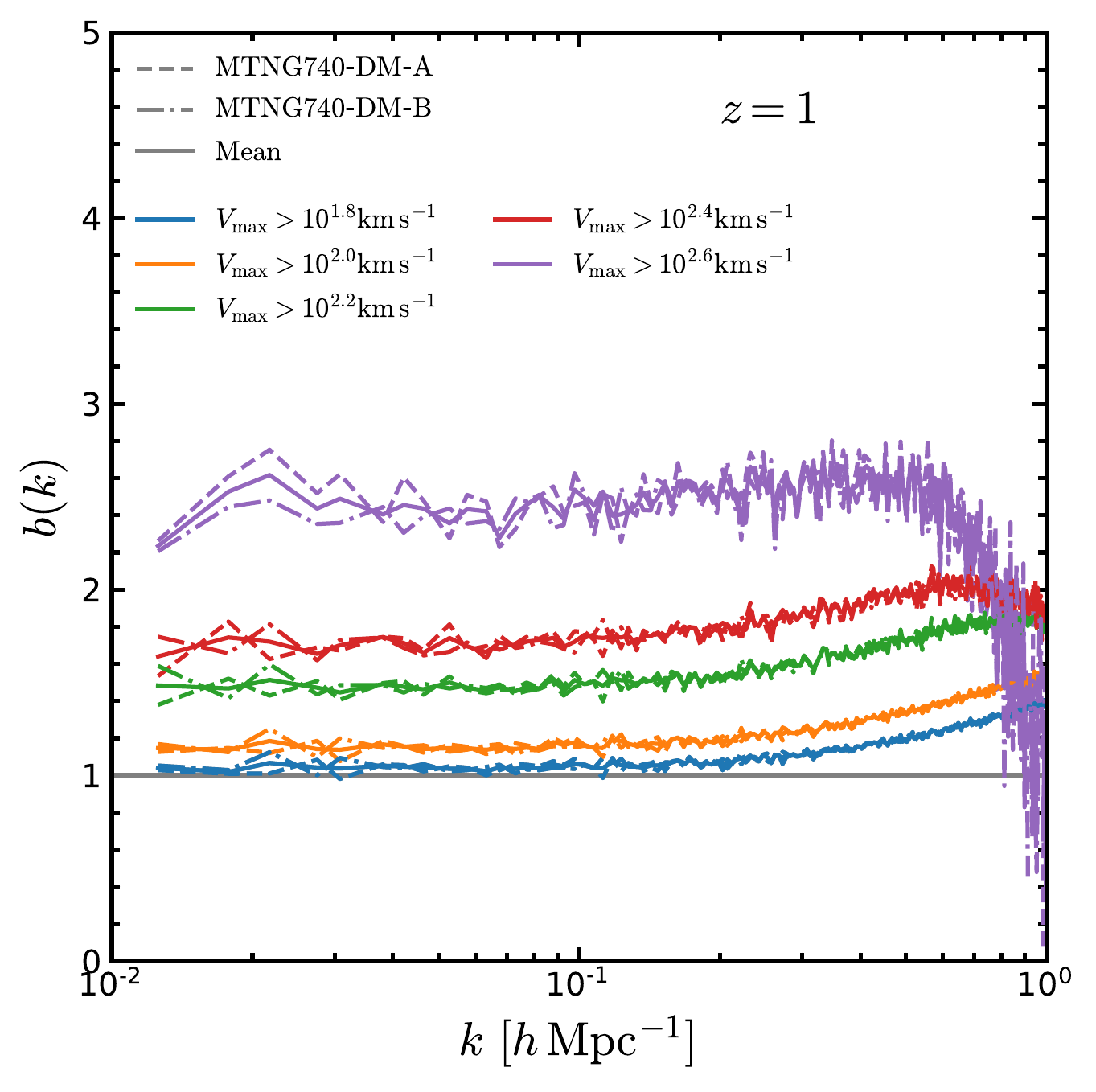}
\caption{Scale dependent halo bias, $b(k)=\sqrt{P_{\rm h}(k)/P_{\rm m}(k)}$, for $M_{200c}-$selected and $V_{\rm max}-$selected haloes at $z=0$ (left panel) and $z=1$ (left panel), for the MTNG740-DM-1 simulations. Different line-styles represent different types of simulations, as labelled.}
\label{fig:halo_bias}
\end{figure*}

%--------- Figure --------------
\begin{figure*}
 \centering
\includegraphics[width=0.45\textwidth]{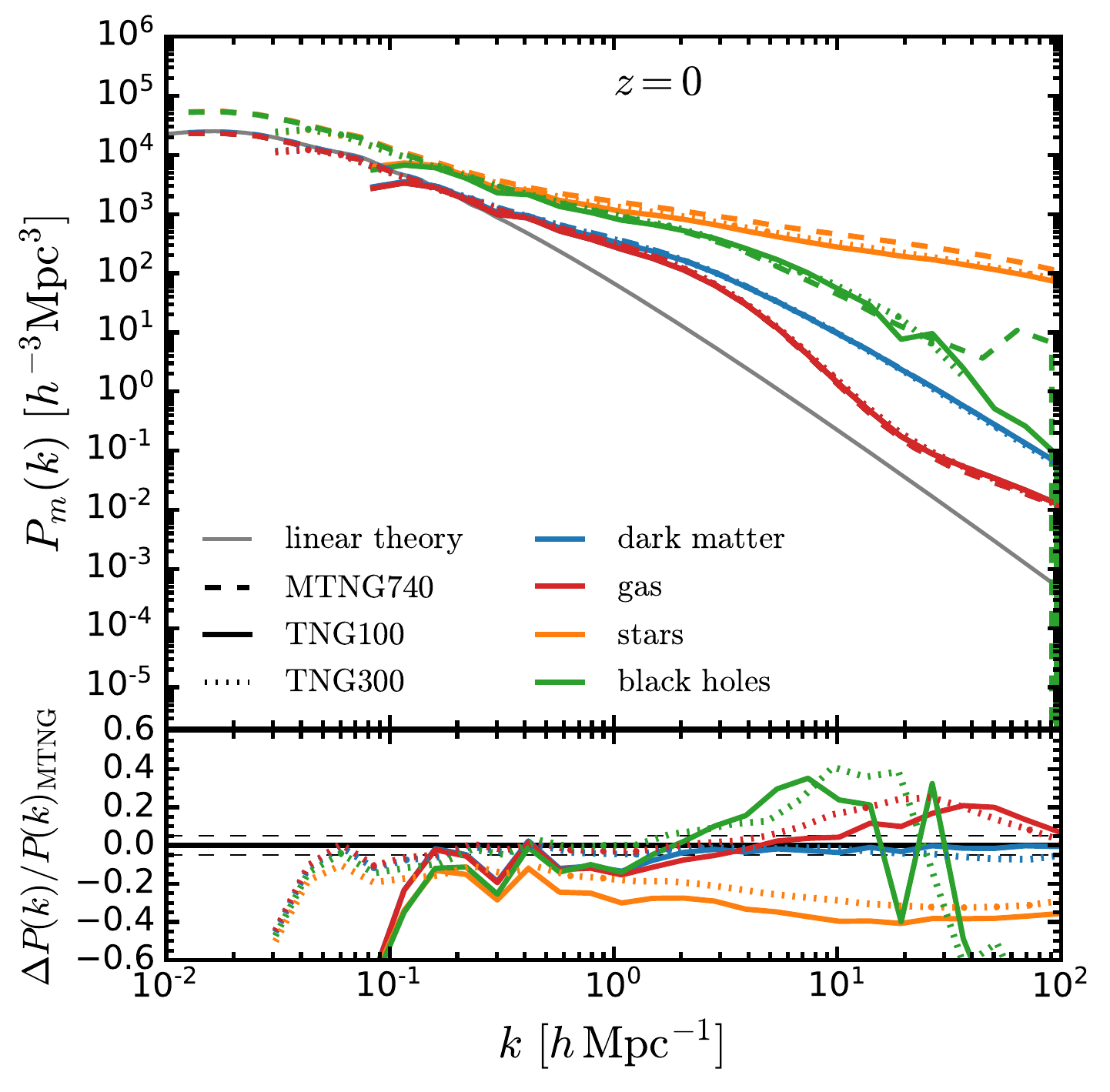}
\includegraphics[width=0.45\textwidth]{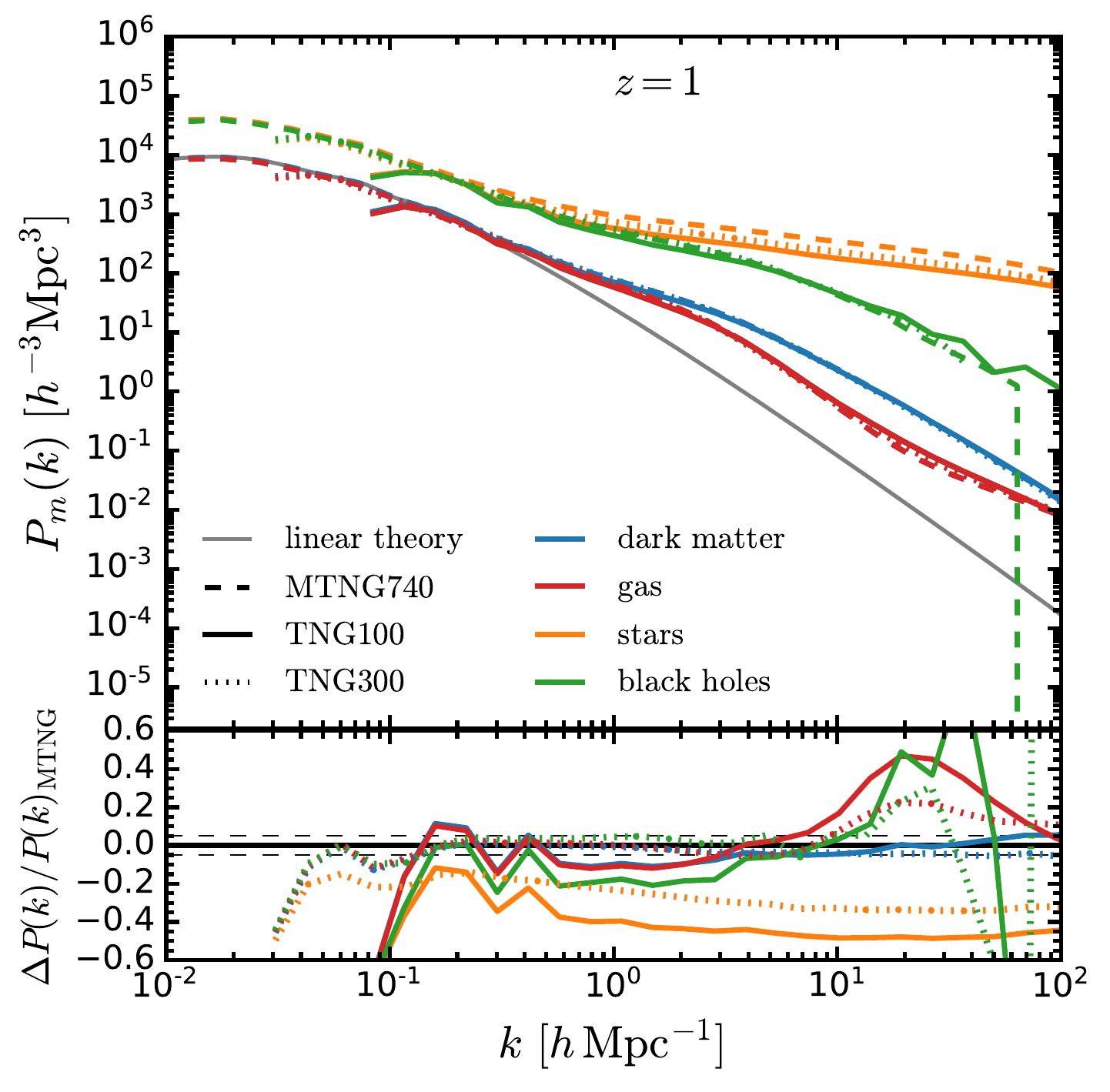}
\caption{Matter power spectra for different mass components of the MTNG740 hydrodynamic simulation (dashed lines) at redshifts $z=0$ (right panel) and $z=1$ (left panel). We include the measurement of the TNG300 (dotted lines) and TNG100 (solid lines) simulations, for comparison. The lower subpanels display the relative difference between the TNG300/TNG100 and MTNG740 measurements.}
\label{fig:Pk_matter}
\end{figure*}

%--------- Figure --------------
\begin{figure}
 \centering
\includegraphics[width=0.5\textwidth]{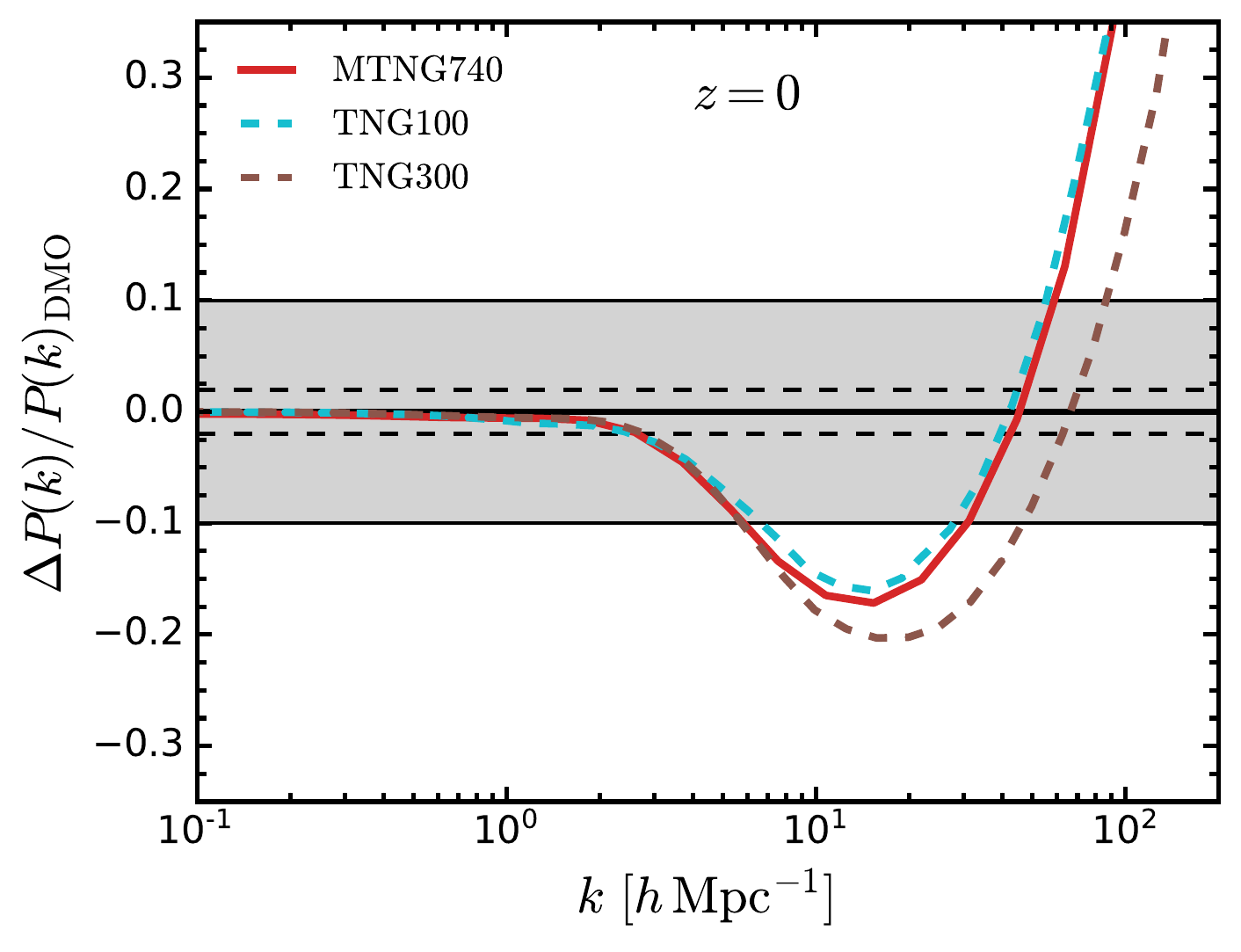}
\caption{Impact of baryonic physics on the total matter spectrum at $z=0$ measured from the MTNG740 runs (red solid line) in comparison with the TNG100 (cyan dashed line) and TNG300 (brown dashed line) simulations. The dashed horizontal lines and the grey shaded region indicate a relative difference of 2 and 10 per cent, respectively.}
\label{fig:Pkm_comp}
\end{figure}

%--------- Figure --------------
\begin{figure}
 \centering
\includegraphics[width=0.5\textwidth]{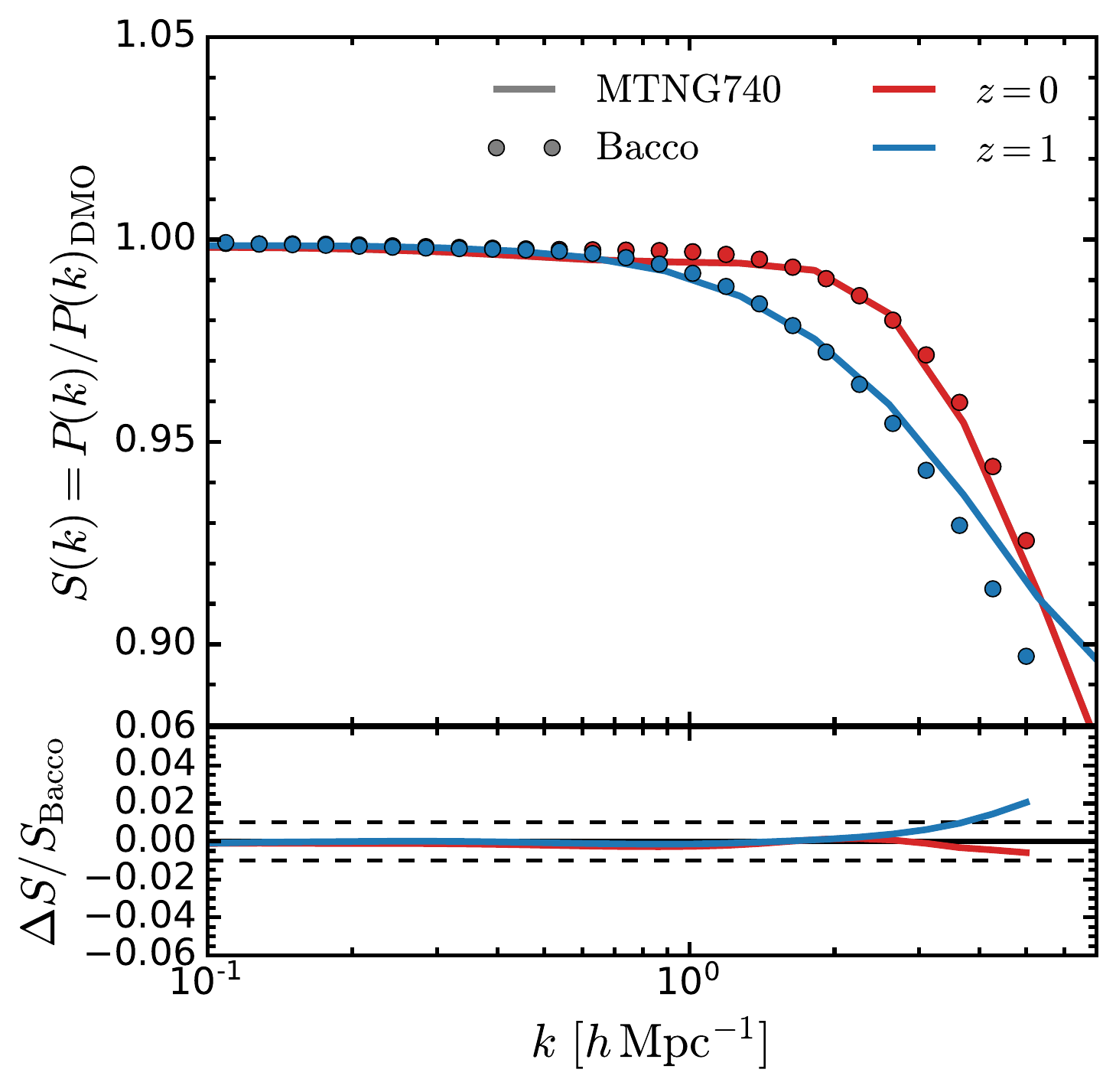}
\caption{Suppression of the matter power spectrum due to baryonic physics in the MTNG740 hydrodynamic simulation relative to MTNG740-DM-1 (solid lines) and the predictions from the {\sc Bacco} emulator (symbols) at $z=0$ and $z=1$. The lower subpanel shows the relative difference between the simulation and the emulator.}
\label{fig:Pkm_bacco}
\end{figure}

%--------- Figure --------------
\begin{figure*}
 \centering
\includegraphics[width=0.46\textwidth]{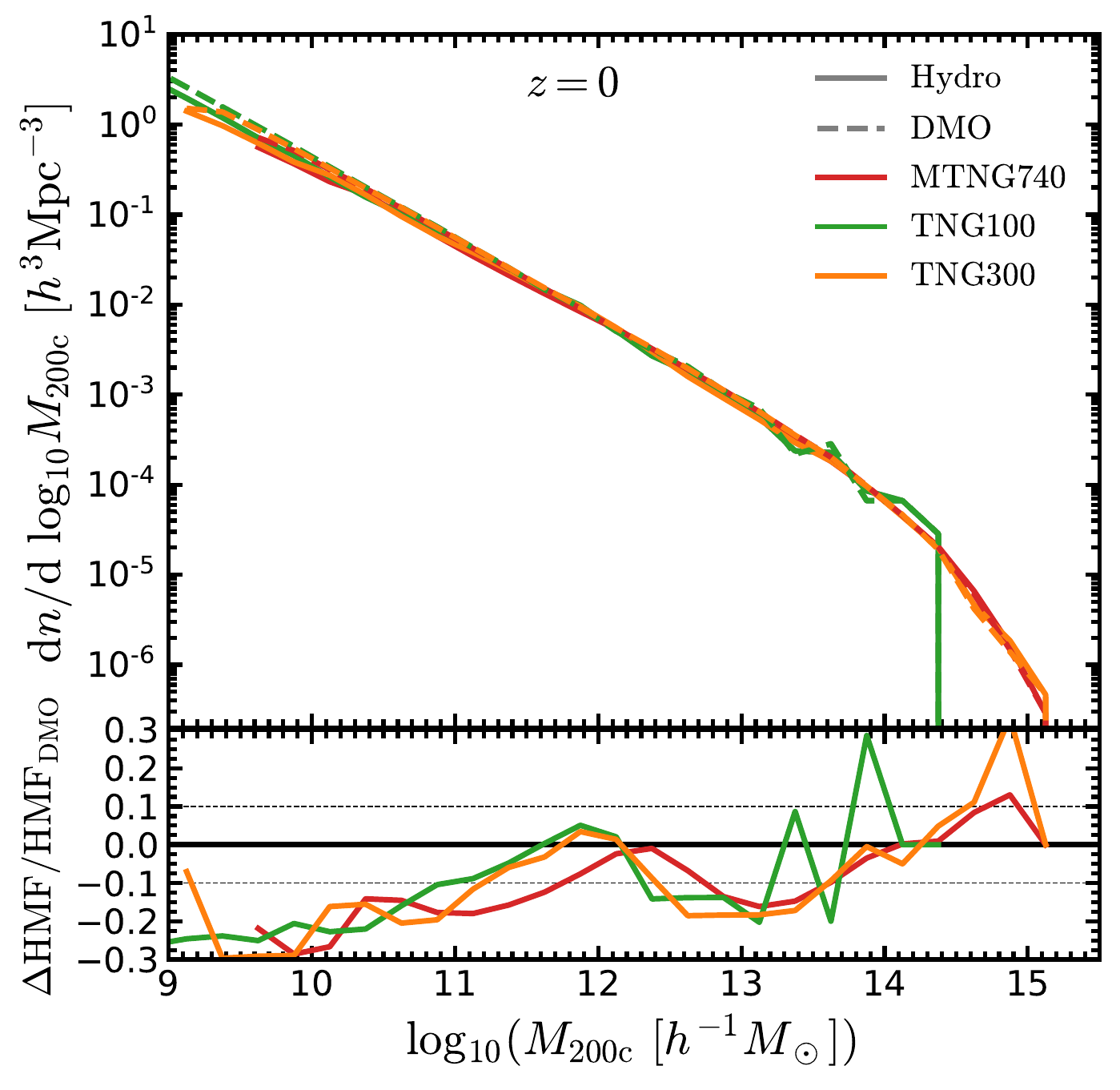}
\includegraphics[width=0.465\textwidth]{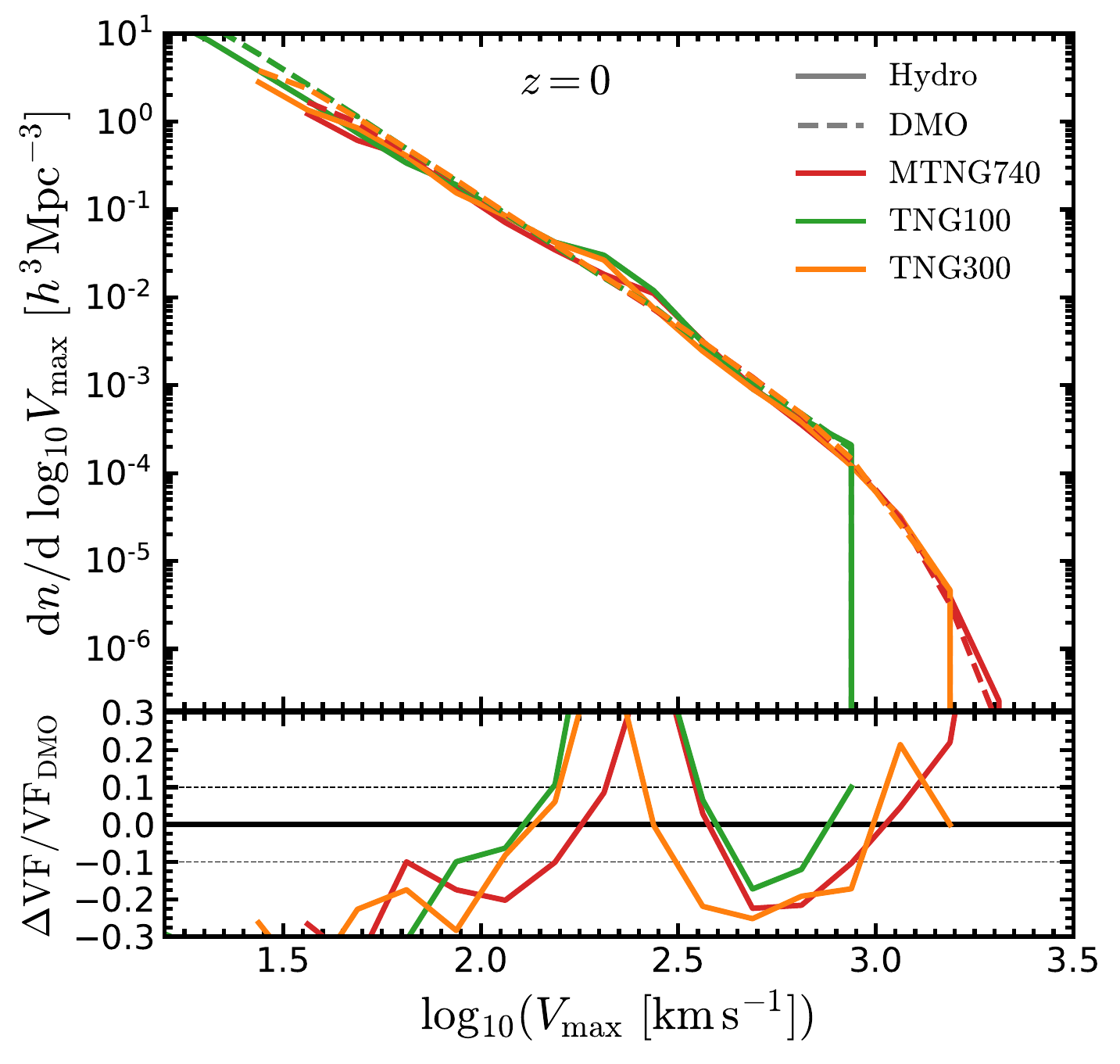}
\caption{The differential halo mass functions as a function of halo mass ($M_{200c}$; left panel), and the main subhalo $V_{\rm max}$ functions (right panel) measured from the MTNG740 hydro and the MTNG740-DM-1 simulations at $z=0$, as labelled. The lower subpanels show the ratio between the hydro halo mass and velocity functions with respect to their dark matter-only versions. We also show the mean of the A and B realisations as a red line.}
\label{fig:dHMF_hydro}
\end{figure*}

%--------- Figure --------------
\begin{figure*}
 \centering
\includegraphics[width=0.47\textwidth]{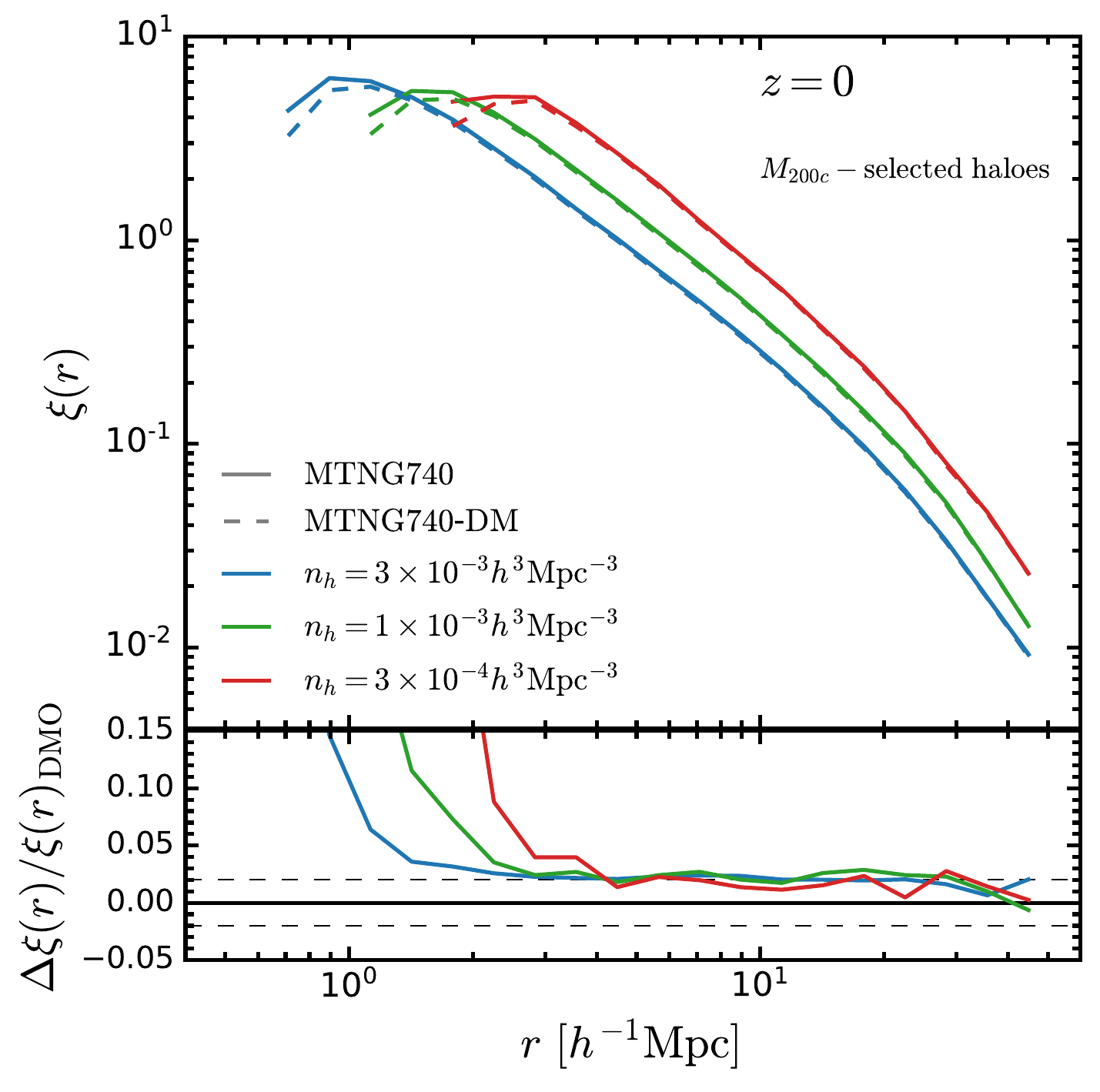}
\includegraphics[width=0.46\textwidth]{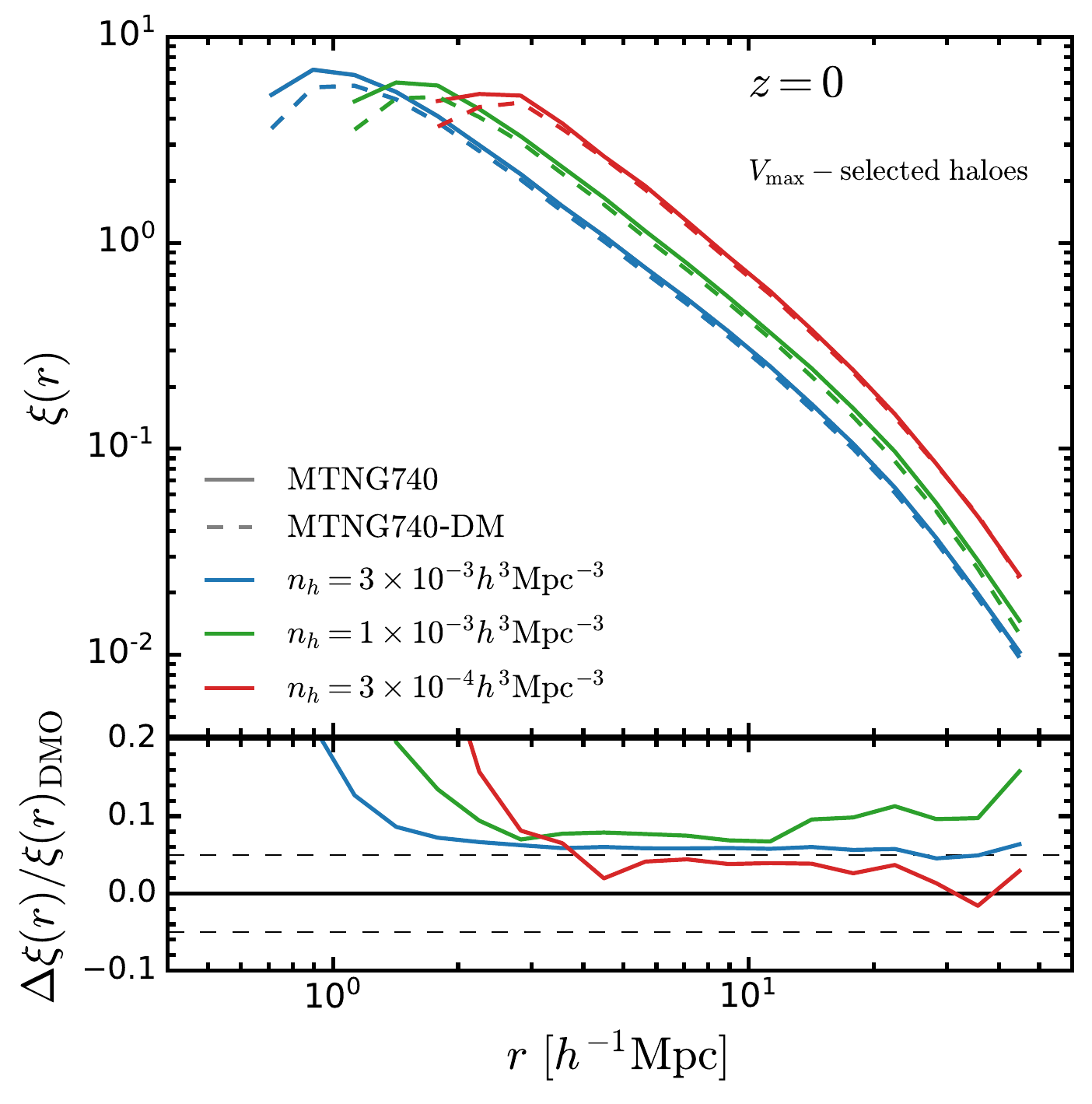}
\includegraphics[width=0.47\textwidth]{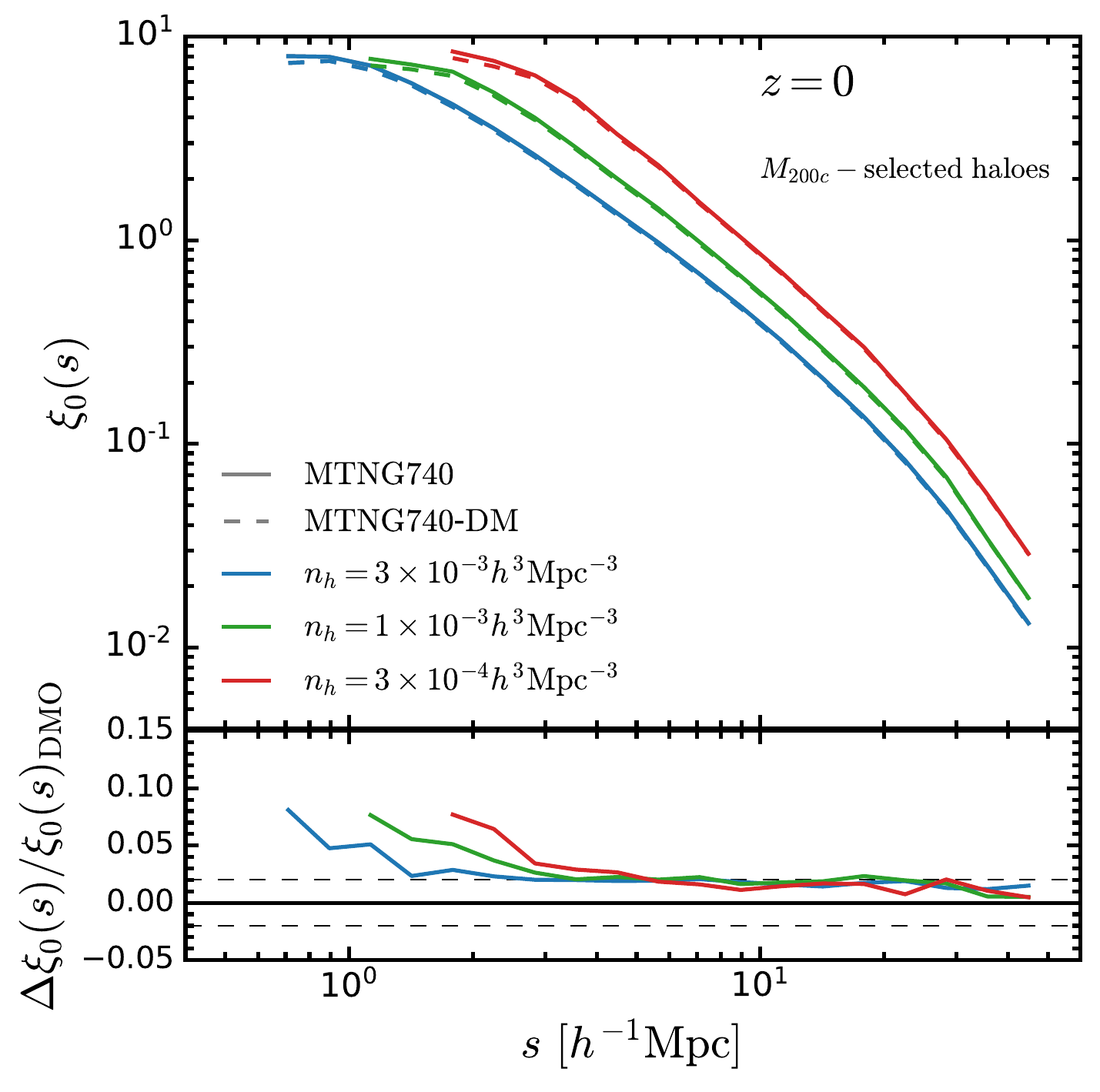}
\includegraphics[width=0.47\textwidth]{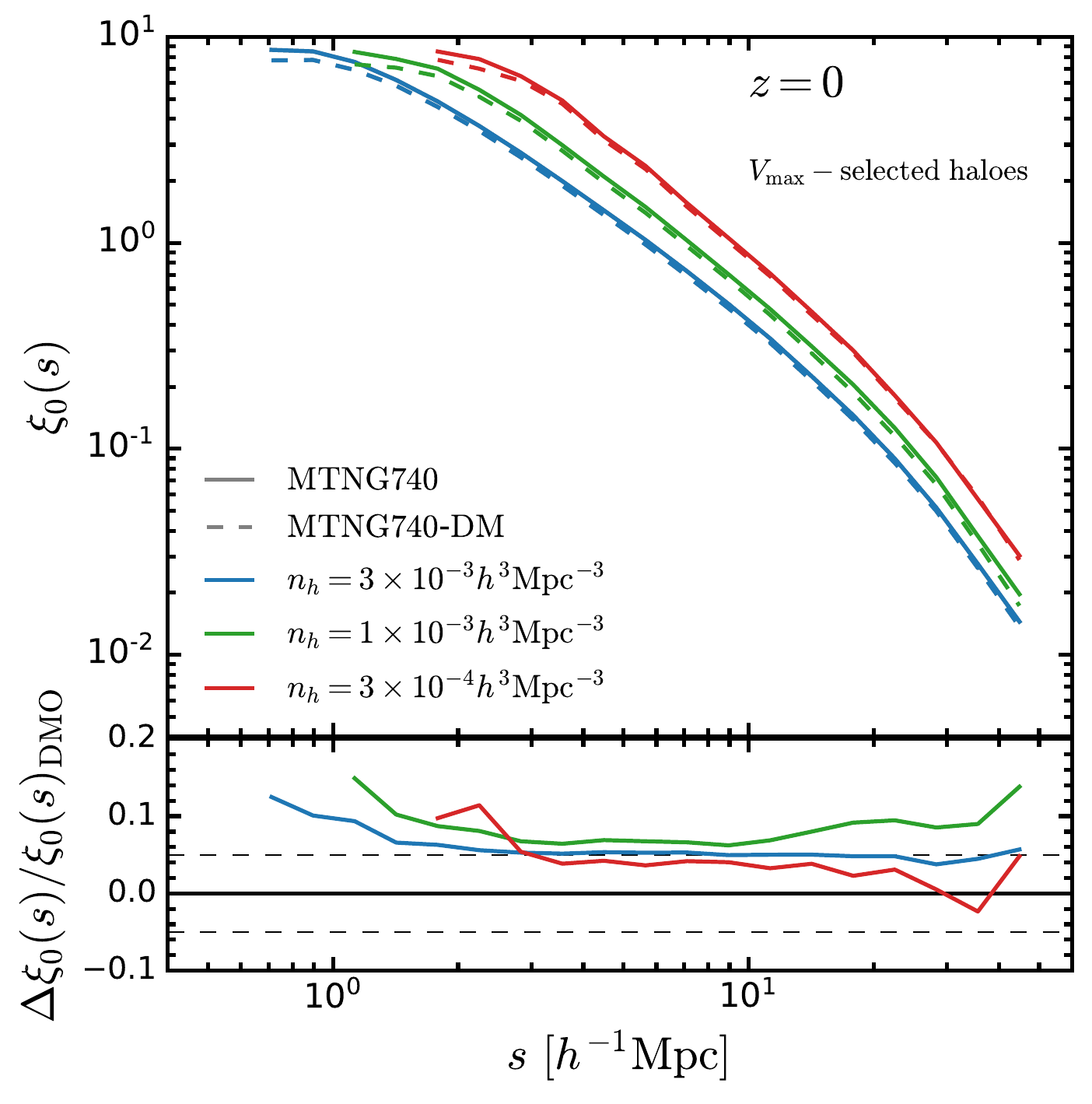}
\caption{The halo two-point correlation function in real-space (upper panels) and redshift-space (lower panels) for three different halo number density cuts, $n_h = 3\times 10^{-3}\hMpcc$ (blue lines), $n_h = 1\times 10^{-3}\hMpcc$ (green lines) and $n_h = 3\times 10^{-4}\hMpcc$ (red lines), at redshift $z=0$ for $M_{200c}-$ and $V_{\rm max}-$selected haloes, as labelled. The lower subpanels show the relative difference between the MTNG740 hydro and the MTNG740-DM measurements.}
\label{fig:xi_halo}
\end{figure*}

Here, we assess the accuracy of cosmological emulators for the non-linear matter power spectrum and the halo mass function when compared with our MTNG simulations. On the one hand, we use predictions of the matter power spectrum from the {\sc CosmicEmu} \citep{Pk:emu_new}, {\sc Bacco} \citep{bacco:emu} and {\sc EuclidEmulator2} \citep{EE2:emu} emulators. All of these emulators were calibrated using a large number of $N$-body simulations $(\mathcal{O}(100))$ with different cosmological parameters\footnote{We refer the reader to the original publications about the emulators for details about the adopted cosmological models and simulations.}.

Instead of comparing the full-shape of the power spectrum, we use the non-linear boost, which is the ratio between the non-linear matter clustering and its linear theory prediction,
\begin{equation}\label{eq:B_Pk}
    B(k) = \frac{P_m(k)}{P_{\rm lin}(k)}\,.
\end{equation}
The comparison of the boost factor, $B(k)$, at $z=0$ and $z=1$, is shown in the left and right panel of Fig.~\ref{fig:Pkm_emu}, respectively. At the present time ($z=0$; lower-left panel), the predictions of the MTNG740-DM simulations and the emulators are consistent within  $1\%$ accuracy for $k>0.2\hMpc$, with  some fluctuations seen at the large-scale modes, but the differences are well within the $5\%$ limit. At early times ($z=1$; lower-right panel), we find slightly larger differences than $1\%$ at the high-$k$ modes, but this is consistent with the claimed accuracy of the emulators.

Making fast predictions for the halo mass function is an important goal for emulators as well. In the left panel of Fig.~\ref{fig:hmf_emu}, we compare the differential halo mass function by \cite{HMF:cosmicEmu} (which is a part of the {\sc CosmicEmu} project) and our MTNG results at $z=0$ (blue lines) and $z=1$ (orange lines). The HMF {\sc CosmicEmu} was built using the $M_{200c}$ halo mass definition and  is limited to the mass range $10^{13} < M_{200c}/[\Msh] < 10^{15.5}$ at $z=0$. The {\sc CosmicEmu} haloes were identified with the FoF algorithm with a linking length parameter of $b=0.168$, and a spherical overdensity ($M_{200c}$) halo catalogue was built using the potential minimum of each FoF group as halo centre, in a similar way to the {\sc Subfind} code. We find an agreement better than $5\%$ for haloes with mass $M_{200c} < 10^{15}\Msh$ ($M_{200c} < 10^{14}\Msh$) at $z=0$ ($z=1$). There are larger differences of $\sim 10\%$ at $z=1$ at the high-mass end.

We also make use of the {\sc DarkQuest} emulator \citep{DarkQuest:emu} to test the accuracy of the HMF. This HMF emulator was calibrated using the $M_{200b}$ halo mass definition, which corresponds to the mass within a sphere with overdensity 200 times the mean background density of the Universe. Note that {\sc DarkQuest} was designed to predict the comoving number density of haloes at a given halo mass; i.e., the cumulative halo mass function, $n(>M_{200b})$, for haloes with masses $M_{200b} > 10^{12}\Msh$. Also, \citet{DarkQuest:emu} applied a correction to the halo mass given by,
\begin{equation}
    M_{\rm 200b}= (1 - N^{-0.55}_p)M^{\rm sim}_{\rm 200b}\,,
\end{equation}
where $N_p$ is the number of particles of a given halo, and $M^{\rm sim}_{\rm 200b}$ is the spherical overdensity mass definition obtained from the halo catalogue. We therefore apply the same correction to our $M_{\rm 200b}$ haloes to have a fair comparison. 

The measured HMFs from our halo catalogues at $z=0$ and $z=1$ are compared against the {\sc DarkQuest} predictions in the right panel of Fig.~\ref{fig:hmf_emu}. In this case, we find a disagreement of $10\%$ for haloes less massive than $M_{\rm 200b} = 10^{13.5}\Msh$. However, the differences are within the $5\%$ level in the mass range $10^{13.5} < M_{\rm 200b}/[\Msh] < 10^{14.5}$. The large deviations at the low-mass end are potentially due to the different halo finder used by \cite{DarkQuest:emu} to construct their halo catalogues. {\sc DarkQuest} was built and validated with {\sc Rockstar} catalogues \citep{Behroozi:2011ju}, while our MTNG haloes were identified by the FoF algorithm combined with {\sc Subfind}. Previously, \cite{Knebe:2011rx} have shown that differences  in the abundance of haloes  of up to $\sim 10\%$ can occur between  {\sc Rockstar} and {\sc Subfind}, which is consistent with our findings in the lower-right panel of Fig.~\ref{fig:hmf_emu}.

%---------------------------------------------------------------
\section{Large-scale matter and halo clustering}
\label{sec:matter_halo_Pk}
%---------------------------------------------------------------
One of the main goals of galaxy surveys is to measure the clustering of matter and galaxies on the largest scales, $r>150\Mpch$ ($k<0.1\hMpc$), to accurately determine the position of the baryonic acoustic oscillations (BAO) peaks with an unprecedented level of precision. In this Section, we present an analysis of the large-scale clustering of matter and dark matter haloes. Note that we are using the average of the MTNG740-DM-A and -B realisations, unless otherwise stated.

%---------------------------------------------------------------
\subsection{Matter and halo clustering at the BAO scale}
\label{sec:Pk_BAO}
%---------------------------------------------------------------
The volume of the MTNG740 simulations allows us to compute the dark matter power spectrum with high accuracy at the BAO scales down to the present time. Fig.~\ref{fig:Pkm_BAO} displays the measured matter power spectrum from our MTNG740-DM simulations divided by the smoothed (no-wiggle) linear theory prediction from $z=30$ (top left panel) to $z=0$ (bottom right panel). At very early times ($z=30$), the measured power spectrum matches perfectly with the linear theory, where we can identify up to six BAO peaks. We start to appreciate small deviations with respect to the linear predictions on small scales, $k>0.3\hMpc$, at $z=15$ of order of 1 percent. At intermediate redshifts, $z=7$ and $z=3$ (middle row of Fig.~\ref{fig:Pkm_BAO}), the non-linear effects are visible beyond the position of the third ($k\sim 0.2\Mpch$) and second ($k\sim 0.13\hMpc$) BAO peak, respectively, showing an enhancement of $5\%$ and $15\%$ in the clustering signal at the smallest scales. At low redshifts ($z\leq 1$; see the bottom row of Fig.~\ref{fig:Pkm_BAO}), we detect negative and positive shifts of the BAO peaks of order $2-3\%$ due to non-linear effects, extending to the largest scales. For comparison, we show the predictions from the {\sc Halofit} model \citep{Halofit} at $z=1$ and $z=0$ (see the blue solid lines in the lower panels of Fig.~\ref{fig:Pkm_BAO}). We find that {\sc Halofit} predicts the same shifts of the BAO peaks as found in the simulation measurements.

We now analyse the large-scale halo clustering in Fourier space. To do so, we select haloes according to their mass ($M_{200c}$) and velocity ($V_{\rm max}$); note that we only use central subhaloes (selected as the largest bound structure identified by {\sc Subfind} in every FoF halo) in our $V_{\rm max}$ samples. The maximum of the circular velocity, $V_{\rm max}$, can be used as a proxy to identify galaxies in dark matter subhaloes and make mock catalogues through the subhalo abundance matching technique \citep[SHAM; see e.g.,][]{Conroy2006,Gerke:2012ru,Klypin:2013rsa,Reddick:2012qy}. In each case, we select two halo samples with different space density, $n_{\rm h} = 1\times10^{-3}\hMpcc$ and $n_{\rm h} = 6.86\times10^{-4}\hMpcc$, at $z=1$. We select samples with these specific number densities at a redshift of unity because this makes them close to the expected spatial density of DESI OII emission line galaxies (ELGs) and Euclid H$\alpha$ emitters \citep{DESI:2016zmz,Euclid:2019clj}, respectively.

We measure the monopole of the redshift-space halo power spectrum because we want to draw a more direct analogy between the halo and galaxy clustering. Also, studying the clustering of $V_{\rm max}$-selected haloes is similar to studying the clustering of mock galaxies from a SHAM catalogue. We here use the distant-observer approximation to shift the positions of haloes from real- to redshift-space, we use the three coordinates, $\hat{x}$, $\hat{y}$ and $\hat{z}$, as the line-of-sight (LOS) to generate three redshift-space catalogues for one real-space catalogue where the new coordinates are,
\begin{equation}\label{eq:coord_z}
{\bf s} = {\bf r} + \frac{(1+z)v_\parallel}{H(z)}\hat{e}_\parallel\,,
\end{equation}
${\bf r}$ is the comoving coordinate vector in real space, ${\bf s}$ is the equivalent of this in redshift-space, and $z$ is the redshift. $H(z)$ is the Hubble parameter, $v_\parallel$ and $\hat{e}_\parallel$ are the components of the peculiar velocity and the unit vector along the LOS direction. So, in total we have six redshift space catalogues. Then, the monopole of the redshift-space power spectrum can be obtained by
\begin{equation}\label{eq:Pk0}
    P_0(k) = \int^1_{0}{P(k,\mu)}~{\rm d}\mu\,,
\end{equation}
where $P(k,\mu)$ is the full two-dimensional power spectrum, and $\mu$ is the cosine of the angle between the separation vector, $\mathbf{s}$ or $\mathbf{k}$, and the line-of-sight in configuration or Fourier space.

We use {\sc Nbodykit} to measure the halo power spectrum from the simulation for wavenumbers $k_{\rm box} < k/[\hMpc] < k_{\rm Ny}$, where $k_{\rm box}$ is the fundamental mode of the box given by Eq.~(\ref{eq:kbox}) and $k_{\rm Ny}$ is the Nyquist frequency computed by Eq.~(\ref{eq:k_ny}) with $N_{\rm grid} = 256$ (which is here the mesh resolution of the power spectrum measurement), using linear $k$-bins with width $\Delta k = 0.005 \hMpc$ and 30 linearly spaced bins between $0$ and $1$ for $\mu$.

The halo clustering measured from the MTNG740-DM simulations is shown in Fig.~\ref{fig:BAO_halo} (blue symbols with error bars), for each halo sample at $z=1$. We focus only on large scales, $k<0.5\hMpc$, where we can clearly see the BAO features in the power spectrum. Also, we can see that samples with a lower number of objects; i.e., $n_h = 6.86\times10^{-4}\hMpcc$, have a higher clustering amplitude than the larger samples ($n_h = 1\times10^{-3}\hMpcc$). This is because the low-density sample is influenced more strongly by massive objects which are more highly clustered. 

In cosmological analyses of the BAOs, it is common to extract the peak position through the dilation parameter, $\alpha$, which is related to the spherical average distance \citep{Eisenstein:2005su}. For $\alpha > 1$ the peak is moved to smaller scales, while $\alpha < 1$ shifts the peak to larger scales \citep{Angulo:2008,Anderson:2013zyy,Ross:2014qpa}. 

As a proof-of-concept example, we estimate the dilation parameter from our halo samples by fitting the monopole of the power spectrum from our simulations to a simple template used by \citet{Springel:2017tpz},
\begin{equation}
    P^{\rm fit}_0(k) = B^2_p(A_1+A_2k+A_3k^2)P_{\rm lin}(k/\alpha)\,,
\end{equation}
where $P_{\rm lin}$ is the linear theory power spectrum, $B_p$ is a bias parameter, and $A_1$, $A_2$ and $A_3$ are further free parameters. We fit our measurements on scales $k<0.5\hMpc$.
The best fit model for each sample is shown together with the measurements in Fig.~\ref{fig:BAO_halo}. In all cases, we can constrain the dilation parameter with a precision of $\sigma_\alpha/\alpha \sim 0.08-1.4$ percent.

%---------------------------------------------------------------
\subsection{Linear halo bias}
\label{sec:halo_bias}
%---------------------------------------------------------------
Haloes are biased tracers of the dark matter density field, hence the relation between the distribution of haloes and matter can be described by the linear halo bias $b$, defined as
\begin{equation}
b \equiv \delta_{\rm h}/\delta\,,
\end{equation}
where $\delta_{\rm h}$ is the halo density contrast and $\delta$ is the density contrast of matter. In terms of a Fourier-space analysis of the clustering, the halo bias can be estimated by,
\begin{equation}\label{eq:b_halo}
b(k) = \sqrt{\frac{P_{\rm h}(k)}{P_{\rm m}(k)}}\,,
\end{equation}
where $P_{\rm h}(k)$ and $P_{\rm m}(k)$ are, respectively, the halo and matter power spectrum in real space.
At sufficiently large (linear) scales, the halo bias is expected to be a scale independent constant, and its amplitude should depend mostly on halo properties, such as the halo mass. This linear halo bias is a basic ingredient of the halo model \citep[see e.g.,][]{Cooray:2002dia}. Since haloes host different galaxy types, it is important to accurately describe the halo bias in order to use it to model galaxy clustering and the galaxy-halo connection \citep[for a review, see][]{Wechsler:2018pic}.

We measure the halo bias for fixed $M_{200c}$ and $V_{\rm max}$ halo samples at $z=0$ and $z=1$, as shown in Fig.~\ref{fig:halo_bias}. We work with haloes with masses above $M_{200c} = 10^{11}\Msh$ (M1), $10^{11.5}\Msh$ (M2), $10^{12}\Msh$ (M3), $10^{12.5}\Msh$ (M4) and $10^{13}\Msh$ (M5), as well as with samples with velocities higher than $V_{\rm max} = 10^{1.8}\,\kms$ (V1), $10^{2}\,\kms$ (V2), $10^{2.2}\,\kms$ (V3), $10^{2.4}\,\kms$ (V4) and $10^{2.6}\,\kms$ (V5). Similarly to the redshift-space power spectrum measurements discussed in Sec.~\ref{sec:Pk_BAO}, we measure the real-space clustering in Fourier space for all halo samples in the range $k_{\rm box} < k/[\hMpc] < 1$ with $k$-bins with width $\Delta k = 0.005 \hMpc$. We additionally show in Fig.~\ref{fig:halo_bias} the individual measurements from the A (dashed lines) and B (dot-dashed lines) realisations, besides the mean of the paired runs (solid lines). In general, more massive haloes and haloes with higher $V_{\rm max}$ velocities display a larger halo bias. 

From the upper panels of Fig.~\ref{fig:halo_bias} we can see that the bias of $M_{200c}$-selected haloes shows an exclusion effect on small scales for the most massive samples (M4 and M5), where the bias drops at $k\sim 0.5\hMpc$ at both output times, $z=0$ and $z=1$. Also, the bias at early times, e.g., $z=1$ (upper right panel), has a higher amplitude for fixed halo mass samples than their present time counterparts (upper left panel). In all cases, we find that the halo bias is constant on very large scales, $k<0.1\hMpc$. However, on mildly non-linear scales, scale-dependent effects set in and become appreciable on scales $k>0.3\hMpc$. 

On the other hand, velocity selected halo samples (bottom panels of Fig.~\ref{fig:halo_bias}) show qualitatively similar results on large-scales to $M_{200c}$-selected haloes.  
Our simulations make very accurate quantitative predictions for these non-linear effects in dark matter-only models, but we caution that they can be affected by baryonic physics, which constitutes the primary systematic uncertainty.

Note that the A- and B- runs produce similar results, but the average of the two realisations tends to cancel out fluctuations on the largest scales, producing a much smoother measurement of the linear halo bias.

%---------------------------------------------------------------
\section{Baryonic effects on basic matter and halo statistics}
\label{sec:baryonic_impact}
%---------------------------------------------------------------

So far we have focused our attention on the results from the MTNG DM-only runs, ignoring baryons, as well as neutrinos. In this section, we briefly discuss the impact of baryons on basic matter and halo statistics. Previously, \cite{Springel:2017tpz} presented a high-level study of the matter and halo clustering results of the full-physics IllustrisTNG simulations, giving some hints of the effects of baryonic feedback of the TNG model on basic matter and halo statistics. In a companion paper, \citet{Pakmor2022} explores the consistency of the MTNG740 full-hydro run with the smaller box TNG simulations. Here we extend this analysis and consider the baryonic impact on basic halo and matter clustering statistics, while a more general study of the baryonic effects also on other observables, such as weak lensing and general galaxy statistics, will be presented in forthcoming work.

%---------------------------------------------------------------
\subsection{Clustering of matter components}
\label{sec:Pk_hydro}
%---------------------------------------------------------------
In Figure~\ref{fig:Pk_matter} we show the clustering power spectra of different matter components in our large MTNG740 hydro run at redshifts $z=0$ and $z=1$; i.e.~separately for dark matter, gas, stellar mass, and black hole mass.
We also compare to equivalent measurements for TNG100 and TNG300 from the IllustrisTNG project \citep{Springel:2017tpz}. We find in general good consistency between the MTNG740 and the TNG100/300 simulations, with some small differences that can be understood as arising from differences in mass resolution, and, in particular, in simulated volume. Reassuringly, there is excellent agreement on large scales, with MTNG740 extending the baryonic results to much larger spatial scales, throughout the BAO region, than possible for IllustrisTNG.

As shown in Fig.~\ref{fig:Pkm_comp} (and in our companion paper by \citet[][their Fig.~4]{Pakmor2022}), the suppression of the total matter clustering predicted by MTNG740 is also in nearly perfect agreement with that found in TNG100, confirming that there is a reduction of power of around $20\%$ at $k\sim10\hMpc$ due to AGN feedback, and a strong enhancement of the clustering on small scales $(k>50\hMpc)$ compared to dark matter-only models as result of the dissipative formation of galaxies. We note that the MTNG740 suppression is almost identical to the TNG300 prediction at $k<5\hMpc$.
On the smallest scales, $k > 10 \hMpc$, the baryonic effects on the total matter power spectrum of MTNG740 agree almost perfectly with the predictions for TNG100 even though TNG300 is closer in mass resolution. We attribute this to the lack of magnetic fields in MTNG740 compared to the TNG simulations, which slightly changes the effectiveness of the AGN feedback and how it couples to star formation  \citep[see the discussion in][]{Pakmor2022}.

This suppression needs to be taken into account, in particular, in weak gravitational lensing analysis, unless one restricts them to comparatively large scales and thus gives up on considerable cosmological information. It is thus important to develop tools to accurately model the suppression. In Figure~\ref{fig:Pkm_bacco}, we consider how well this suppression can be described by the {\sc Bacco} emulator \citep{Bacco:emu_b} at $z=0$ and $z=1$. The emulator matches our numerical results with impressive accuracy, albeit limited to the large-scale regime, $k < 5\hMpc$, for which the emulator has been calibrated. Note that we use the {\sc Bacco} parameters that match with the TNG300 suppression obtained by \citet{Bacco:emu_b}, which  seems adequate as MTNG740 and TNG300 give almost identical results \citep{Pakmor2022} in the $k$-range of the {\sc Bacco} emulator. The level of agreement is nevertheless encouraging and suggests that parametrisations like {\sc Bacco}  can be of great help to marginalise over complicated astrophysical effects such as AGN feedback.

%---------------------------------------------------------------
\subsection{Halo abundance and clustering}
\label{sec:halo_abundance}
%---------------------------------------------------------------
Previously, it has already been shown that the presence and evolution of gas and stars in hydrodynamical simulations affects the mass of haloes \citep{Vogelsberger2014, Springel:2017tpz}. This is because gas can be expelled from the inner regions of haloes thanks to the baryonic feedback processes, directly modifying the mass of a halo and affecting its ability to grow through the attraction of further matter. This mass change is expected to not only affect the abundance of dark matter haloes as a function of mass, but also their clustering.

In the left panel of Fig.~\ref{fig:dHMF_hydro}, we show the differential halo mass function (HMF) measured from the full-hydro (solid lines) and DM-only (dashed lines) runs of the MTNG740 (red lines), TNG100 (green lines) and TNG300 (orange lines) simulations at $z=0$, for comparison. In the bottom panels we show the residuals in both cases. Compared to dark matter only, we can see that the MTNG740 hydro run predicts 20$\%$ fewer low-mass haloes $(M_{200c} < 10^{11}\Msh)$, an additional 10$\%$ deficit of haloes with masses $M_{200c} \sim 10^{13}\Msh$, and also an enhancement of the most massive haloes $(M_{200c} > 10^{14}\Msh)$ of $\sim 5\%$. These results reaffirm the findings of \cite{Springel:2017tpz} for the TNG physics models. 

Full hydrodynamical simulations also offer the possibility to explore  predictions for the number of dark matter substructures. In the right panel of Fig.~\ref{fig:dHMF_hydro}, we show the differential subhalo $V_{\rm max}$ function at $z=0$. We use only the most massive (central) subhaloes to measure such velocity functions from the full-physics and DM-only MTNG740 runs, as well as for TNG100 and TNG300.

We find that there is a lack of \emph{low-velocity} subhaloes, $V_{\rm max} \sim 125\,\kms$, of around 20$\%$ in the hydro simulation; this is consistent with what we showed in the right panel of Fig.~\ref{fig:dHMF_hydro} for low-mass haloes. Also, we can see that there is a peak in the MTNG740-hydro velocity functions at $V_{\rm max} \sim 310\,\kms$ producing $30\%$ more subhaloes with respect to DM-only simulations. This peak divides the subhalo population into two samples. On one hand, we have low-velocity subhaloes that are affected strongly by supernovae feedback and which are related to low-mass haloes, and on the other hand, there are high-velocity subhaloes that occupy cluster-like haloes, and which are more strongly influenced by AGN feedback.

Of particular relevance observationally is how the structural changes in the properties of haloes are reflected in their clustering properties. Note that changes of halo properties invariably introduce differences in sample selection, which can in turn influence the clustering signal. To obtain a first quantitative assessment of this we consider in Figure~\ref{fig:xi_halo} halo samples of different space densities ($n_h = 3\times10^{-3}\hMpcc$, $1\times10^{-3}\hMpcc$ and $3\times10^{-4}\hMpcc$), either selected as a function of halo mass, or as a function of the halo maximum circular velocity. Note that the latter can be strongly affected by effects such as adiabatic compression due to baryonic physics, even if the halo virial mass itself does not change (as we showed in the left panel of Fig.~\ref{fig:dHMF_hydro}). We give measurements at $z=0$ both for real-space (upper panels) and redshift-space (bottom panels).

Interestingly, we find substantial differences in the clustering amplitudes on all scales.  While they are comparatively small at the level of 2 percent for mass selected samples in both real- and redshift-space, they become larger ($>5\%$) and sample-density dependent on small-scales. Our results are consistent with previous findings by \cite{Beltz-Mohrmann:2021try} for the TNG300 simulation. For circular velocity selected samples, the differences are substantial, amounting to $\sim 5\%$ at large scales. For distances of order $1\Mpch$ and smaller, strong halo exclusion effects that depend on the sample density become highly noticeable in real space whereas they are partially alleviated in redshift space. Overall, these results stress once more the strong impact of the sample selection on any clustering analysis \citep[see also][]{Angulo2012}. 

%---------------------------------------------------------------
\section{Summary and conclusions}
\label{sec:conc}
%---------------------------------------------------------------
In this paper, we have introduced the {\sc MillenniumTNG} simulation project. Our flagship runs consist of two fixed-and-paired DM-only simulations with $4320^3$ dark matter particles in a volume of $500^3\Mpchc$ \citep[which is the same volume as in the iconic {\sc Millennium} run;][]{Springel:2005nw}, a single full-physics hydrodynamical simulation which evolves $2\times 4320^3$ dark matter and gaseous resolution elements in the same cosmological volume from $z=63$ to the present time, and an extremely large simulation with more than a trillion dark matter particles which also includes massive neutrinos $(\Sigma m_\nu = 0.1\,{\rm eV})$ in a volume of $(3\,{\rm Gpc})^3$. The full-hydro simulation employs the {\sc IllustrisTNG} model of galaxy formation \citep{Weinberger:2017MNRAS,Pillepich:2017jle}.  In addition, we have complemented the simulation set with a number of smaller runs to investigate resolution dependence, box-size dependence, as well as the impact of different neutrino masses.

In this paper, we have introduced the primary technical aspects of the runs (see Table~\ref{tab:sims}) and describe their data products. However, our analysis only focuses on the matter clustering and halo statistics of the DM-only and full-physics MTNG740 runs. An in-depth analysis of the MTNG-neutrino simulations will be presented in a forthcoming paper.

We have performed a series of convergence tests by comparing the measured non-linear matter power spectrum (see Fig.~\ref{fig:MTNG_Pk_res}), the halo mass functions for two halo mass definitions ($M_{\sc 200c}$ and $M_{\rm FoF}$; shown in Figs.~\ref{fig:MTNG_HMF_res} and \ref{fig:MTNG_Mfof_res}, respectively), and the real-space halo clustering for three halo mass selected samples (Fig.~\ref{fig:xih_res}) at $z=0$ and $z=1$, of our level-1 run with the lower resolution simulations. We find sub percent agreement between the $4320^3$ and $2160^3$ resolutions for all measurements over a large range of scales and halo masses. Additionally, we have assessed the convergence of our numerical calculations by comparing the non-linear matter clustering and halo abundances of the MTNG740-DM and MTNG185-DM boxes (see Fig.~\ref{fig:MTNG_box}). The two simulated volumes agree very well ($\lesssim 2\%$) and the largest differences are directly associated with effects due to the limited volume of the MTNG185-DM run.

We have shown that computing two realisations of a given model using fixed-and-paired $N$-body simulations allows us to make accurate cosmological predictions at large scales despite the still quite limited volume  (see Fig.~\ref{fig:diff_Pkm}), and that this technique yields results comparable to those predicted by the mean of one hundred independent realisations (Fig.~\ref{fig:R_test}). This confirms that the fixed-and-paired technique effectively boosts the statistical power that is achievable with a given simulation volume.  

We have compared the non-linear boost factor, Eq.~\eqref{eq:B_Pk}, and the halo mass functions from a set of recent cosmological emulators ({\sc CosmicEmu}, {\sc Bacco}, {\sc EuclidEmulator2}, and {\sc DarkQuest}) to measurements from our simulations, finding reasonably good agreement in all the cases over the range where the emulators were calibrated (Figs.~\ref{fig:Pkm_emu} and \ref{fig:hmf_emu}). This can be viewed both as a reassuring confirmation of the high-quality calibration reached by these emulators, as well as an independent test of our simulation results if one already trusts these emulators.

We have also given a first impression of the large-scale clustering of matter and haloes around the BAO scale in Fourier space (Figs.~\ref{fig:Pkm_BAO} and \ref{fig:BAO_halo}). We have shown that the MTNG740 simulations have enough statistical power and volume to extract the BAO features through the dilation parameter, $\alpha$. We also show the linear halo bias for $M_{200c}$ and $V_{\rm max}$-selected haloes in Fig.~\ref{fig:halo_bias}.

We have validated the clustering of different matter components (dark matter, gas, stars and black holes) of the full-physics MTNG740 simulation by comparing with results from the TNG100 and TNG300 simulations (see Figs.~\ref{fig:Pk_matter} and \ref{fig:Pkm_comp}), finding good agreement between MTNG and its TNG predecessors. We also compared the  suppression of the total matter power spectrum due to baryonic physics between MTNG and the {\sc Bacco} emulator (see Fig.~\ref{fig:Pkm_bacco}), finding sub percent agreement on scales $k<5\hMpc$. The halo and substructure abundances (Fig.~\ref{fig:dHMF_hydro}) are also in good agreement between MTNG740 and TNG100/300.

Finally, we have explored baryonic effects on halo clustering in real- and redshift-space at $z=0$ in Fig.~\ref{fig:xi_halo}. We find that baryons affect the clustering of $M_{200c}$-selected haloes between $\sim 2\%$ in real-space up to $>20\%$ in redshift-space. The clustering of $V_{\rm max}$-selected haloes is modified even more, here the inclusion of baryons causes differences of $\sim 50\%$ in both real- and redshift-space.

Numerical calculations of the size of the MTNG simulations are needed to assist modern galaxy surveys in achieving their goals to constrain cosmological parameters at the sub percent level. With the MTNG project, we aim to produce realistic mock galaxy catalogues using the physically motivated TNG model in combination with semi-analytical models of galaxy formation. However, before presenting galaxy mock catalogues constructed in this manner, we need to carefully assess the reliability of the model foundations; i.e.~whether the MTNG simulations pass tests of numerical accuracy.  In this paper, we have examined this question and arrived at an affirmative assessment, making us confident that the MTNG simulations are a powerful tool to study a number of cosmological questions, some of them we address both in our companion papers and in our forthcoming work.

%---------------------------------------------------------------
\section*{Acknowledgements}
%---------------------------------------------------------------
We thank the anonymous referee for a constructive report that helped to improve the paper.
We wish to thank Giovanni Aric\`o and Raul Angulo for providing the TNG300 baryonic parameters of the {\sc Bacco} emulator.
CH-A acknowledges support from the Excellence Cluster ORIGINS which is funded by the Deutsche Forschungsgemeinschaft (DFG, German Research Foundation) under Germany's Excellence Strategy -- EXC-2094 -- 390783311. 
VS and LH acknowledge support by the Simons Collaboration on ``Learning the Universe''. 
LH is supported by NSF grant AST-1815978.  
SB is supported by the UK Research and Innovation (UKRI) Future Leaders Fellowship [grant number MR/V023381/1]. 
The authors gratefully acknowledge the Gauss Centre for Supercomputing (GCS) for providing computing time on the GCS Supercomputer SuperMUC-NG at the Leibniz Supercomputing Centre (LRZ) in Garching, Germany, under project pn34mo. 
This work used the DiRAC@Durham facility managed by the Institute for Computational Cosmology on behalf of the STFC DiRAC HPC Facility, with equipment funded by BEIS capital funding via STFC capital grants ST/K00042X/1, ST/P002293/1, ST/R002371/1 and ST/S002502/1, Durham University and STFC operations grant ST/R000832/1. 

%---------------------------------------------------------------
\section*{Data Availability}
%---------------------------------------------------------------
The {\sc MillenniumTNG} simulations will be made fully publicly available at \url{https://www.mtng-project.org} in 2024. The data underlying this article will be shared upon reasonable request to the corresponding authors.

\bibliographystyle{mnras}
\bibliography{ref} 

%%%%%%%%%%%%%%%%% APPENDICES %%%%%%%%%%%%%%%%%%%%%

%\appendix

% Don't change these lines
\bsp	% typesetting comment
\label{lastpage}
\end{document}